\documentclass[a4paper,11pt]{article}
\usepackage{jheppub} % for details on the use of the package, please
\usepackage{comment}
\usepackage[T1]{fontenc}
\usepackage{amssymb}
\usepackage{amsmath}
\usepackage{color}
\usepackage{xspace}
\usepackage{appendix}
\usepackage{float}
\usepackage{subcaption}
\usepackage{graphicx}
\usepackage{hyperref}
\usepackage{chngcntr}
\usepackage{diagbox}
\usepackage{wasysym}
\usepackage{multirow}
\usepackage{placeins}
\usepackage{booktabs}        
\usepackage{multirow}        
\usepackage{array}           
\usepackage{xcolor}          
\usepackage{amsmath}   
\usepackage{comment}
\usepackage{slashbox}
\usepackage[utf8x]{inputenc} 
\newcounter{JW}

\newcommand{\bqa}{\begin{eqnarray}}
\newcommand{\eqa}{\end{eqnarray}}

\def\as{\alpha_s}

\def\mglong{{\tt MadGraph5\_aMC@NLO}}
\def\mgshort{{\tt MG5\_aMC}}
\def\geneva{{\tt GENEVA}}
\def\nnlojet{{\tt NNLOJET}}
\def\ihixs{{\tt iHixs2}}

\def\ggxy{{\tt ggxy}}

\chardef\MyArticleWithColor=\pdfcolorstackinit page direct{0 g}

\preprint{CPTNP-2026-004}
\title{Fully differential Higgs boson pair production at N$^3$LO with top quark mass effects}
\author[a]{Xuan Chen,}
\author[a]{Yuesheng Dai,}
\author[a]{Hai Tao Li,}
\author[a]{Shi-Yuan Li,}
\author[b]{Hua-Sheng Shao,}
\author[a,c]{and Jian Wang}

\affiliation[a]{School of Physics, Shandong University, Jinan, Shandong 250100, China}

\affiliation[b]{Laboratoire de Physique Th\'eorique et Hautes Energies (LPTHE), UMR 7589, Sorbonne Universit\'e et CNRS, 4 place Jussieu, 75252 Paris Cedex 05, France}

\affiliation[c]{Center for High Energy Physics, Peking University, Beijing 100871, China}

\emailAdd{xuan.chen@sdu.edu.cn}
\emailAdd{yueshengdai@mail.sdu.edu.cn}
\emailAdd{haitao.li@sdu.edu.cn}
\emailAdd{lishy@sdu.edu.cn}
\emailAdd{huasheng.shao@lpthe.jussieu.fr}
\emailAdd{j.wang@sdu.edu.cn}

\abstract{
    Higgs-boson pair production is of fundamental importance for probing the Higgs potential. At hadron colliders, the dominant production channel proceeds via gluon-gluon fusion (ggF) mediated by a top-quark loop. We report the first fully differential predictions for Higgs-boson pair production through ggF at next-to-next-to-next-to-leading order (N$^3$LO) in the strong coupling $\alpha_s$ in the heavy-top-quark limit (HTL). Fiducial cross section and selected differential distributions are presented at a center-of-mass energy of $\sqrt{s}$ = 14 TeV, under realistic experimental selection cuts. The N$^3$LO QCD corrections reduce the scale uncertainties of the next-to-next-to-leading order fiducial and differential predictions by approximately a factor of three, bringing the theoretical uncertainty to the percent level in the HTL. After incorporating top-quark-mass effects at next-to-leading order in $\alpha_s$, we provide one of the most precise parton-level differential predictions to date for ongoing experimental searches for Higgs-boson pair production at the LHC.
}

\begin{document}
\maketitle
\flushbottom

%%%%%%%%%%%%%%%%%%%%%%%%%%
\section{Introduction\label{sec:introduction}}
One of the most peculiar and least understood sectors of the Standard Model (SM) of elementary particles is the Higgs potential $V(H)$, which governs electroweak symmetry breaking, the generation of particle masses, the nature of the electroweak phase transition in the early Universe, and the stability of the vacuum. Here, $H$ denotes an SU(2)$_L$ Higgs doublet. The specific form of the Higgs potential realized in the SM has motivated numerous attempts to embed it into more fundamental ultraviolet theories. What is currently known about the Higgs potential can be summarized by two facts:
\begin{itemize}
\item The potential has a local minimum at
\begin{equation}
H = H_0 = \left(0,\frac{v}{\sqrt{2}}\right)^T\,,
\end{equation}
where the vacuum expectation value $v$ is determined by the Fermi constant $G_F$ via
\begin{equation}
v = (\sqrt{2} G_F)^{-1/2}\approx 246.22~\mathrm{GeV}\,.
\end{equation}
The Fermi constant $G_F$ is currently determined from the muon lifetime with extremely high precision at the parts-per-million (ppm) level~\cite{ParticleDataGroup:2024cfk}.
\item The second derivative of the potential with respect to the Higgs field $h$, defined through
\begin{equation}
H = \left(0,\frac{v+h}{\sqrt{2}}\right)^T\,,
\end{equation}
evaluated at the local minimum, corresponds to the squared mass of the Higgs boson $m_h^2$, i.e.,
\begin{equation}
m_h^2=\left.\frac{\partial^2}{\partial h^2}V(H)\right|_{h=0}\,.
\end{equation}
The current experimental determination of $m_h$ has a relative precision of less than one permille~\cite{ParticleDataGroup:2024cfk}.
\end{itemize}
The trilinear and quartic Higgs self-interaction couplings, $\lambda_{3h}$ and $\lambda_{4h}$, are defined as
\begin{equation}
\lambda_{3h}=\frac{1}{3!\,v}\left.\frac{\partial^3}{\partial h^3}V(H)\right|_{h=0}\,,\quad \lambda_{4h}=\frac{1}{3!}\left.\frac{\partial^4}{\partial h^4}V(H)\right|_{h=0}\,.
\end{equation}
In the SM, they are not free parameters but  fully determined by other SM parameters. In particular, at tree level, both couplings are equal
\begin{equation}
\lambda_{3h}^{\mathrm{SM}}=\lambda_{4h}^{\mathrm{SM}} = \lambda^{\mathrm{SM}} \equiv \frac{m_h^2}{2 v^2}\,.\label{eq:HiggsselfcoupinSM}
\end{equation}
These two Higgs self-couplings can be directly probed through double- and triple-Higgs boson production processes, respectively.

%Beyond tree level, the trilinear and quartic Higgs self-couplings receive radiative corrections from fermionic, scalar, and gauge-boson loops. In the SM, the dominant next-to-leading order (NLO) contributions arise from top-quark loops and induce corrections of order $\mathcal{O}(10\%)$ for the trilinear coupling and up to $\mathcal{O}(20\%)$ for the quartic coupling, depending on the renormalization scheme and kinematic definition. 

The quartic Higgs self-coupling $\lambda_{4h}$ is very weakly constrained to date, since experimental searches for triple-Higgs production at the LHC have only recently been initiated. The first measurements were performed by the ATLAS~\cite{ATLAS:2024xcs} and CMS~\cite{CMS:2025jkb,CMS:2025gos} collaborations using Run~2 data in the $6b$ and $4b2\gamma$ final states. The current allowed ranges of $\kappa_{4h}\equiv\lambda_{4h}/\lambda_{4h}^{\mathrm{SM}}$ at 95\% confidence level (CL) are $-230<\kappa_{4h}<240$ (ATLAS)~\cite{ATLAS:2024xcs} and $-190<\kappa_{4h}<190$ (CMS)~\cite{CMS:2025gos}, assuming all other couplings take their SM values.

In contrast, searches for di-Higgs production started soon after the discovery of the Higgs boson~\cite{ATLAS:2012yve,CMS:2012qbp} using LHC Run~1 data~\cite{ATLAS:2015sxd,CMS:2017yfv,CMS:2016cma}. Analyses of Run~2 data by the ATLAS and CMS collaborations have covered a wide range of decay channels. For example, ATLAS has studied final states such as $b\bar{b}\tau^{+}\tau^{-}$, $b\bar{b}\gamma\gamma$, $b\bar{b}b\bar{b}$, $b\bar{b}WW^{*}$, $WW^{*}\gamma\gamma$, and $WW^{*}WW^{*}$~\cite{ATLAS:2018fpd,ATLAS:2018hqk,ATLAS:2018ili,ATLAS:2021ifb,ATLAS:2022xzm,ATLAS:2023qzf,ATLAS:2024pov}, while CMS has considered $b\bar{b}\tau^{+}\tau^{-}$, $b\bar{b}\gamma\gamma$, $b\bar{b}b\bar{b}$, $b\bar{b}ZZ$, and multi-lepton final states~\cite{CMS:2020tkr,CMS:2022hgz,CMS:2022cpr,CMS:2022gjd,CMS:2022kdx,CMS:2022omp}. The first di-Higgs analyses using Run~3 data have also been performed in refs.~\cite{ATLAS:2025hhd,CMS:2025ero}. The current best constraints on the trilinear Higgs self-coupling $\kappa_{3h}\equiv\lambda_{3h}/\lambda^{\mathrm{SM}}_{3h}$ at 95\% CL are $-0.4<\kappa_{3h}<6.3$ from ATLAS~\cite{ATLAS:2022jtk} and $-1.24<\kappa_{3h}<6.49$ from CMS~\cite{CMS:2022dwd}, assuming SM values for all other couplings. With the full HL-LHC dataset, the projected 95\% CL sensitivity on the self-coupling modifier is expected to reach $0.5<\kappa_{3h}<1.7$~\cite{ATLAS:2025eii}, and the di-Higgs production process can be observed with a significance of more than $5\sigma$.

\begin{figure}[hbt!]
\centering
 \includegraphics[width=0.3\textwidth]{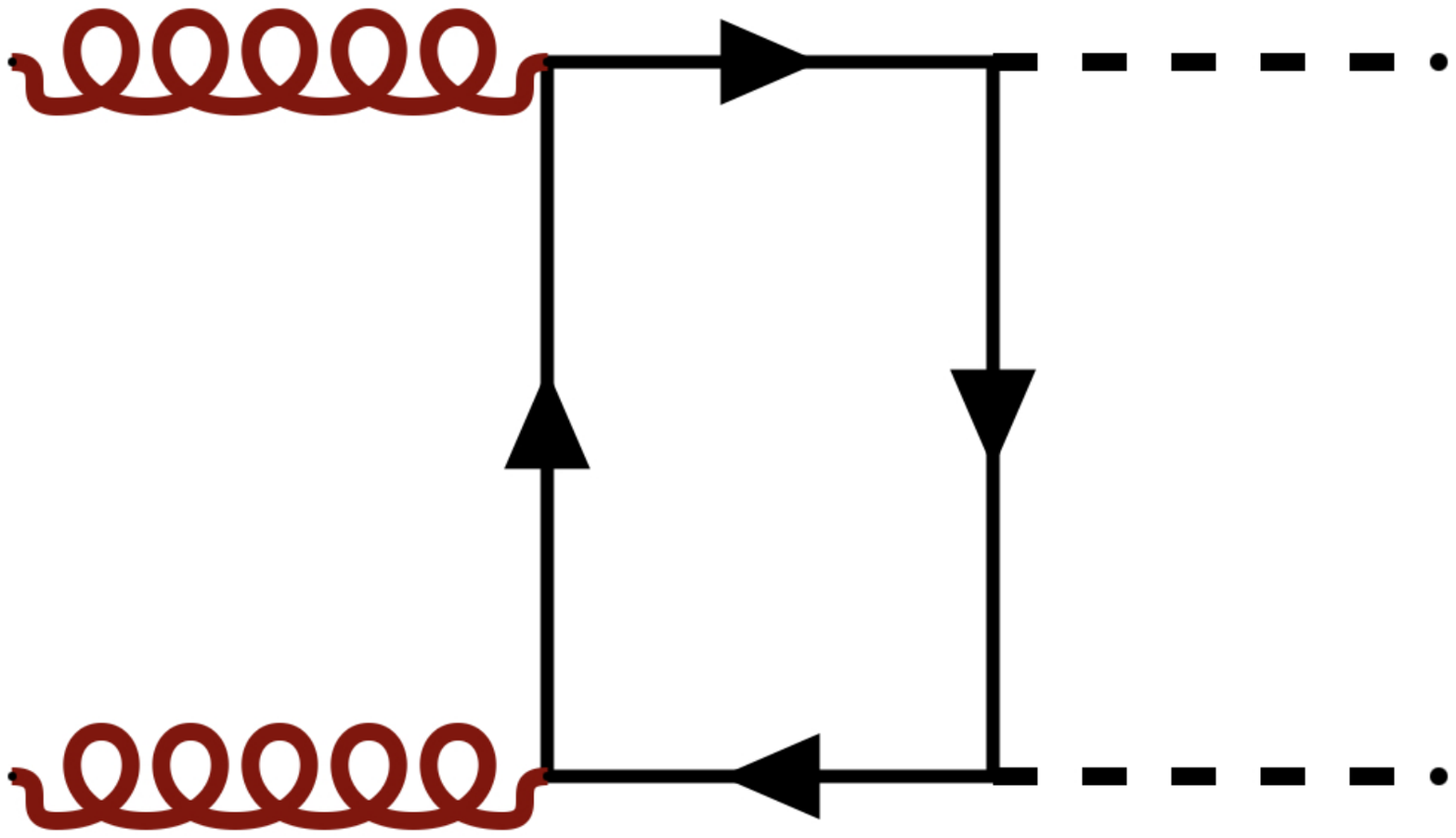}
 \hspace{0.5cm}
 \includegraphics[width=0.3\textwidth]{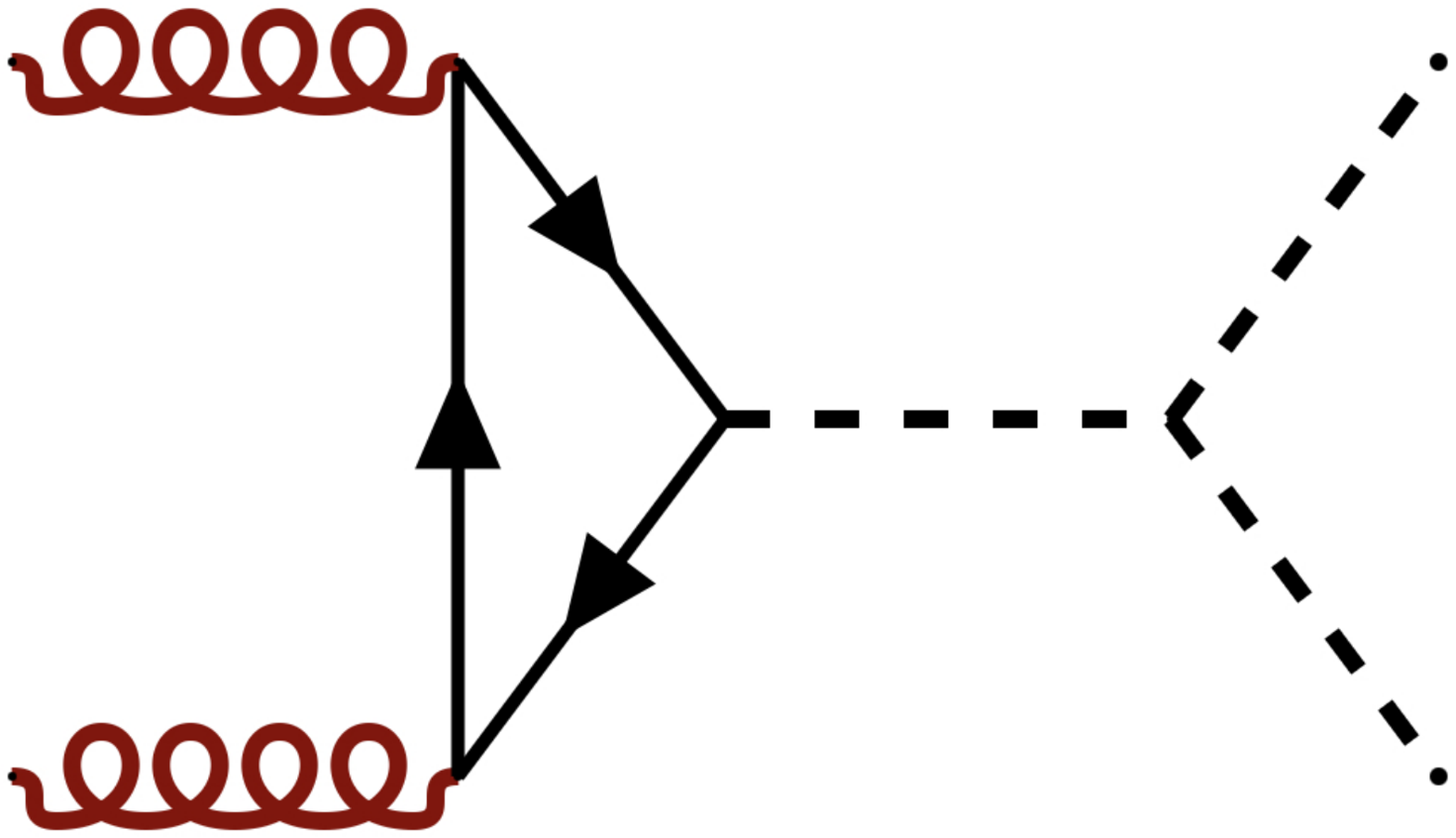}
  \caption{ LO representative Feynman diagrams for the Higgs boson pair production in the SM.}
  \label{fig:LOFeyDia-SM}
\end{figure}

In this paper, we focus on improving the theoretical predictions for Higgs boson pair production via gluon-gluon fusion (ggF), $gg\to hh$, which accounts for more than 90\% of the total di-Higgs yield at the LHC. At leading order (LO) in the SM, the process proceeds via a top-quark loop, as shown in figure~\ref{fig:LOFeyDia-SM}. 
Due to the presence of massive propagators, perturbative calculations for this process are particularly challenging. 
In fixed-order computations, only next-to-leading order (NLO) QCD corrections are currently available~\cite{Borowka:2016ypz,Borowka:2016ehy,Baglio:2018lrj,Davies:2019dfy,Baglio:2020wgt,Baglio:2020ini,Bagnaschi:2023rbx,Davies:2025qjr},
which increase the LO predictions~\cite{Glover:1987nx,Plehn:1996wb} by about $66\%$ and reduce the scale uncertainties to approximately $\pm 13\%$ at $\sqrt{s}=$ 14 TeV.
Moreover, the choice of top-quark mass renormalization scheme introduces an additional significant theoretical uncertainty, which can reach up to $30\%$ for certain differential distributions~\cite{Baglio:2018lrj,Baglio:2020ini}.
Recently, the full NLO electroweak (EW) corrections were computed, leading to a reduction of the total LO cross section by about $4\%$~\cite{Bi:2023bnq}.
Independent calculations including partial EW contributions have also been reported in refs.~\cite{Muhlleitner:2022ijf,Davies:2022ram,Davies:2023npk,Heinrich:2024dnz,Davies:2025wke,Bonetti:2025vfd,Bhattacharya:2025egw}. The NLO QCD predictions can be further improved through soft-gluon resummation~\cite{Ferrera:2016prr,deFlorian:2018tah} or by matching to parton showers~\cite{Heinrich:2017kxx,Jones:2017giv,Heinrich:2019bkc,Bagnaschi:2023rbx,Alioli:2025xcu}. Complete next-to-NLO (NNLO) QCD corrections with full top-quark mass dependence are still unavailable, despite ongoing dedicated efforts~\cite{Davies:2023obx,Davies:2024znp,Davies:2025ghl}. Within the narrow-width approximation, NLO QCD corrections to fiducial cross sections have been studied for $gg\to hh\to b\bar{b}\gamma\gamma$~\cite{Li:2024ujf} and $gg\to hh\to b\bar{b}\tau^+\tau^-$~\cite{Li:2025gbx}, where they are found to be sizable. These large corrections predominantly originate from the strong sensitivity to soft and collinear QCD radiation at fixed order, an effect that can be largely mitigated once parton-shower effects are included~\cite{Braun:2025pmf}.

\begin{figure}[hbt!]
\centering
 \includegraphics[width=0.275\textwidth]{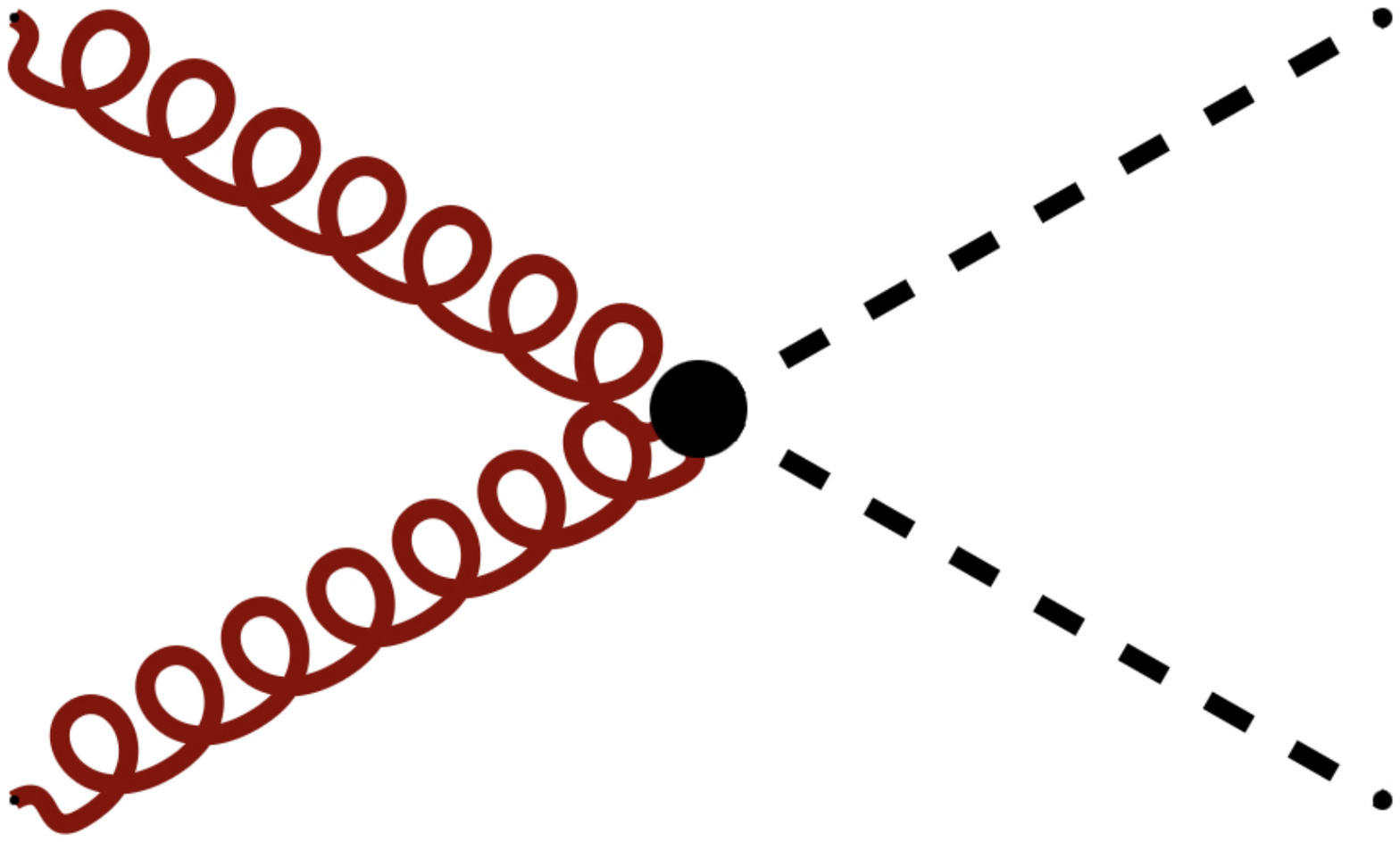}
 \hspace{0.65cm}
 \includegraphics[width=0.275\textwidth]{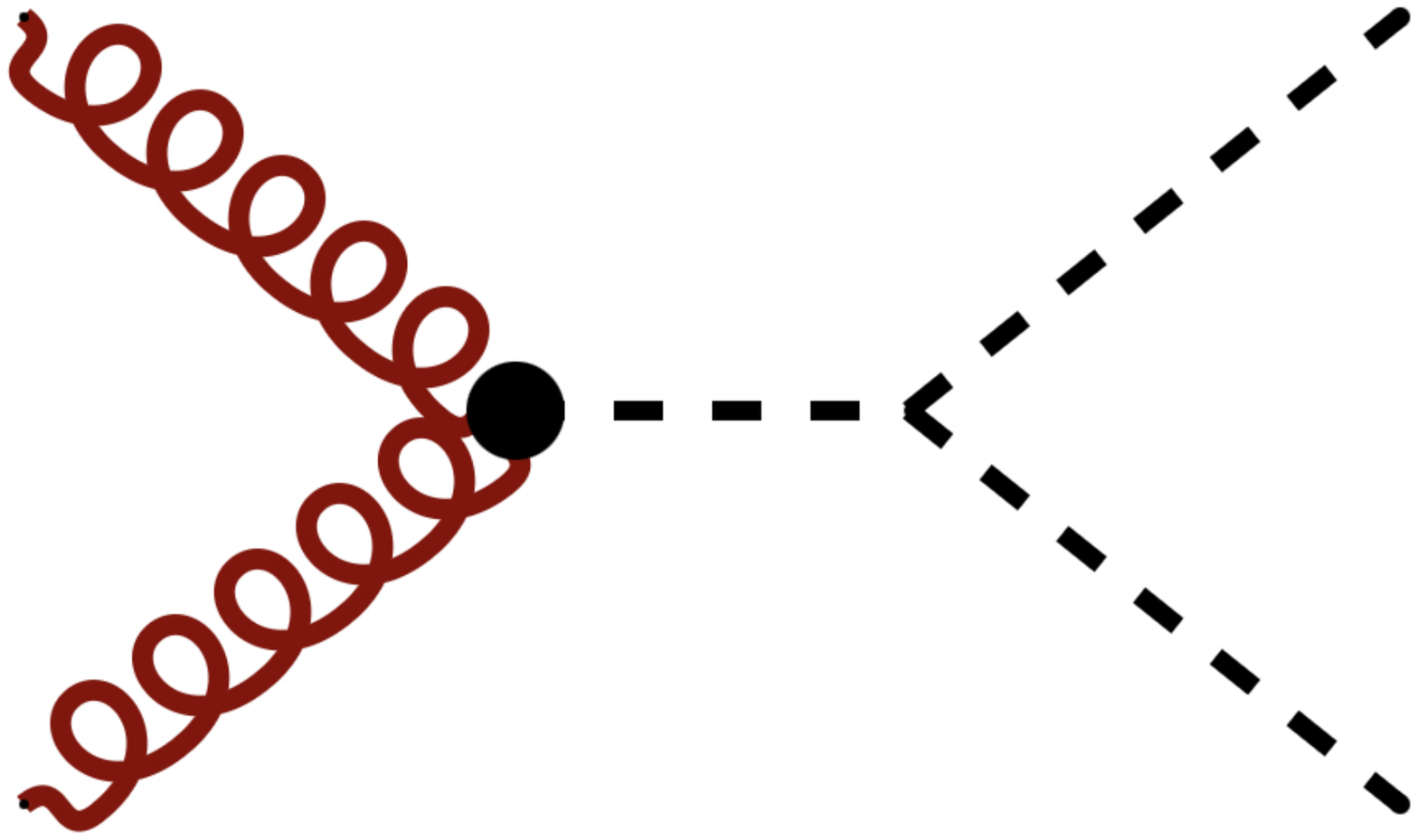}
  \caption{ LO representative Feynman diagrams for the Higgs pair production in the HTL.}
  \label{fig:LOFeyDia-HTL}
\end{figure}

An efficient way to improve the theoretical predictions for the cross sections is to employ the heavy top-quark limit (HTL), corresponding to $m_t\to\infty$. In this approximation, the top-quark loops are integrated out and reduce to local contact interactions, represented by the black dots in figure~\ref{fig:LOFeyDia-HTL}. This effective description is obtained by assuming \mbox{$m_t\gg m_h$} at the amplitude level.
This simplification renders higher-order perturbative QCD corrections considerably more tractable. In the HTL, NLO, NNLO, and next-to-NNLO (N$^3$LO) QCD corrections have been computed in ref.~\cite{Dawson:1998py}, refs.~\cite{deFlorian:2013uza,deFlorian:2013jea,Grigo:2014jma,deFlorian:2016uhr}, and refs.~\cite{Chen:2019lzz,Chen:2019fhs}, respectively. At N$^3$LO accuracy, however, only the inclusive total cross section and the di-Higgs invariant-mass distribution are known in a fully differential manner, while other kinematic distributions are available only approximately. The goal of this paper is to overcome this limitation at N$^3$LO QCD accuracy. Beyond fixed-order calculations, soft-gluon resummation has been investigated at next-to-next-to-next-to-leading logarithmic (N$^3$LL) accuracy in ref.~\cite{AH:2022elh} and at next-to-next-to-leading logarithmic (NNLL) accuracy in earlier works~\cite{Shao:2013bz,deFlorian:2015moa,DeFlorian:2018eng}. In addition, NNLO QCD predictions matched to parton showers in the HTL are available within the \geneva\ event generator~\cite{Alioli:2022dkj}. The HTL is generally expected to be valid when the partonic center-of-mass energy $\sqrt{\hat{s}}$ satisfies $\sqrt{\hat{s}}\ll 2m_{t}\simeq 345~\mathrm{GeV}$. However, the production of an on-shell Higgs boson pair requires $\sqrt{\hat{s}}\geq 2m_{h}\simeq 250~\mathrm{GeV}$. As a result, finite top-quark mass effects cannot be neglected in a reliable theoretical prediction. Indeed, numerous studies in the literature~\cite{Maltoni:2014eza,Frederix:2014hta,Grigo:2013rya,Grigo:2015dia,Degrassi:2016vss,Grazzini:2018bsd,Davies:2019xzc} have investigated how to incorporate finite top-quark mass effects in calculations based on the HTL.

In this work, we present the first fully differential QCD corrections to $gg\to hh$ at N$^3$LO in the HTL, combined with NLO QCD predictions including full top-quark mass dependence. The remainder of the paper is organized as follows. In section~\ref{sec:n3lo}, we introduce the theoretical and computational framework in the HTL and provide technical details of our validation. In section~\ref{sec:results}, we present our numerical results. We conclude in section~\ref{sec:conclusion}. Additional plots illustrating the dependence on the top-quark mass renormalization scheme at NLO QCD accuracy are collected in appendix~\ref{app:top-quark-mass-scheme}.

\section{Theoretical and computational framework\label{sec:n3lo}}

\begin{comment}
In this section, we briefly review the framework and results of refs.~\cite{Chen:2019fhs,Chen:2019lzz}. For further details, the reader is referred to those works. We also present the new method to compute the fully differential cross section in this section, as well as the numerical validations.
\end{comment}

As mentioned above, the fully differential N$^3$LO QCD calculations for $gg\to hh$ in this paper are performed in the HTL. The effective Lagrangian describing the coupling of the Higgs field $h$ to the gluon field-strength tensors is given by
\begin{align}\label{eq:effL}
 \mathcal{L}_{\rm eff}&=-\frac{1}{4}G_{\mu\nu}^aG^{a\,\mu\nu}\left(C_h\frac{h}{v}-C_{hh}\frac{h^2}{2v^2}\right)\,,
\end{align}
where $G^{a\,\mu\nu}$ denotes the gluon field-strength tensor. The Wilson coefficients $C_h$ and $C_{hh}$ are obtained by matching the full SM onto the low-energy effective theory~\cite{Chetyrkin:1997un}, in which the top-quark degree of freedom has been integrated out. These coefficients start at $\mathcal{O}(\alpha_s)$ and are analytically known up to $\mathcal{O}(\alpha_s^4)$~\cite{Kataev:1981gr, Inami:1982xt, Chetyrkin:1997iv, Chetyrkin:1997un, Schroder:2005hy, Chetyrkin:2005ia, Kniehl:2006bg,Baikov:2016tgj,Spira:2016zna,Gerlach:2018hen}. We refrain from presenting their explicit expressions here; they can be found in section 2.1 of ref.~\cite{Chen:2019fhs}, using the on-shell (OS) top-quark mass scheme.

\begin{figure}[hbt!]
\centering
\begin{subfigure}{0.3\textwidth}
  \centering
  \includegraphics[width=\textwidth, height=0.2\textheight, keepaspectratio]{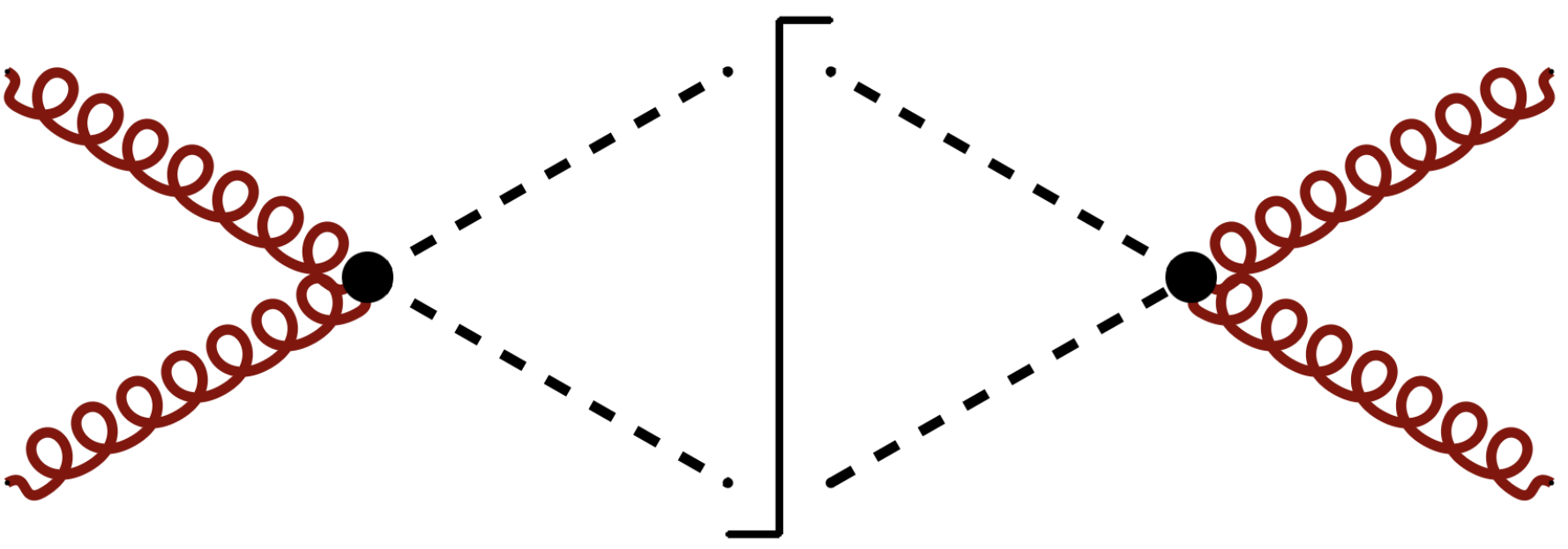}
  \caption{} 
  \label{subfig:class-a}
\end{subfigure}
\begin{subfigure}{0.3\textwidth}
  \centering
  \includegraphics[width=\textwidth, height=0.2\textheight, keepaspectratio]{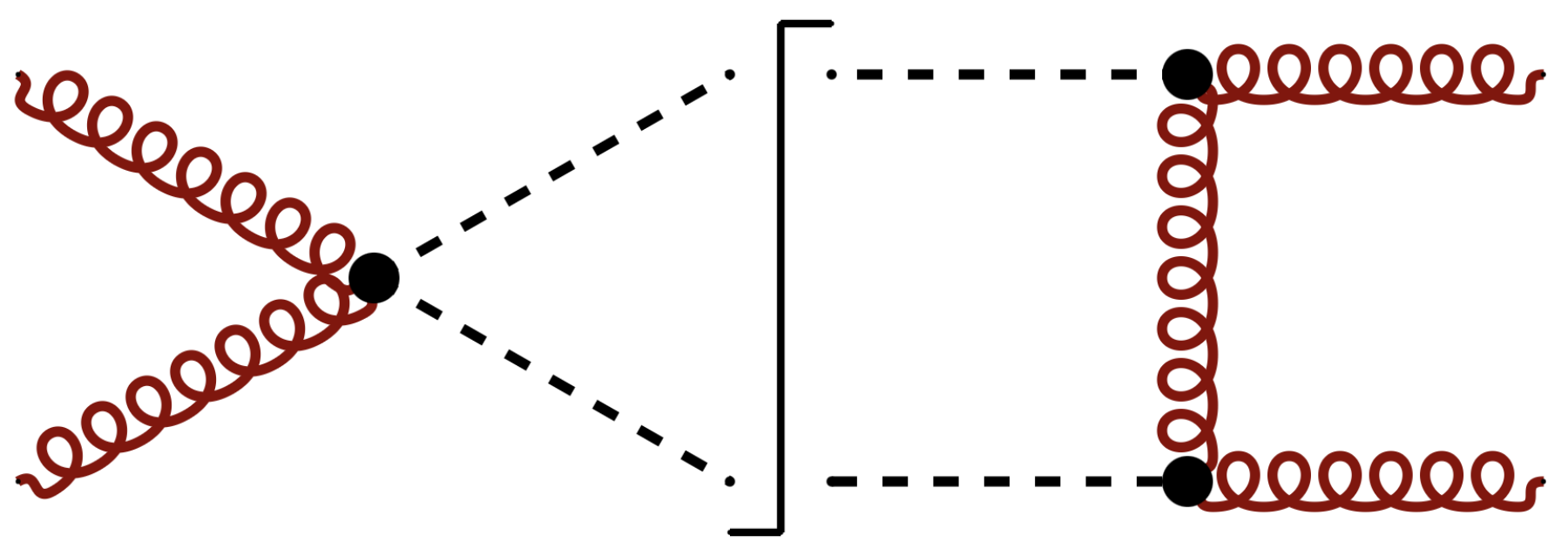}
  \caption{}
  \label{subfig:class-b}
\end{subfigure}
\begin{subfigure}{0.3\textwidth}
  \centering
  \includegraphics[width=\textwidth, height=0.2\textheight, keepaspectratio]{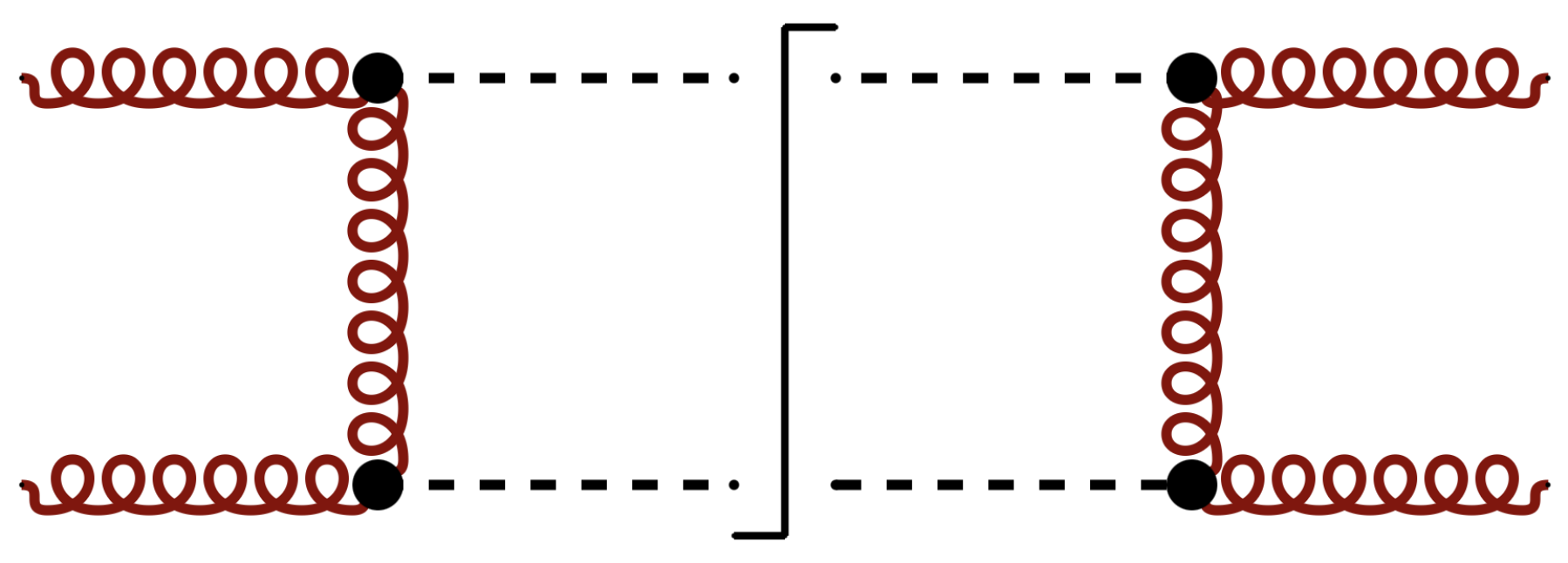}
  \caption{}
  \label{subfig:class-c}
\end{subfigure}
  \caption{Representative Born-level Cutkosky-cut diagrams for the three classes of Higgs-pair production in the HTL.}
  \label{fig:FeynDia-classes}
\end{figure}

As suggested in refs.~\cite{Chen:2019lzz,Chen:2019fhs}, the perturbative QCD calculations in the HTL can be conveniently organized into three classes according to the number of effective vertices at the squared-amplitude level. Specifically, we denote contributions with two, three, and four effective vertex insertions as class-$a$, $b$, and $c$, respectively. Representative Born-level cut diagrams for each class are shown in figure~\ref{fig:FeynDia-classes}. 
Accordingly, the (differential) di-Higgs cross section in the HTL can be decomposed as
\begin{align}
    d\sigma_{hh} = d\sigma^{a}_{hh} + d\sigma^{b}_{hh} + d\sigma^{c}_{hh}\,.
\end{align}
Since each effective vertex contributes at least at $\mathcal{O}(\alpha_s)$, the lowest perturbative orders of class-$a$, $b$, and $c$ are $\mathcal{O}(\alpha_s^2)$, $\mathcal{O}(\alpha_s^3)$, and $\mathcal{O}(\alpha_s^4)$, respectively. Their contributions at each perturbative order in $\alpha_s$ are summarized in table~\ref{tab:classes}.
\begin{table}[hbt!]
\centering
\begin{tabular}{lcccc}
\toprule
\multirow{2}{*}{Class} & \multicolumn{4}{c}{Perturbative Order} \\
\cmidrule{2-5}
 & LO & NLO & NNLO & N$^3$LO \\
\midrule
Sum & $\mathcal{O}(\alpha_s^2)$ & $\mathcal{O}(\alpha_s^3)$ & $\mathcal{O}(\alpha_s^4)$ & $\mathcal{O}(\alpha_s^5)$ \\
Class-$a$ & $\mathcal{O}(\alpha_s^2)$ & $\mathcal{O}(\alpha_s^3)$ & $\mathcal{O}(\alpha_s^4)$ & $\mathcal{O}(\alpha_s^5)$ \\
Class-$b$ & --- & $\mathcal{O}(\alpha_s^3)$ & $\mathcal{O}(\alpha_s^4)$ & $\mathcal{O}(\alpha_s^5)$ \\
Class-$c$ & --- & --- & $\mathcal{O}(\alpha_s^4)$ & $\mathcal{O}(\alpha_s^5)$ \\
\bottomrule
\end{tabular}
\caption{Contributions of the different classes at each perturbative order in $\alpha_s$.}
\label{tab:classes}
\end{table}
In summary, up to N$^k$LO ($k\in\mathbb{N}$), one needs to compute the N$^k$LO, N$^{k-1}$LO, and N$^{k-2}$LO corrections to $d\sigma^{a}_{hh}$, $d\sigma^{b}_{hh}$, and $d\sigma^{c}_{hh}$, respectively. In the following, we describe the different techniques employed to compute the differential distributions in each class up to N$^3$LO.

\subsection{Calculation of the class-$a$ contribution\label{sec:classa}}

The class-$a$ differential cross section can be directly related to the single-Higgs differential cross section $d\sigma_h$ via~\cite{deFlorian:2013jea,Chen:2019lzz,Chen:2019fhs} 
\begin{align}
    d \sigma_{h h}^a=\frac{dm_{hh}^2}{2\pi}d\Phi(p_{hh}\to p_{h,1} p_{h,2})\frac{1}{2v^2}\left|\frac{C_{h h}}{C_h}-\frac{6 \lambda_{3h} v^2}{m_{h h}^2-m_h^2}\right|^2 \left(\left.d\sigma_{h}\right|_{p_h\to p_{hh}}\right)\,,
    \label{eq:htohh}
\end{align}
where $p_{h,1}$ and $p_{h,2}$ denote the four-momenta of the two final-state Higgs bosons, and $p_{hh}$ is the four-momentum of the Higgs-boson pair, with the invariant mass $p_{hh}^2=m_{hh}^2$. The two-body phase-space measure $d\Phi(p_{hh}\to p_{h,1} p_{h,2})$ is defined as
\begin{equation}
d\Phi(p_{hh}\to p_{h,1} p_{h,2})=(2\pi)^4\delta^{(4)}\left(p_{hh}-p_{h,1}-p_{h,2}\right)\prod_{i=1}^{2}{\left[\frac{d^3\vec{p}_{h,i}}{(2\pi)^32E_{h,i}}\right]}\,,
\end{equation}
where $E_{h,i}$ and $\vec{p}_{h,i}$ are the energy and three-momentum components of $p_{h,i}$.
In this paper, we fix the trilinear Higgs coupling to its SM value, $\lambda_{3h}=\lambda_{3h}^{\mathrm{SM}}$, as given in eq.~\eqref{eq:HiggsselfcoupinSM}. The single-Higgs production cross section $d\sigma_h$ is known up to N$^3$LO in QCD~\cite{Anastasiou:2015ema,Anastasiou:2016cez,Mistlberger:2018etf,Dulat:2018bfe,Cieri:2018oms,Chen:2021isd,Billis:2021ecs}. The substitution $p_h\to p_{hh}$ in eq.~\eqref{eq:htohh} implies that the Higgs four-momentum in the single-Higgs cross section is replaced by the four-momentum of the Higgs-boson pair. The factor inside the absolute-value brackets corresponds to the ratio of the squared amplitude for Higgs-pair production to that for single-Higgs production, and it should also be expanded in $\alpha_s$ when providing the corresponding fixed-order predictions.

To account for the renormalization- and factorization-scale dependence of class-$a$ at N$^3$LO, we employ the renormalization-group evolution given in appendix C of ref.~\cite{Chen:2019fhs} (cf. eqs.(C.4) and (C.11) therein). Specifically, the scale dependence of the N$^3$LO differential cross section of class-$a$ reads
\begin{align}
\frac{\partial}{\partial \ln \mu_R} d\sigma_{h h}^{a,\mathrm{N}^3\mathrm{LO}}\left(\mu_R, \mu_F\right)
&=2\left(\frac{\alpha_{s}(\mu_R)}{4\pi}\right)^{3}\chi \frac{dm_{hh}^2}{2\pi}d\Phi(p_{hh}\to p_{h,1}p_{h,2})\frac{1}{2v^2}
\nonumber\\ & \times 
\left(\left.\sigma_h^{\mathrm{LO}}\left(\mu_R, \mu_F\right)\right|_{p_h \to p_{h h}}\right) 
\left(1-\frac{6 \lambda_{3h} v^2}{m_{h h}^2-m_h^2}\right)+\mathcal{O}\left(\alpha_s^6\right)\\
&=2\left(\frac{\alpha_{s}(\mu_R)}{4\pi}\right)^{3}\chi d\sigma_{hh}^{a,\mathrm{LO}}(\mu_R,\mu_F)\left(1-\frac{6 \lambda_{3h} v^2}{m_{h h}^2-m_h^2}\right)^{-1}+\mathcal{O}(\alpha_s^6)\,,\nonumber
\end{align}
where 
\begin{equation}
\chi\equiv\frac{16}{9}(32n_{f}^{2}-420n_{f}-1461)\,,
\end{equation}
and $n_f$ denotes the number of light-quark flavors.
This term induces an additional contribution compared to the single-Higgs case when reconstructing the scale dependence.

The N$^3$LO QCD corrections to the inclusive single-Higgs production cross section can be computed using the public code \ihixs~\cite{Dulat:2018rbf}, which has been employed in refs.~\cite{Chen:2019fhs,Chen:2019lzz}. However, according to eq.~\eqref{eq:htohh}, knowledge of the inclusive single-Higgs cross section $\sigma_h$ alone allows the computation only of the total cross section and the di-Higgs invariant-mass distribution, while other differential distributions at N$^3$LO accuracy were available only in approximate form in ref.~\cite{Chen:2019fhs}. 
Subsequently, the fully differential single-Higgs production cross section was computed at N$^3$LO, with infrared divergences handled using the $q_T$-slicing method~\cite{Cieri:2018oms,Billis:2021ecs} and the projection-to-Born method~\cite{Chen:2021isd}.~\footnote{The projection-to-Born method was originally proposed in ref.~\cite{Cacciari:2015jma}.} These developments pave the way toward achieving a fully differential prediction for Higgs-boson pair production at N$^3$LO in QCD.

In this work, we calculate the class-$a$ contribution to $gg\to hh$ in the \nnlojet\ framework~\cite{Huss:2025iov} using eq.~\eqref{eq:htohh}. The class-$a$ di-Higgs squared amplitude $|\mathcal{M}^a_{hh}|^2$ is obtained from its single-Higgs counterpart $|\mathcal{M}_h|^2$ through the relation
\begin{align}
\label{eq:class-matrix}
    |\mathcal{M}_{hh}^{a}|^{2}=\frac{1}{2v^{2}}\left|\frac{C_{h h}}{C_h}-\frac{6 \lambda_{3h} v^2}{m_{h h}^2-m_h^2}\right|^2 \times\left(\left. |\mathcal{M}_{h}|^{2} \right|_{p_h \to p_{h h}}\right)\,,
\end{align}
where both the Wilson coefficients, $C_{hh}$ and $C_h$,
and the single-Higgs squared amplitude must be expanded as a series in $\alpha_s$. This allows us to obtain the class-$a$ di-Higgs squared amplitude $|\mathcal{M}_{hh}^{a}|^{2}$ up to N$^3$LO. Both Higgs pair production and the associated Higgs-pair-plus-jet ($hhj$) production have been implemented in \nnlojet\ up to NNLO in QCD. For both processes at NNLO, we employ the antenna subtraction method~\cite{Gehrmann-DeRidder:2004ttg,Gehrmann-DeRidder:2005svg,Gehrmann-DeRidder:2005alt,GehrmannDeRidder:2005cm,Daleo:2006xa,Daleo:2009yj,Gehrmann:2011wi,Boughezal:2010mc,Gehrmann-DeRidder:2012too,Currie:2013vh,Chen:2022ktf,Gehrmann:2023dxm}, as implemented in \nnlojet, to handle infrared divergences in real-radiation contributions.

To verify our implementation in \nnlojet\ for the class-$a$ contribution, we compare the NNLO total cross sections for di-Higgs production with the results of ref.~\cite{Chen:2019fhs}. In addition, we compare the LO and NLO total and fiducial cross sections for di-Higgs plus $n$ jets, with $0\leq n\leq 3$, against the \mglong\ (\mgshort\ hereafter) generator~\cite{Alwall:2014hca,Frederix:2018nkq}, using the NLO Universal Feynman Output (UFO) format~\cite{Degrande:2011ua,Darme:2023jdn} model~\cite{Chen:2019fhs} corresponding to the effective Lagrangian in eq.~\eqref{eq:effL}. We find perfect agreement in all cases. Table~\ref{tab:Higgs_plus_jets_validation} summarizes our validation.
\begin{table}[hbt!]
    \centering
    \begin{tabular}{|c|c|c|c|c|}
    \hline
        & $hh$ & $hhj$  & $hhjj$  & $hhjjj$   
    \\
    \hline
    LO   & \mgshort & \mgshort & \mgshort & \mgshort  
     \\
    \hline
     NLO  & \mgshort & \mgshort&\mgshort& ---
     \\
    \hline
     NNLO & ref.~\cite{Chen:2019fhs} & --- & --- & --- 
     \\
    \hline
    \end{tabular}
    \caption{Validation summary for Higgs-pair and Higgs-pair plus jets productions in \nnlojet.}
    \label{tab:Higgs_plus_jets_validation}
\end{table}

\begin{comment}
In addition, a comparison of the total class-$a$ cross section for di-Higgs production obtained with \nnlojet\ and ref.~\cite{Chen:2019fhs} is shown in table~\ref{tab:class-a-LO-NNLO}. {\color{red} XC: Do we need to show this table?}
\begin{table}[hbt!]
    \centering
    \begin{tabular}{|c|c|c|c|}
    \hline
        &  $\sigma_{hh}^{a,\mathrm{LO}}$  & $\sigma_{hh}^{a,\mathrm{NLO}}$  & $\sigma_{hh}^{a,\mathrm{NNLO}}$   \\
    \hline
    \nnlojet    &13.804(2)& 26.1751(6) &31.250(2)  
     \\
    \hline
     ref.~\cite{Chen:2019fhs}   &13.80&26.16&31.24 
     \\
    \hline
    \end{tabular}
    \caption{Total cross sections (in units of fb) for the class-$a$ contribution to Higgs-pair production in $pp$ collisions at $\sqrt{s}=13~\rm{TeV}$. Results in the second row are obtained with \nnlojet; those in the last row are taken from ref.~\cite{Chen:2019fhs}.}
    \label{tab:class-a-LO-NNLO}
\end{table}
\end{comment}

To compute the fully differential N$^3$LO di-Higgs cross section for class-$a$, we use the $q_T$-slicing method~\cite{Catani:2007vq,Cieri:2018oms} within \nnlojet. In this approach, the class-$a$ differential cross section can be decomposed into two contributions:
\begin{align}
d\sigma^a_{hh} = d\sigma^a_{hh}\Big|_{p_{T,hh}<p_T^{\rm veto}} + d\sigma^a_{hh}\Big|_{p_{T,hh}>p_T^{\rm veto}}\,,
\label{eq:cutoffa}
\end{align}
where $p_{T,hh}$ is the transverse momentum of the Higgs-pair system. The first and second terms on the right-hand side of eq.~\eqref{eq:cutoffa} correspond to the regions with transverse-momentum cuts $p_{T,hh}<p_T^{\rm veto}$ and $p_{T,hh}>p_T^{\rm veto}$, respectively. 

The first term with sufficiently small $p_T^{\rm veto}$ can be computed using the $q_T$-resummation formalism developed in soft-collinear effective field theory (SCET)~\cite{Bauer:2000ew,Bauer:2000yr,Bauer:2001ct,Bauer:2001yt,Beneke:2002ph}. Explicitly,
\begin{align}\label{eq:qt_smalla}
   \left.d\sigma^a_{hh}\right|_{p_{T,hh}<p_T^{\mathrm{veto}}} \propto H^a\otimes B_g \otimes B_g \otimes S \times \left[ 1+  \mathcal{O}\left(\left(\frac{p_T^{\mathrm{veto}}}{m_{hh}}\right)^{n}\right)\right]\,,
\end{align}
where power-suppressed terms have been neglected. The power $n$ appearing in the leading power-suppressed terms of eq.~\eqref{eq:qt_smalla} is $2$ for inclusive observables, such as total cross sections. In contrast, in the presence of fiducial cuts, $n$ can be reduced to $1$ (i.e., linear power corrections)~\cite{Salam:2021tbm,Grazzini:2017mhc,Ebert:2019zkb,Alekhin:2021xcu}. We will discuss the linear power corrections later. The formalism in eq.~\eqref{eq:qt_smalla} involves a convolution of a hard function $H^a$, two gluon transverse-momentum-dependent (TMD) beam functions $B_g$, and a TMD soft function $S$. The hard function $H^a$, which is process-dependent, is given in appendix A of ref.~\cite{Chen:2019fhs} for di-Higgs production, while the gluon TMD beam functions $B_g$ and the TMD soft function $S$ are process-independent. Two-loop analytic results for the quark and gluon TMD beam functions can be found in refs.~\cite{Gehrmann:2012ze, Gehrmann:2014yya,Luebbert:2016itl,Echevarria:2016scs,Luo:2019hmp, Luo:2019bmw},
and the three-loop results are given in refs.~\cite{Luo:2019szz,Ebert:2020yqt,Luo:2020epw}. The soft function $S$ is also known up to three loops~\cite{Li:2016ctv}. 

Due to the non-vanishing transverse momentum of the Higgs-pair system in the second term of eq.~\eqref{eq:cutoffa}, events with at least one additional jet are selected. In other words, an N$^3$LO computation of the class-$a$ contribution requires the calculation of the NNLO QCD cross section for Higgs-boson pair production in association with a jet, which has been implemented and validated in \nnlojet, as described previously. In the small transverse-momentum region, the numerical stability of \nnlojet\ has been optimized for Higgs plus jet production at NNLO QCD in refs.~\cite{Chen:2018pzu,Cieri:2018oms}. Further validations of the extension to Higgs-boson pair plus jet production are presented below.

\begin{figure}[hbt!]
%\hspace{-0.3cm}
 \includegraphics[width=0.5\textwidth]{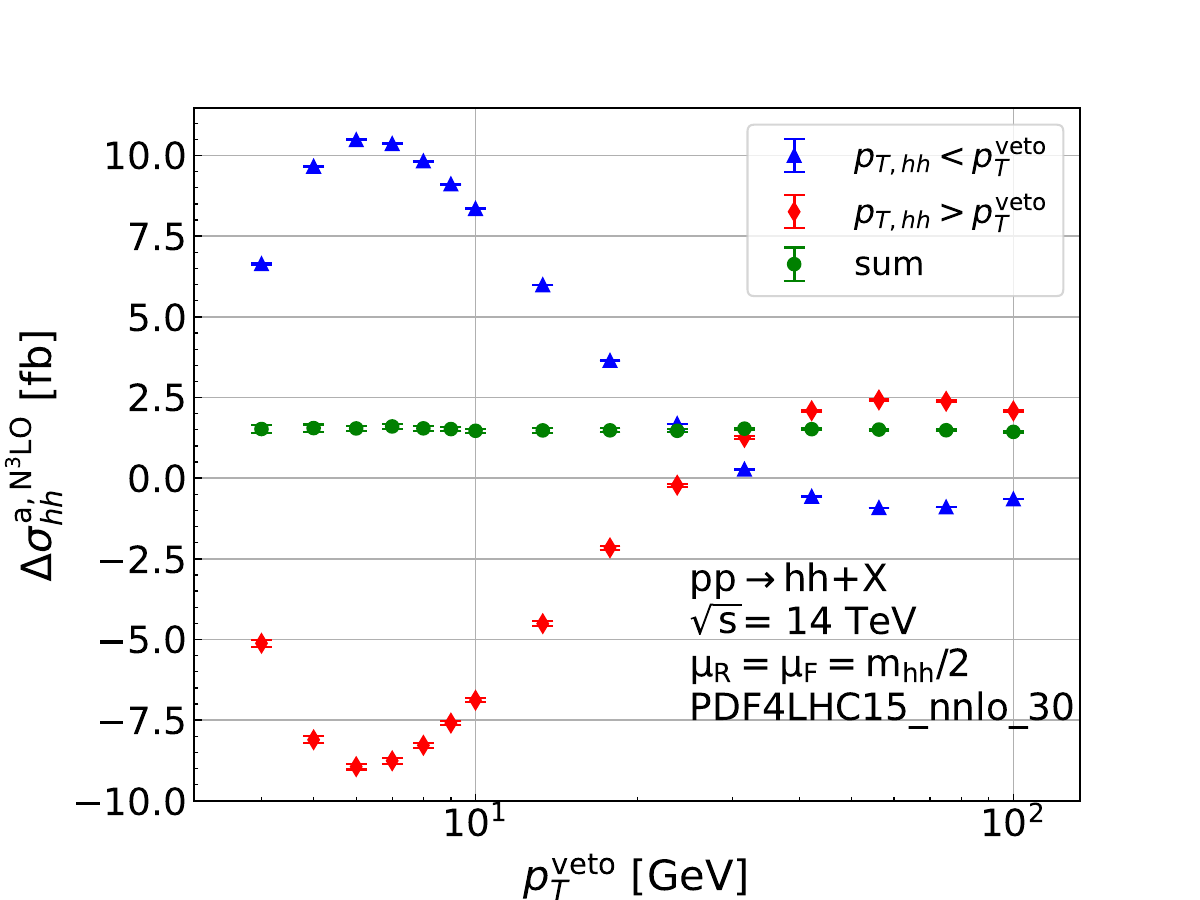}
 %\hspace{-0.8cm}
 \includegraphics[width=0.5\textwidth]{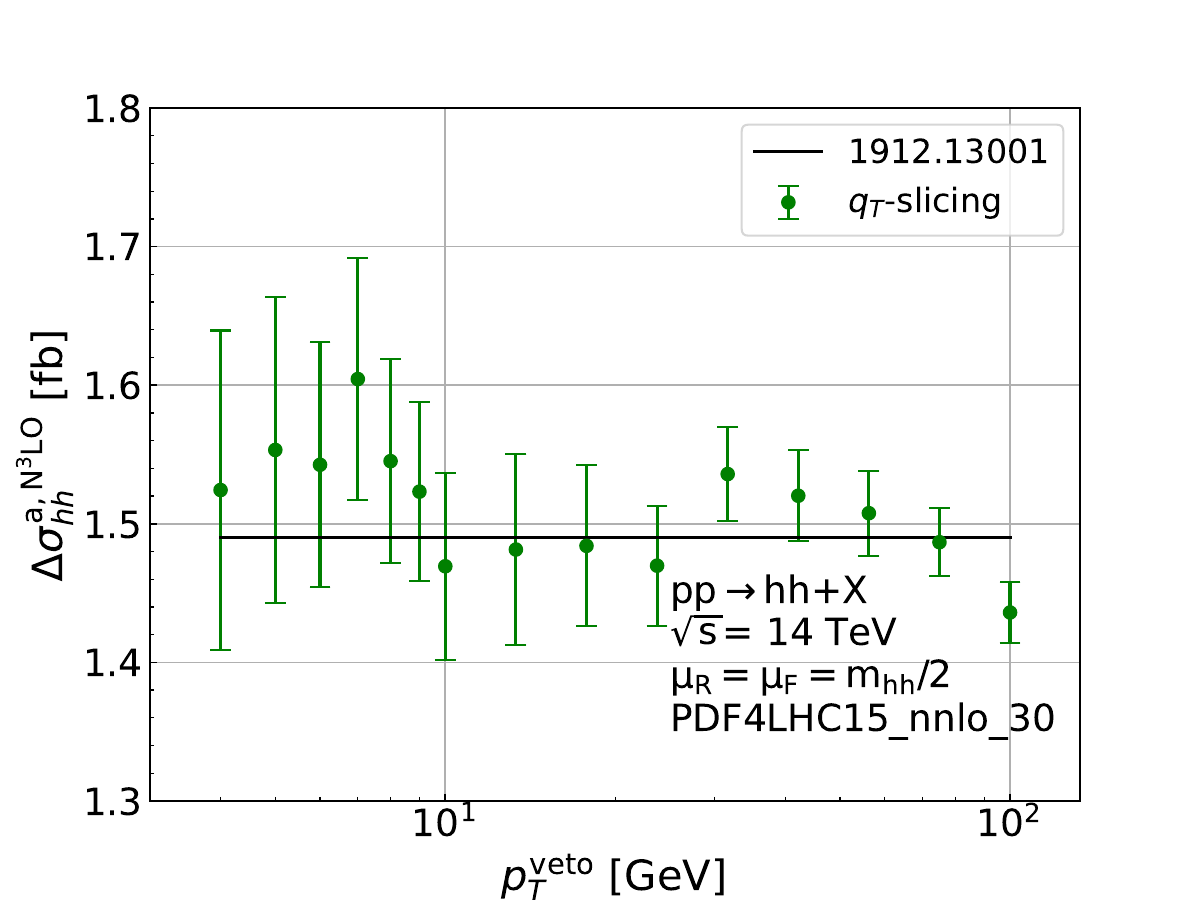}
  \caption{The $p_T^{\mathrm{veto}}$ dependence of the N$^3$LO ($\mathcal{O}(\alpha_s^5)$) corrections $\Delta\sigma_{hh}^{a,\mathrm{N}^3\mathrm{LO}}$ (green circles) to the total cross section of the class-$a$ contribution in $pp$ collisions at $\sqrt{s}=14$ TeV. The left panel displays the breakdown of $\Delta\sigma_{hh}^{a,\mathrm{N}^3\mathrm{LO}}$ into $p_{T,hh}<p_T^{\mathrm{veto}}$ (blue triangles) and $p_{T,hh}>p_T^{\mathrm{veto}}$ (red diamonds) contributions. In the right panel, our $q_T$-slicing result (green circles) is compared with the inclusive cross section from ref.~\cite{Chen:2019fhs}, obtained using \ihixs\ and eq.~\eqref{eq:htohh} (black line). Error bars indicate Monte Carlo integration uncertainty.}
  \label{fig:HH-a-N3LO_only}
\end{figure}

For sufficiently small $p_T^{\mathrm{veto}}$, such that power-suppressed terms can be safely neglected, the sum of the two contributions in eq.~\eqref{eq:cutoffa} is expected to be independent of the choice of $p_T^{\mathrm{veto}}$. This independence also serves as a strong cross-check of the implementation, since the two terms on the right-hand side of eq.~\eqref{eq:cutoffa} are computed using completely different techniques. We now discuss the $p_T^{\mathrm{veto}}$ (in)dependence in our results.
For the total cross section, we explicitly check the $p_T^{\mathrm{veto}}$ dependence of the N$^3$LO ($\mathcal{O}(\alpha_s^5)$) corrections $\Delta\sigma_{hh}^{a,\mathrm{N}^3\mathrm{LO}}$, as shown in figure~\ref{fig:HH-a-N3LO_only}. The calculation is performed for $pp$ collisions at $\sqrt{s}=14$ TeV using the parton distribution function (PDF) set {\tt PDF4LHC15\_nnlo\_30}~\cite{Butterworth:2015oua}. The left panel displays the decomposition of the corrections into $p_{T,hh}<p_T^{\mathrm{veto}}$ (blue triangles) and $p_{T,hh}>p_T^{\mathrm{veto}}$ (red diamonds) parts. Although the two separate contributions depend strongly on the cutoff, their sum exhibits only a very weak dependence on $p_T^{\mathrm{veto}}$ when $p_T^{\mathrm{veto}}\ll m_{hh}$. As a second cross-check, we compare our $\Delta\sigma_{hh}^{a,\mathrm{N}^3\mathrm{LO}}$ result obtained using $q_T$-slicing with the inclusive cross section from ref.~\cite{Chen:2019fhs}, which was computed using \ihixs\ and eq.~\eqref{eq:htohh}. This comparison, shown in the right panel of figure~\ref{fig:HH-a-N3LO_only}, demonstrates that for $p_T^{\rm veto}\simeq 10$ GeV the two computations agree within numerical uncertainties.

\begin{figure}[hbt!]
 \centering
 \includegraphics[width=0.75\textwidth]{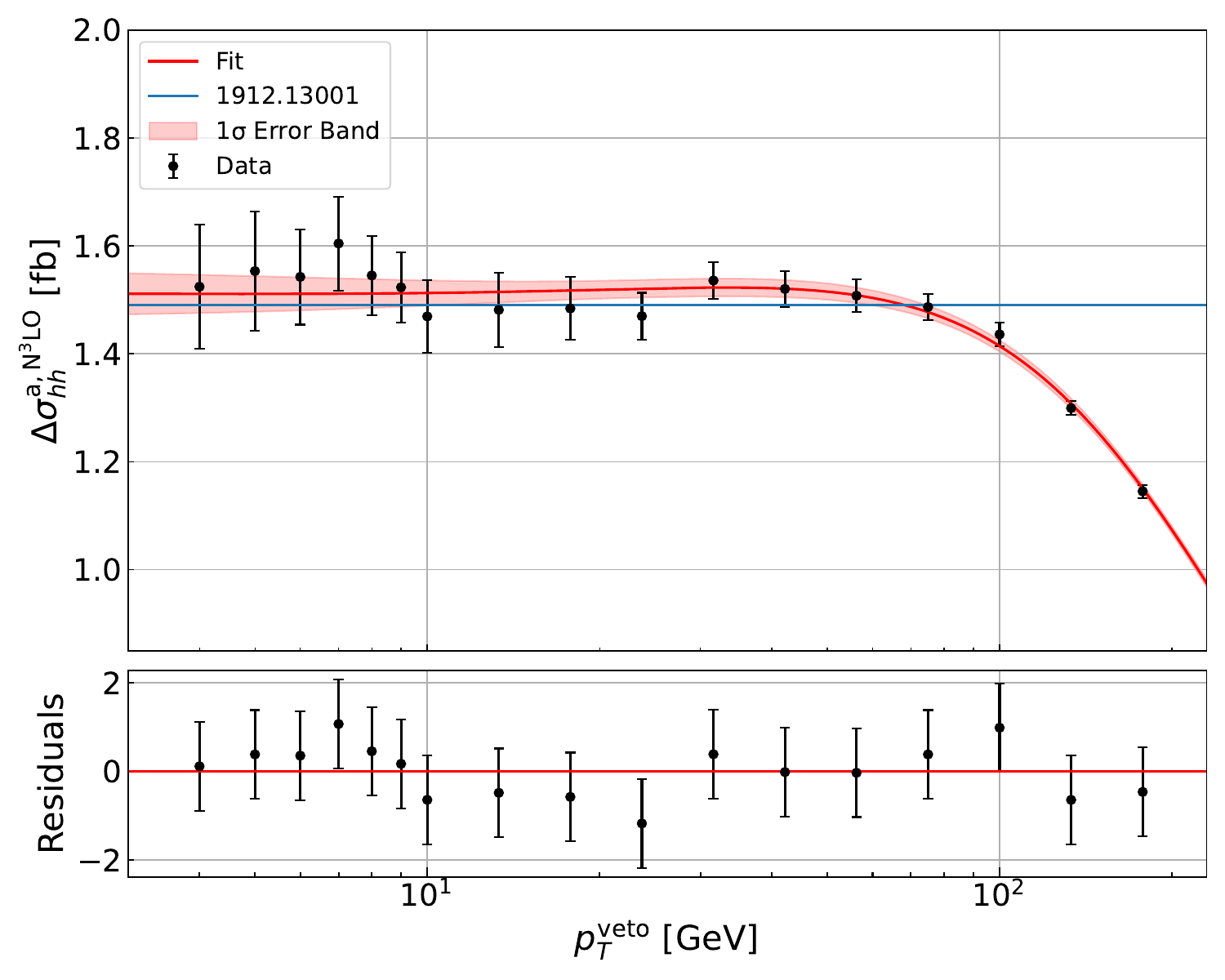}
  \caption{Fit of $\Delta\sigma^{\mathrm{a,N^{3}LO}}_{hh}(p_{T}^{\rm veto})$ using the ansatz in eq.~\eqref{eq:DN3LOaansatz}. The lower panel shows the residuals of the fit. The numerical data points are shown as black dots, with error bars indicating the Monte Carlo integration statistical uncertainties. The red line represents the central value of the fit curve, and the light red band denotes the fit error. The blue line corresponds to the result in ref.~\cite{Chen:2019fhs}.}
  \label{fig:HH-a-N3LO_only-fitting}
\end{figure}

To estimate the systematic error due to the choice of $p_T^{\rm veto}$ in the N$^3$LO corrections, we first assume the following functional form of the slicing power corrections at the total cross section level:
\begin{align}\label{eq:DN3LOaansatz}
    \Delta \sigma_{hh}^{a,\mathrm{N}^3 \mathrm{LO}}\left(p_T^{\mathrm{veto}}\right) = \Delta \sigma_{hh}^{a,\mathrm{N}^3\mathrm{LO}}(0)+\left(\frac{\alpha_s}{2 \pi}\right)^3 \left(\frac{p_T^{\mathrm{veto}}}{m_{hh}}\right)^2 \sum_{n=0}^5 a_n^{(3)} \ln ^n {\left(\frac{m_{hh}}{p_T^{\mathrm{veto}}}\right)}\,, 
\end{align}
and perform a numerical fit to extract $\Delta \sigma_{hh}^{a,\mathrm{N}^3\mathrm{LO}}(0)$ and the coefficients $a_n^{(3)}$. Our fitted results for the inclusive total cross section are shown in figure~\ref{fig:HH-a-N3LO_only-fitting}. We obtain $\Delta \sigma_{hh}^{a,\mathrm{N}^3 \mathrm{LO}}(0)=1.513(47)~\mathrm{fb}$ with 3\% uncertainty on the N$^3$LO coefficient. Our fitted result is in perfect agreement with the $\Delta \sigma_{hh}^{a,\mathrm{N}^3 \mathrm{LO}}$ result in ref.~\cite{Chen:2019fhs} using \ihixs\ and eq.~\eqref{eq:htohh}, which amounts to $1.4904~\mathrm{fb}$. The difference between $\Delta \sigma_{hh}^{a,\mathrm{N}^3\mathrm{LO}}(0)$ and $\Delta \sigma_{hh}^{a,\mathrm{N}^3\mathrm{LO}}$ can be taken as an estimate of our systematic uncertainty, which amounts to approximately $1.5\%$ of the N$^3$LO correction and is negligible for the total cross section.

\begin{figure}[hbt!]
%\hspace{-0.3cm}
 \includegraphics[width=0.5\textwidth]{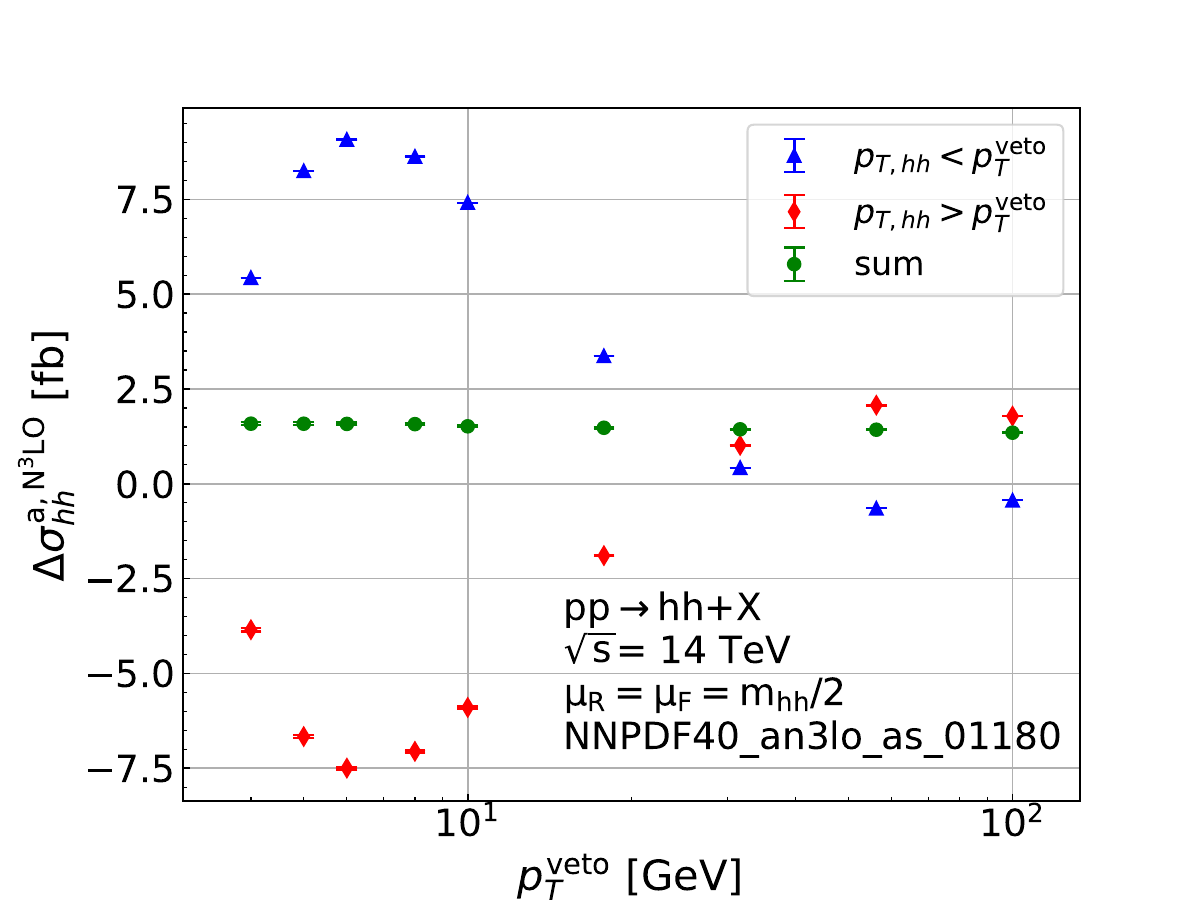}
 %\hspace{-0.8cm}
 \includegraphics[width=0.5\textwidth]{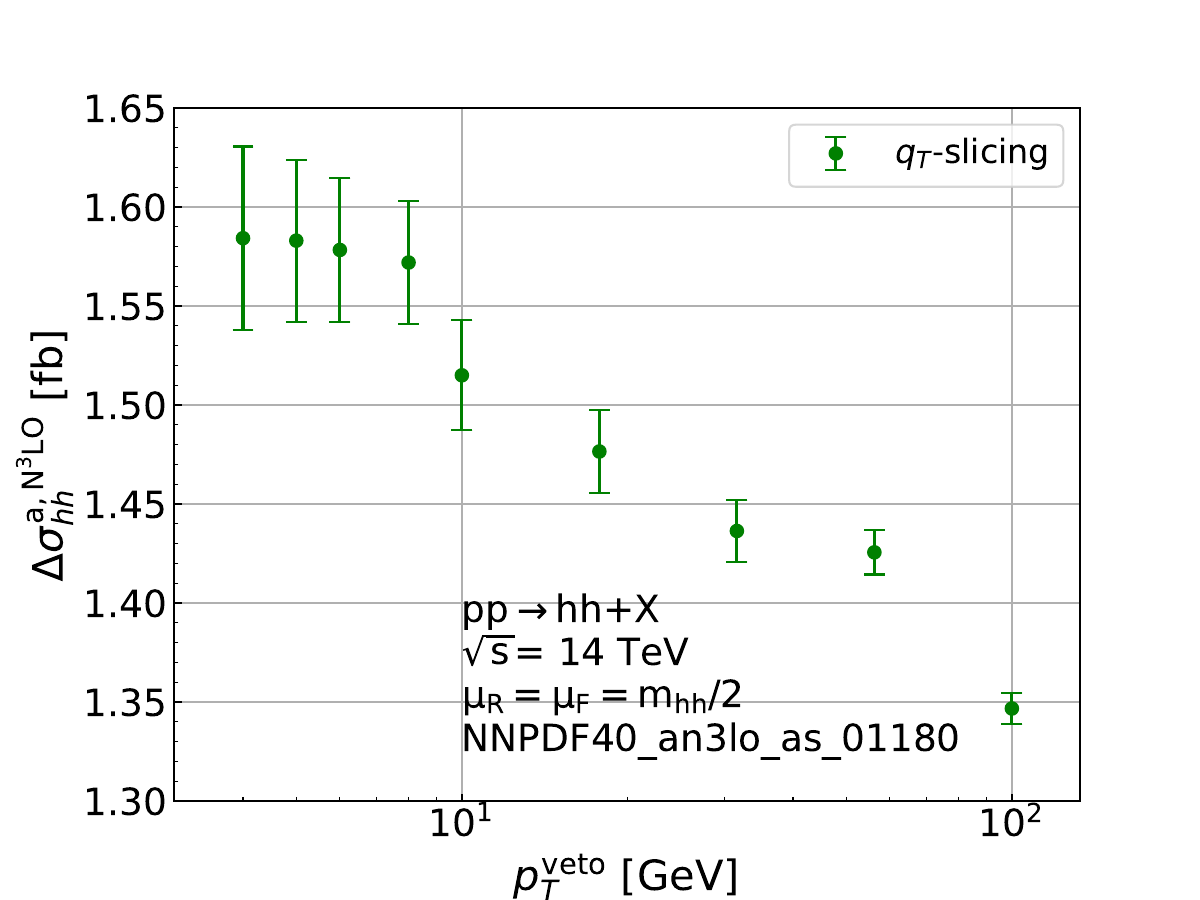}
  \caption{
  The $p_T^{\mathrm{veto}}$ dependence of the N$^3$LO ($\mathcal{O}(\alpha_s^5)$) corrections $\Delta\sigma_{hh}^{a,\mathrm{N}^3\mathrm{LO}}$ (green circles) to the fiducial cross section of the class-$a$ contribution in $pp$ collisions at $\sqrt{s}=14$ TeV. The fiducial region is defined in eq.~\eqref{eq:fiducialcuts}. The left panel displays the breakdown of $\Delta\sigma_{hh}^{a,\mathrm{N}^3\mathrm{LO}}$ into $p_{T,hh}<p_T^{\mathrm{veto}}$ (blue triangles) and $p_{T,hh}>p_T^{\mathrm{veto}}$ (red diamonds) contributions. Error bars indicate Monte Carlo integration uncertainty.}
  \label{fig:HH-a-N3LO_only-Fid}
\end{figure}

The $p_T^{\mathrm{veto}}$ dependence can be further examined for the fiducial cross section with the fiducial cuts defined in eq.~\eqref{eq:fiducialcuts}. The presence of a fiducial phase-space volume generally leads to linear power corrections~\cite{Salam:2021tbm,Grazzini:2017mhc,Ebert:2019zkb,Alekhin:2021xcu} in eq.~\eqref{eq:qt_smalla}, which we compensate using a simple recoil prescription~\cite{Catani:2015vma,Ebert:2020dfc,Becher:2020ugp,Buonocore:2021tke,Camarda:2021jsw,Chen:2022cgv}.
Upon applying the recoil scheme, the $p_T^{\mathrm{veto}}$ dependence of the N$^3$LO corrections $\Delta \sigma_{hh}^{a,\mathrm{N}^3\mathrm{LO}}$ is shown in figure~\ref{fig:HH-a-N3LO_only-Fid}. 
The calculation is performed for $pp$ collisions at $\sqrt{s}=14$ TeV using the PDF set {\tt NNPDF40\_an3lo\_as\_01180}~\cite{NNPDF:2024nan}.
We again observe a strong cancellation of the $p_T^{\mathrm{veto}}$ dependence in the left panel of figure~\ref{fig:HH-a-N3LO_only-Fid} when summing the $p_{T,hh}<p_T^{\mathrm{veto}}$ (blue triangles) and $p_{T,hh}>p_T^{\mathrm{veto}}$ (red diamonds) contributions. Meanwhile, as shown in the right panel, the fiducial N$^3$LO corrections $\Delta\sigma^{a,\mathrm{N}^3\mathrm{LO}}_{hh}$ reach a plateau for $p_T^{\mathrm{veto}}\lesssim 8$ GeV.  

To estimate the systematic uncertainty associated with $p_T^{\mathrm{veto}}$ for fiducial cross sections, we take $p_{T}^{\rm veto}=5~\mathrm{GeV}$ as a reference value and vary it to $4~\mathrm{GeV}$ and $6~\mathrm{GeV}$ for all differential predictions. This procedure mimics the $p_T^{\mathrm{veto}}$ extrapolation uncertainty of the inclusive cross section shown in figure~\ref{fig:HH-a-N3LO_only-fitting}, yielding a comparable numerical effect.

\subsection{Calculation of the class-$b$ contribution\label{sec:classb}}

In this subsection, we discuss the methods employed to calculate the class-$b$ contribution. As shown in table~\ref{tab:classes}, obtaining the N$^3$LO QCD ($\mathcal{O}(\alpha_s^5)$) corrections to di-Higgs production in the HTL requires the computation of the NNLO QCD corrections to the class-$b$ contribution, whose LO is $\mathcal{O}(\alpha_s^3)$. 
The NLO QCD corrections for the class-$b$ contribution are computed using the automated \mgshort\ framework, which employs the local FKS infrared-divergence subtraction scheme~\cite{Frixione:1995ms,Frixione:1997np,Frederix:2009yq}, with the NLO UFO model {\tt HEFT\_DH\_UFO}~\footnote{\href{https://feynrules.irmp.ucl.ac.be/wiki/HEFT_DH}{https://feynrules.irmp.ucl.ac.be/wiki/HEFT\_DH}} generated according to eq.~(\ref{eq:effL}). Further details of the UFO model can be found in appendix B of ref.~\cite{Chen:2019fhs}.

For the NNLO corrections to this contribution, we again employ the $q_T$-slicing method, as in eqs.~\eqref{eq:cutoffa} and \eqref{eq:qt_smalla}, with the superscript $a$ replaced by $b$.
For the term with \mbox{$p_{T,hh}<p_T^{\mathrm{veto}}$}, the only difference compared to the class-$a$ contribution at leading power in $p_T^{\mathrm{veto}}/m_{hh}$ is the hard function $H^b$, whose explicit expressions can be found in appendix A of ref.~\cite{Chen:2019fhs}, with the two-loop amplitudes taken from ref.~\cite{Banerjee:2018lfq}.
The term with $p_{T,hh}>p_T^{\mathrm{veto}}$ corresponds to the NLO QCD corrections to Higgs-pair production in association with a jet, restricted to three effective-vertex insertions at the squared-amplitude level. This calculation can be performed using \mgshort\ with the NLO UFO model {\tt HEFT\_DH\_UFO}.

\begin{figure}[hbt!]
%\hspace{-0.3cm}
 \includegraphics[width=0.5\textwidth]{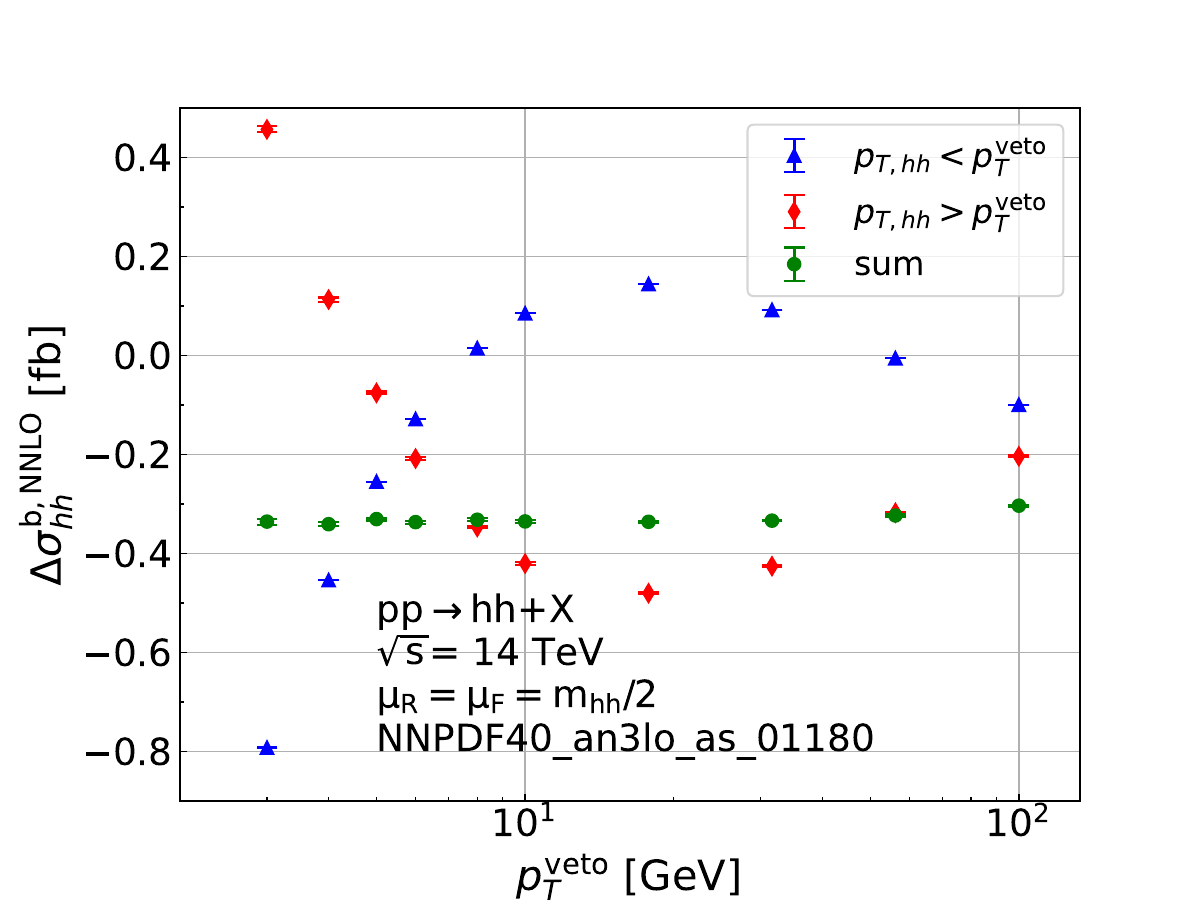}
 %\hspace{-0.8cm}
 \includegraphics[width=0.5\textwidth]{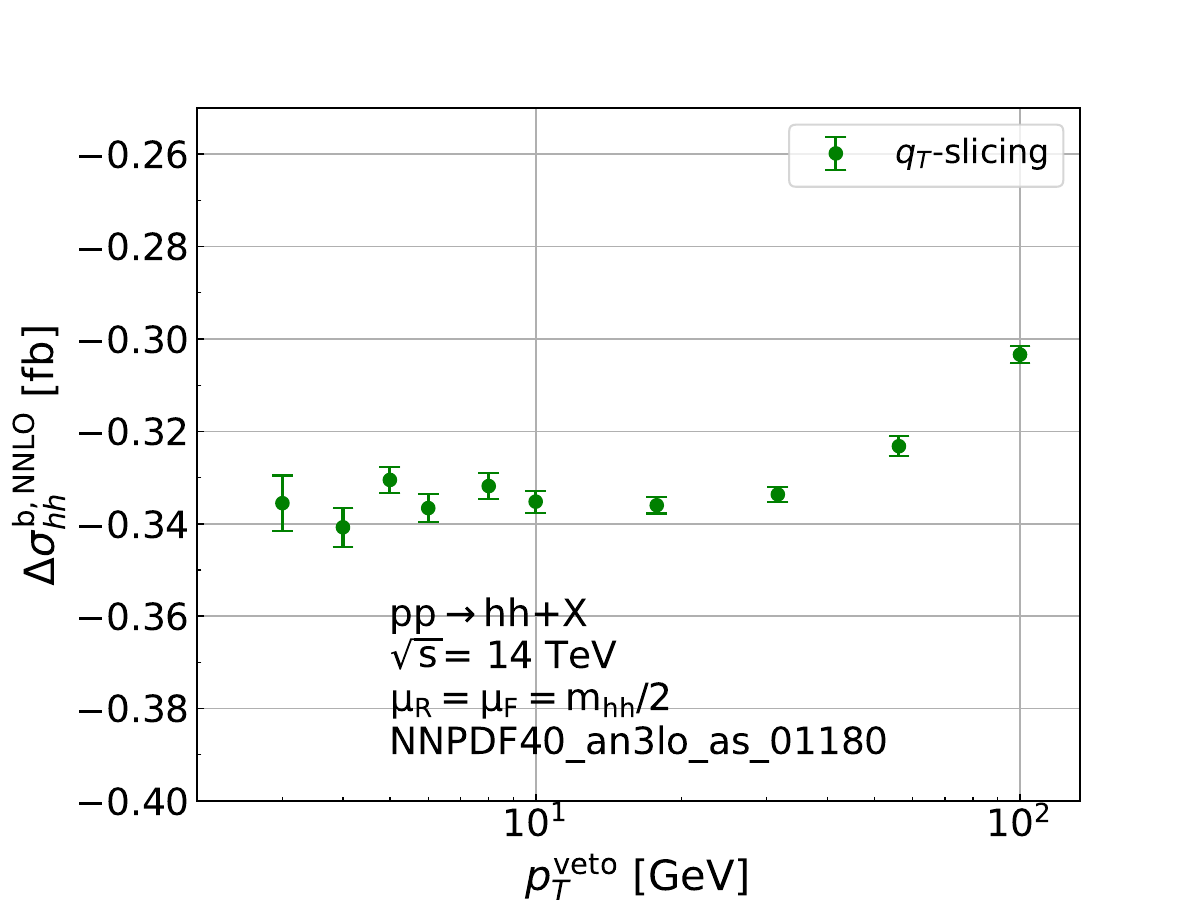}
  \caption{
  The $p_T^{\mathrm{veto}}$ dependence of the NNLO ($\mathcal{O}(\alpha_s^5)$) correction $\Delta\sigma_{hh}^{b,\mathrm{NNLO}}$ (green circles) to the fiducial cross section of the class-$b$ contribution in $pp$ collisions at $\sqrt{s}=14$ TeV. The fiducial region is defined in eq.~\eqref{eq:fiducialcuts}. The left panel displays the breakdown of $\Delta\sigma_{hh}^{b,\mathrm{NNLO}}$ into $p_{T,hh}<p_T^{\mathrm{veto}}$ (blue triangles) and $p_{T,hh}>p_T^{\mathrm{veto}}$ (red diamonds) contributions. Error bars indicate Monte Carlo integration uncertainty.}
  \label{fig:HH-b-NNLO_only}
\end{figure}

The $p_{T}^{\rm veto}$ dependence of the NNLO ($\mathcal{O}(\alpha_s^5)$) correction $\Delta \sigma^{b,\mathrm{NNLO}}_{hh}$ within the fiducial region defined in eq.~\eqref{eq:fiducialcuts} is illustrated in figure~\ref{fig:HH-b-NNLO_only}. 
A strong cancellation of the $p_T^{\mathrm{veto}}$ dependence between the two contributions from $p_{T,hh}<p_T^{\mathrm{veto}}$ (blue triangles) and $p_{T,hh}>p_T^{\mathrm{veto}}$ (red diamonds) is observed in the left panel.
The right panel provides a zoomed-in view of the $p_T^{\mathrm{veto}}$ region relevant for our study using the $q_{T}$-slicing method. In general, the fiducial cross section shows good convergence for $p_{T}^{\rm veto}\lesssim 8$ GeV, as the values of $\Delta \sigma^{b,\mathrm{NNLO}}_{hh}$ at different $p_{T}^{\rm veto}$ agree within their Monte Carlo statistical uncertainties.

\subsection{Calculation of the class-$c$ contribution\label{sec:classc}}

Finally, as shown in table~\ref{tab:classes}, we discuss the class-$c$ contribution. For this class, NLO QCD calculations are sufficient to reach $\mathcal{O}(\alpha_s^5)$. In this work, we compute this contribution using \mgshort\ with the FKS subtraction method. As in the evaluation of $\Delta \sigma_{hh}^{b,\mathrm{NLO}}$ -- which was not explicitly mentioned before -- the infrared-finite real partonic subprocesses $q\bar{q}\to hh g$ are not automatically included in the NLO run for $gg\to hh$ in \mgshort. These subprocesses must
be incorporated as a separate LO run, because the organization of real-emission contributions is based on the underlying Born processes~\cite{Alwall:2014hca}.

The infrared subtraction and slicing methods, along with the event generators employed to obtain the fully differential N$^3$LO cross sections for Higgs-pair production in the HTL, are summarized in table~\ref{tab:calcmethod}.
\begin{table}[hbt!]
    \centering
    \begin{tabular}{|c|c|c|c|}
    \hline
    \multirow{2}{*}{Order}    &  NLO  & NNLO  & $\rm N^{3}LO$   \\
    \cline{2-4}
  &   $\mathcal{O}(\alpha_s^3)$ &   $\mathcal{O}(\alpha_s^4)$  & $\mathcal{O}(\alpha_s^5)$\\ \hline
    $a$   &  antenna (\nnlojet)  &   antenna (\nnlojet) &   $q_T$-slicing (\nnlojet) \\ \hline
    $b$   &  - &   FKS (\mgshort) &   $q_T$-slicing (\mgshort+a private code) \\ \hline
    $c$           & - &  - &   FKS (\mgshort) \\ \hline
    \end{tabular}
    \caption{Summary of the infrared subtraction and slicing methods and event generators (in parentheses) used for the various contributions to the N$^3$LO differential cross sections in the HTL.}
    \label{tab:calcmethod}
\end{table}
The $q_T$-slicing calculations of the $\mathcal{O}(\alpha_s^5)$ contributions in classes $a$ and $b$ are carried out using a combination of \nnlojet, \mgshort, and in-house codes.

\section{Results\label{sec:results}}

In this section, we present our predictions for the fiducial and differential cross sections of Higgs-pair production at the LHC. After specifying our calculational setup in subsection \ref{sec:setup}, we first discuss our results in the HTL in subsection \ref{sec:resHTL}, and then combine these HTL results with the full top-quark-mass-dependent NLO QCD results in subsection \ref{sec:n3lomt}.

\subsection{Calculational setup\label{sec:setup}}

In our numerical calculations, we consider proton-proton collisions at the LHC with a center-of-mass energy of $\sqrt{s}=14~\mathrm{TeV}$. For the SM input parameters, we take $v=246.2~\mathrm{GeV}$, the Higgs mass $m_{h}=125~\mathrm{GeV}$, the OS top-quark mass $m_{t}=173~\mathrm{GeV}$, and the trilinear Higgs self-coupling $\lambda_{3h}=\lambda_{3h}^{\mathrm{SM}}$ (cf. eq.~\eqref{eq:HiggsselfcoupinSM}). We work in five massless quark flavors ($n_f=5$) and include full color contributions. The final-state Higgs are treated as on-shell particles. We use the {\tt NNPDF40\_an3lo\_as\_01180} PDF set~\cite{NNPDF:2024nan} and the corresponding $\alpha_s$ running provided by the {\tt LHAPDF6} library~\cite{Buckley:2014ana}.
The central values of the renormalization and factorization scales are set to half of the di-Higgs invariant mass,
\begin{align}
\mu_{R}=\mu_{F}=\mu_{0}\equiv \frac{m_{hh}}{2}\,.
\end{align}
To estimate the uncertainty due to missing higher-order QCD corrections, we vary $\mu_{R}$ and $\mu_{F}$ according to the standard $7$-point scale variation,
\begin{align}
\label{scale}
   \begin{aligned}
\mu_{R}=\xi_{R}\mu_{0}\,,\quad \mu_{F}=\xi_{F}\mu_{0}\,,
    \end{aligned}
\end{align}
with
\begin{align}
    \bigg(\xi_{R},\xi_{F}\bigg)\in\Bigg\{\bigg(1,1\bigg),\bigg(1,\frac{1}{2}\bigg),\bigg(1,2\bigg),\bigg(\frac{1}{2},1\bigg),\bigg(2,1\bigg),\bigg(\frac{1}{2},\frac{1}{2}\bigg),\bigg(2,2\bigg)\Bigg\}\,.
\end{align}
For contributions evaluated using the $q_{T}$-slicing method (cf. table~\ref{tab:calcmethod}), we choose \mbox{$p_{T}^{\rm veto}=5~\mathrm{GeV}$} as the central value and vary the $p_{T}^{\mathrm{veto}}$ value across the three points of $4$, $5$, and $6$ GeV to estimate the uncertainty associated with the choice of the slicing parameter.

As a showcase of our analysis, we employ the following fiducial cuts: 
\begin{equation}
\label{eq:fiducialcuts}
p_{T, h_{1}}>30~\mathrm{GeV}\,, \quad p_{T, h_{2}}>20~\mathrm{GeV}\,, \quad |y_{h}|<2.4\,,
\end{equation}
where $h_1$ and $h_2$ denote the leading and subleading Higgs bosons, ordered by their transverse momenta $p_{T,h_i}$.
This fiducial volume is designed according to ATLAS and CMS detectors. In particular, asymmetric transverse-momentum cuts are widely adopted to regulate back-to-back configurations at LO, and were originally motivated by their usefulness in stabilizing fixed-order predictions for di-jet production~\cite{Klasen:1995xe,Harris:1997hz,Frixione:1997ks,Alioli:2010xa,Frederix:2016ost}. Recent studies~\cite{Salam:2021tbm,Buonocore:2021tke}, however, have shown that such asymmetric cuts are still insufficient to fully stabilize the fixed-order perturbative series, and have proposed alternative cuts, such as product cuts and staggered cuts,~\footnote{For example, one may impose staggered $p_T$ cuts on the Higgs bosons, ordered by rapidity.} which remain less commonly used choices. Nevertheless, in this paper we adopt the widely used asymmetric cuts as a representative example, leaving the exploration of alternative cuts for future work.

\subsection{Heavy top-quark limit\label{sec:resHTL}}

In this subsection, we discuss our results in the HTL. Both the phase-space-integrated fiducial cross sections and differential distributions are presented.

\subsubsection{Phase-space-integrated fiducial cross sections}

The phase-space-integrated fiducial cross sections, together with their decomposition into the three classes in the HTL from LO to N$^3$LO, are given in table~\ref{tab:integrated-fiducial-cross-sections}.
\begin{table}[hbt!]
    \centering
    \begin{tabular}{|c|c|c|c|c|}
    \hline
        &  $\sigma_{hh}^{\mathrm{LO}}$ [fb]  & $\sigma_{hh}^{\mathrm{NLO}}$ [fb]  & $\sigma_{hh}^{\mathrm{NNLO}}$ [fb]  & $\sigma_{hh}^{\mathrm{N}^3\mathrm{LO}}$ [fb] \\
    \hline
    Sum  &  $14.85_{-22 \%}^{+30 \%}$  &   $27.87_{-15 \%}^{+18 \%}$ &   $32.74_{-7.3 \%}^{+5.2 \%}$  & $34.00(4)_{-2.9 \%}^{+1.4 \%}\pm0.03\%$
     \\
    \hline
     Class-$a$  &  $14.85_{-22 \%}^{+30 \%}$  &   $28.25_{-15 \%}^{+18 \%}$ &   $33.62_{-7.7 \%}^{+5.7 \%}$  & $35.20(4)_{-3.4 \%}^{+1.7 \%}\pm0.02\%$
     \\
    \hline
     Class-$b$   &  0 &   $-0.3758(2)^{+42\%}_{-28\%}$ &   $-0.8854(2)^{+29\%}_{-12\%}$  & $-1.216(3)^{+13\%}_{-14\%}\pm0.8\%$
     \\
    \hline
     Class-$c$   & 0 &  0 &   $0.003065(3)^{+55\%}_{-34\%}$  & $0.008865(3)^{+41\%}_{-28\%}$
     \\
     \hline
    \end{tabular}
    \caption{Phase-space-integrated fiducial cross sections (in units of fb) from LO to N$^3$LO at $\sqrt{s}=14~\rm TeV$. The first percentages indicate the systematic uncertainties arising from scale variations according to eq.~\eqref{scale}. The second percentages, when applicable, correspond to the uncertainties associated with the choice of $p_T^{\mathrm{veto}}$. Values given in parentheses denote the Monte Carlo statistical uncertainties.}
    \label{tab:integrated-fiducial-cross-sections}
\end{table}
We observe that the main contribution arises from class-$a$, which accounts for $+103.5\%$ of the N$^3$LO cross section $\sigma_{hh}^{\mathrm{N}^3\mathrm{LO}}$. 
The class-$b$ contribution amounts to only $-3.6\%$ of $\sigma_{hh}^{\mathrm{N}^3\mathrm{LO}}$, while class-$c$ contributes $+0.03\%$ and is therefore almost negligible. This hierarchy can be understood from the structure of the effective Lagrangian in eq.~\eqref{eq:effL}: the inclusion of an additional effective-vertex insertion leads to a suppression factor of $\alpha_s/3\pi\sim1\%$. 

Similar to the case of inclusive cross sections studied in ref.~\cite{Chen:2019fhs}, the NLO QCD corrections enhance the LO fiducial cross section by $88\%$, the NNLO QCD corrections further increase the NLO result by $17\%$, and the N$^3$LO QCD corrections provide additional $3.8\%$ improvement. For $k\geq2$, the size of the QCD corrections at N$^k$LO is well captured by the scale-variation uncertainties at the lower order N$^{k-1}$LO. This pattern indicates a well-behaved perturbative expansion and suggests asymptotic convergence of the $\alpha_s$ series.

The inclusion of higher-order QCD corrections also leads to a significant reduction of the scale uncertainties. As shown in table~\ref{tab:integrated-fiducial-cross-sections}, the scale uncertainty is reduced by a factor of $2.6$ from NLO to NNLO, and is further reduced by a factor of $2.9$ from NNLO to N$^3$LO. The fractional scale uncertainty at N$^3$LO is ${}^{+1.4\%}_{-2.9\%}$, which is again close to, though slightly larger than, that in the inclusive case (cf. table 3 of ref.~\cite{Chen:2019fhs}). We also observe a mild cancellation of scale uncertainties between class-$a$ and class-$b$ contributions at N$^3$LO. In particular, the scale-uncertainty band for class-$a$ alone, ${}^{+1.7\%}_{-3.4\%}$, is about $19\%$ larger than that of the sum. This behavior can be attributed to operator mixing effects, as discussed in ref.~\cite{Zoller:2016iam}.

Finally, we comment on the theoretical uncertainty associated with the variation of the slicing parameter $p_T^{\mathrm{veto}}$. This uncertainty arises in the $\mathcal{O}(\alpha_s^5)$ corrections to both the class-$a$ and class-$b$ contributions and is quoted as the second percentage in table~\ref{tab:integrated-fiducial-cross-sections}. It exceeds the Monte Carlo statistical uncertainty only for the class-$b$ contribution, corresponding to roughly $3.6$ times the standard deviation. In all cases, it is more than an order of magnitude smaller than the remaining scale uncertainties and can therefore be considered negligible.

\subsubsection{Differential distributions\label{sec:mhhnomt}}
\begin{figure}[hbt!]
%\hspace{-0.3cm}
  \begin{subfigure}[b]{0.5\textwidth}
    \includegraphics[width=\textwidth]{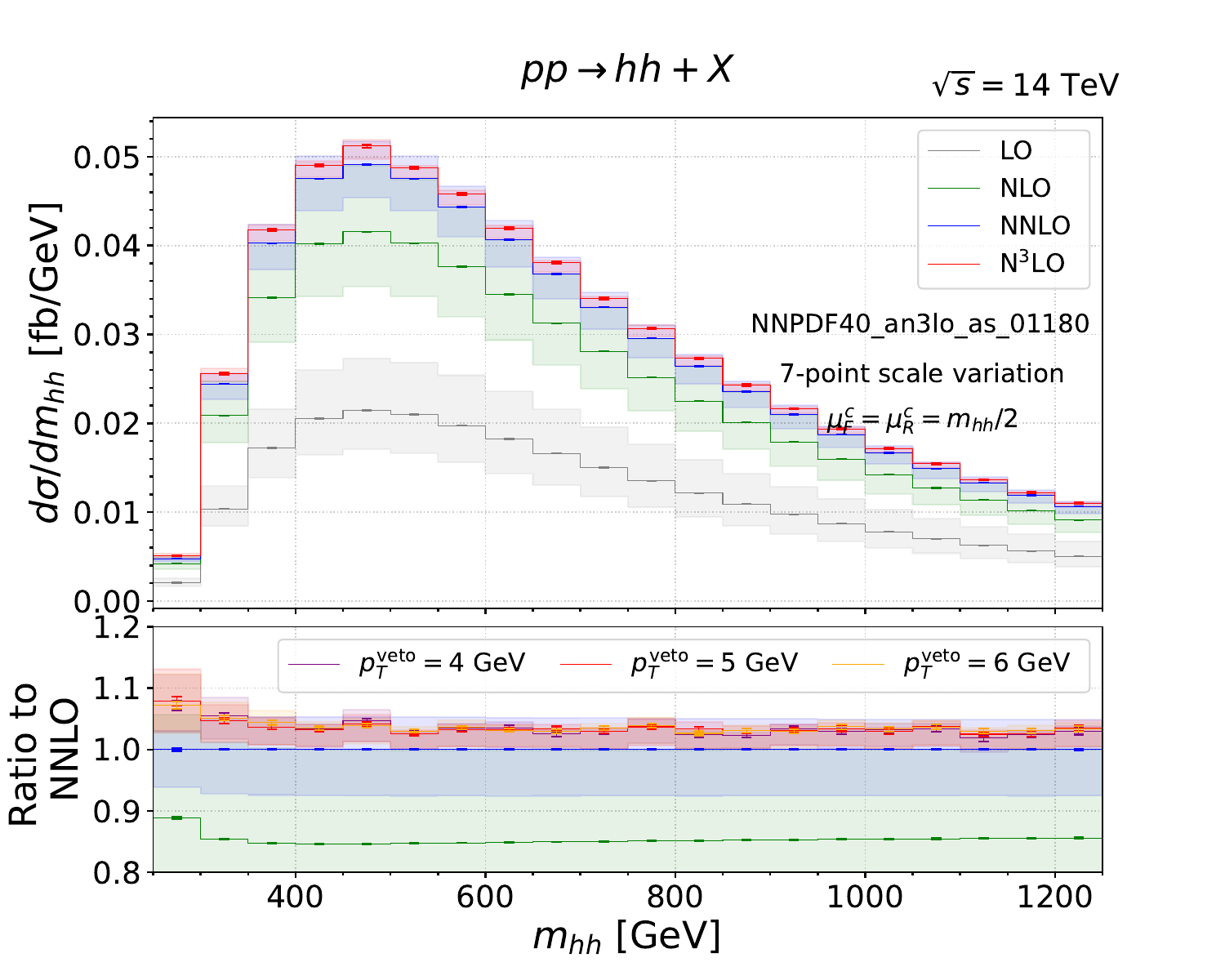}
    \caption{}
    \label{fig:m_hh-a+b+c}
  \end{subfigure}%\hspace{-0.8cm}
  \begin{subfigure}[b]{0.5\textwidth}
    \includegraphics[width=\textwidth]{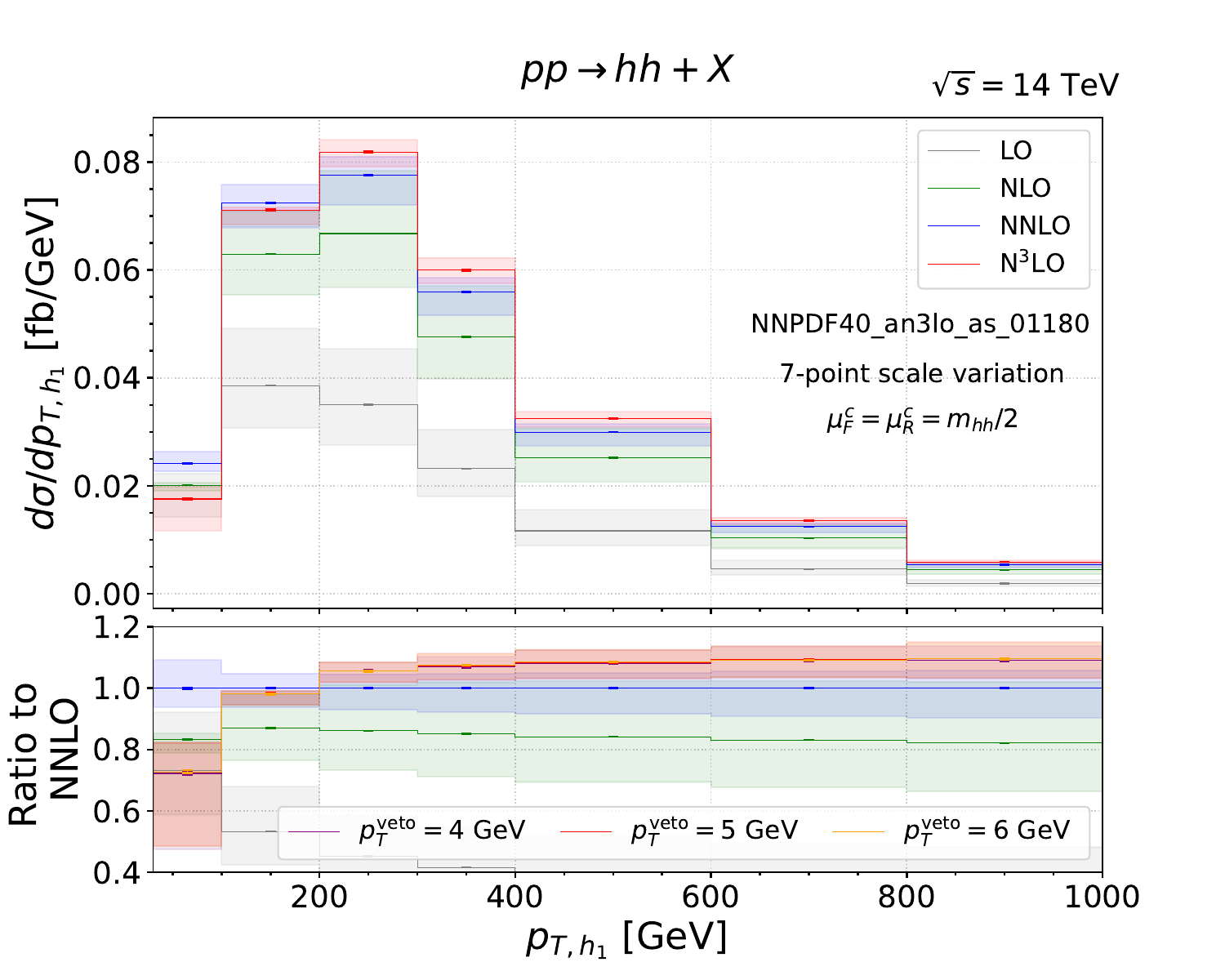}
    \caption{}
    \label{fig:pt_h1st-a+b+c}
  \end{subfigure}
  \begin{subfigure}[b]{0.5\textwidth}
    \includegraphics[width=\textwidth]{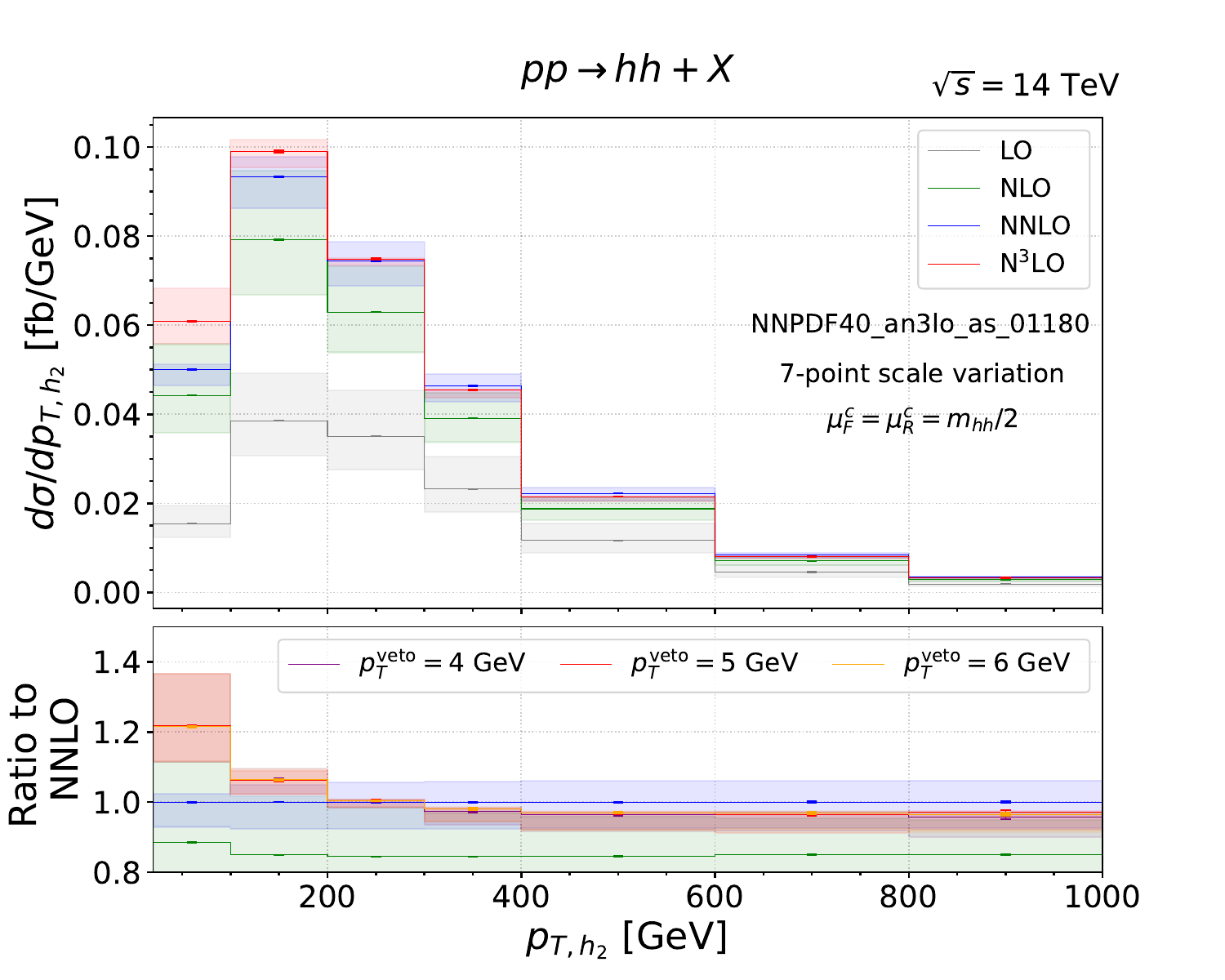}
    \caption{}
    \label{fig:pt_h2nd-a+b+c}
  \end{subfigure}%\hspace{-0.8cm}
  \begin{subfigure}[b]{0.5\textwidth}
    \includegraphics[width=\textwidth]{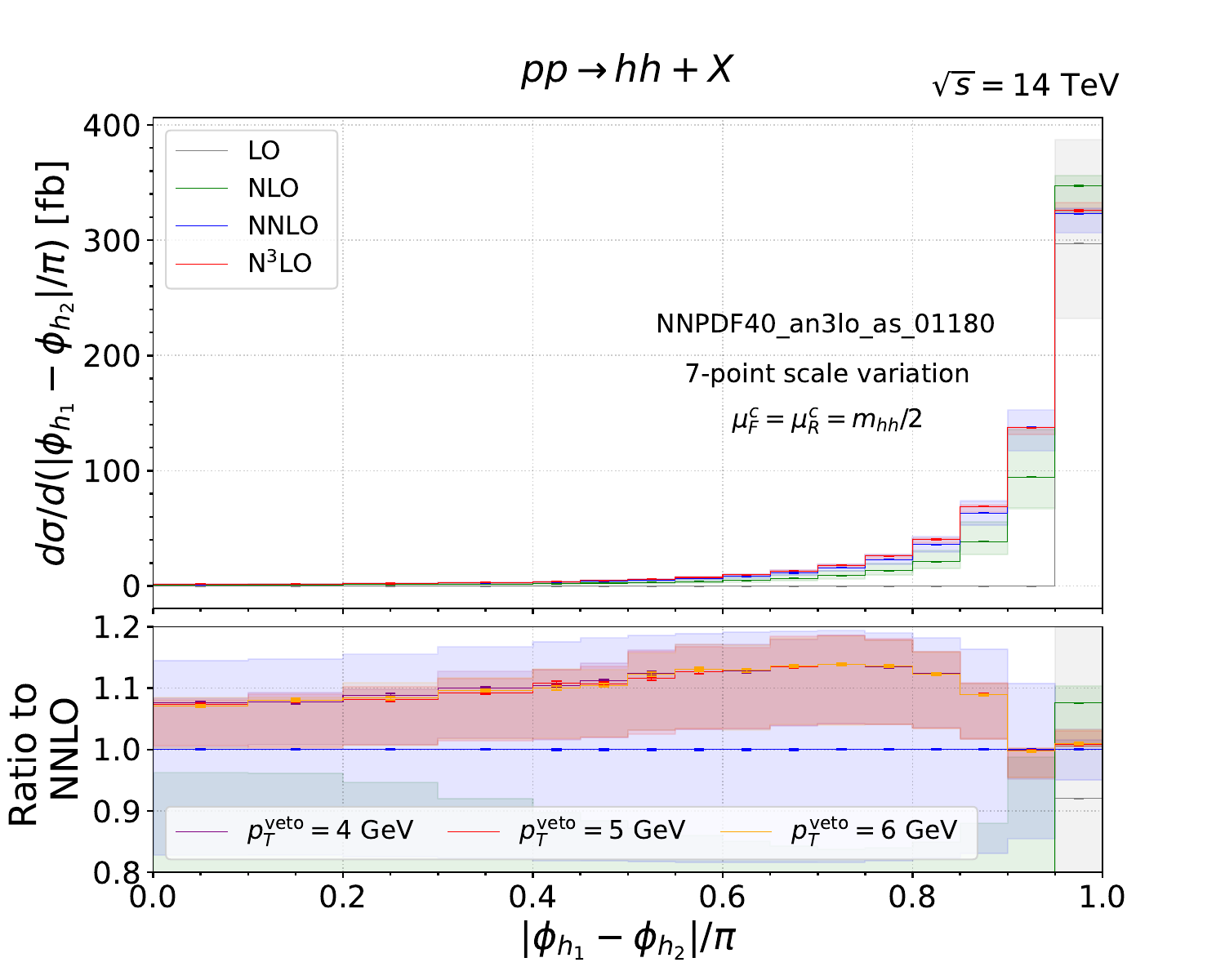}
    \caption{}
    \label{fig:abs(phi_h1st-phi_h2nd)_over_pi-a+b+c}
  \end{subfigure}
  \caption{Differential distributions for Higgs-boson pair production in the HTL from LO to $\mathrm{N^{3}LO}$ accuracy: 
(a) The invariant mass of the Higgs pair, $m_{hh}$.
(b) The transverse momentum of the leading-$p_{T}$ Higgs, $p_{T,h_1}$. 
(c) The transverse momentum of the subleading-$p_{T}$ Higgs, $p_{T,h_2}$.
(d) The azimuthal angle difference between the Higgs bosons, $|\phi_{h_1}-\phi_{h_2}|/\pi$.
The colored bands represent theoretical uncertainties from $7$-point scale variations. The bottom panels show the ratios with respect to NNLO for three different values of $p_{T}^{\rm veto}$.}
  \label{fig:HH-a+b+c-1}
\end{figure}

\begin{figure}[hbt!]
  %\centering
%  \hspace{-0.3cm}
  \begin{subfigure}[b]{0.5\textwidth}
    \includegraphics[width=\textwidth]{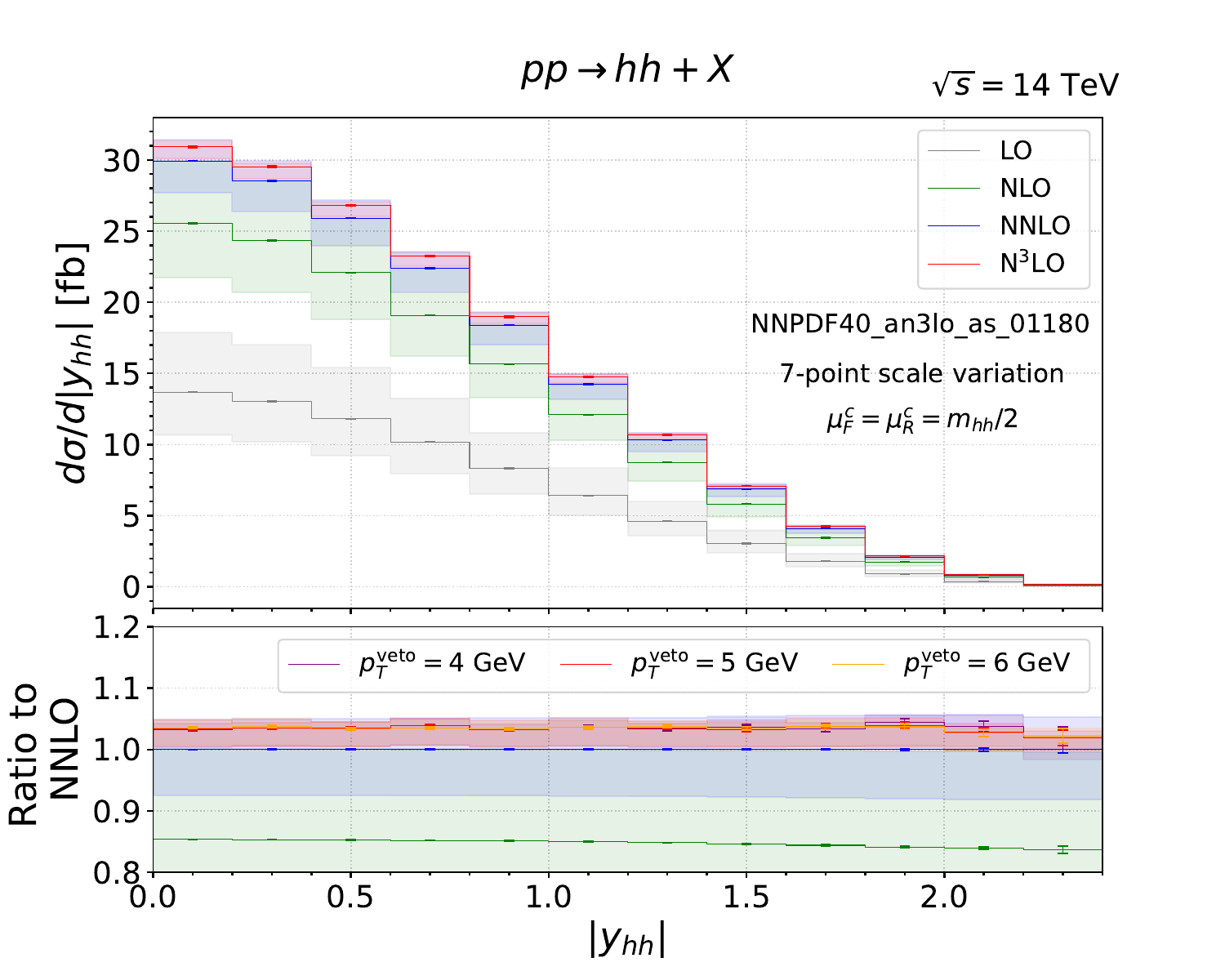}
    \caption{}
    \label{fig:abs(y_hh)-a+b+c}
  \end{subfigure}%\hspace{-0.8cm}
  \begin{subfigure}[b]{0.5\textwidth}
    \includegraphics[width=\textwidth]{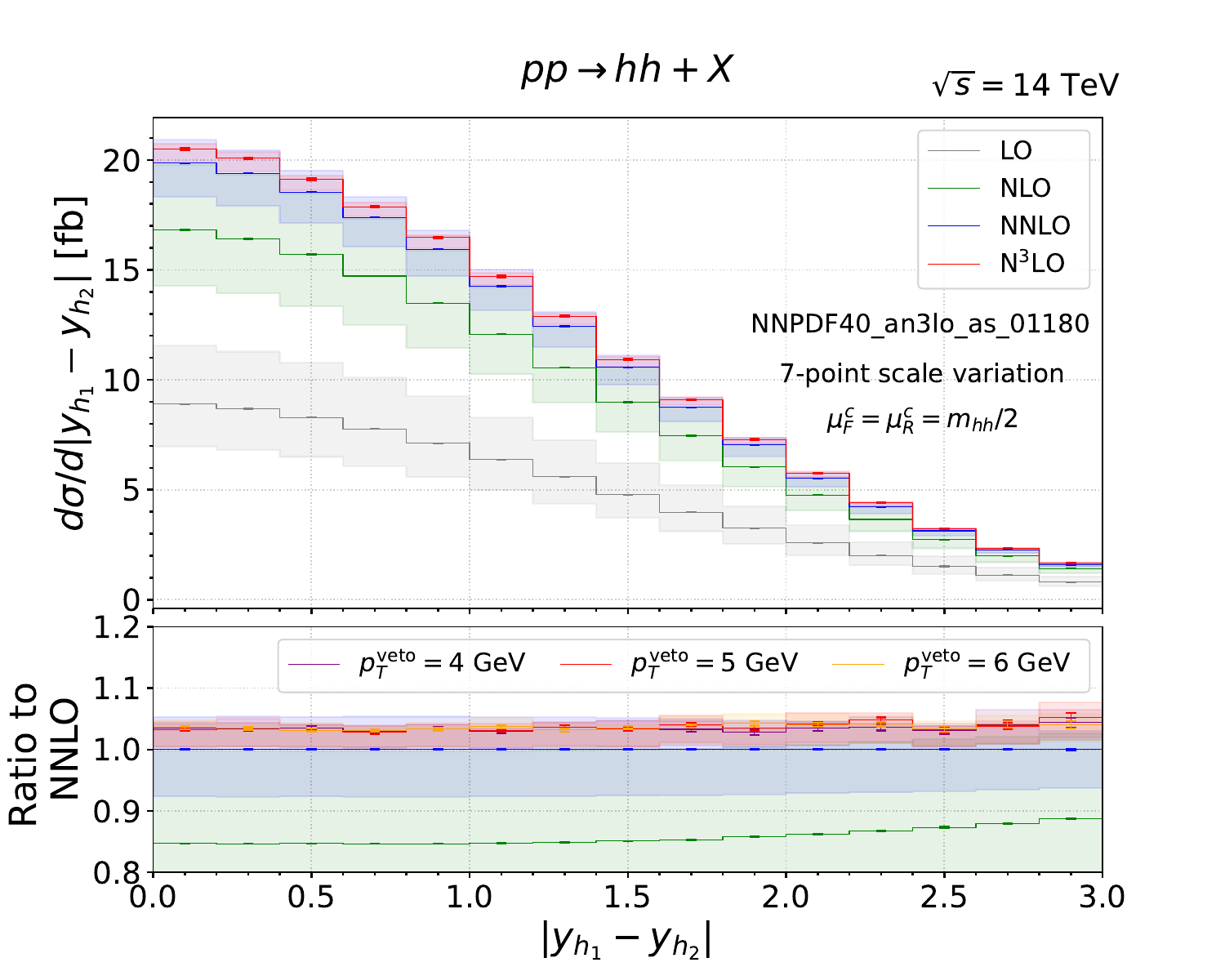}
    \caption{}
    \label{fig:abs(y_h1st-y_h2nd)-a+b+c}
  \end{subfigure}
  \begin{subfigure}[b]{0.5\textwidth}
    \includegraphics[width=\textwidth]{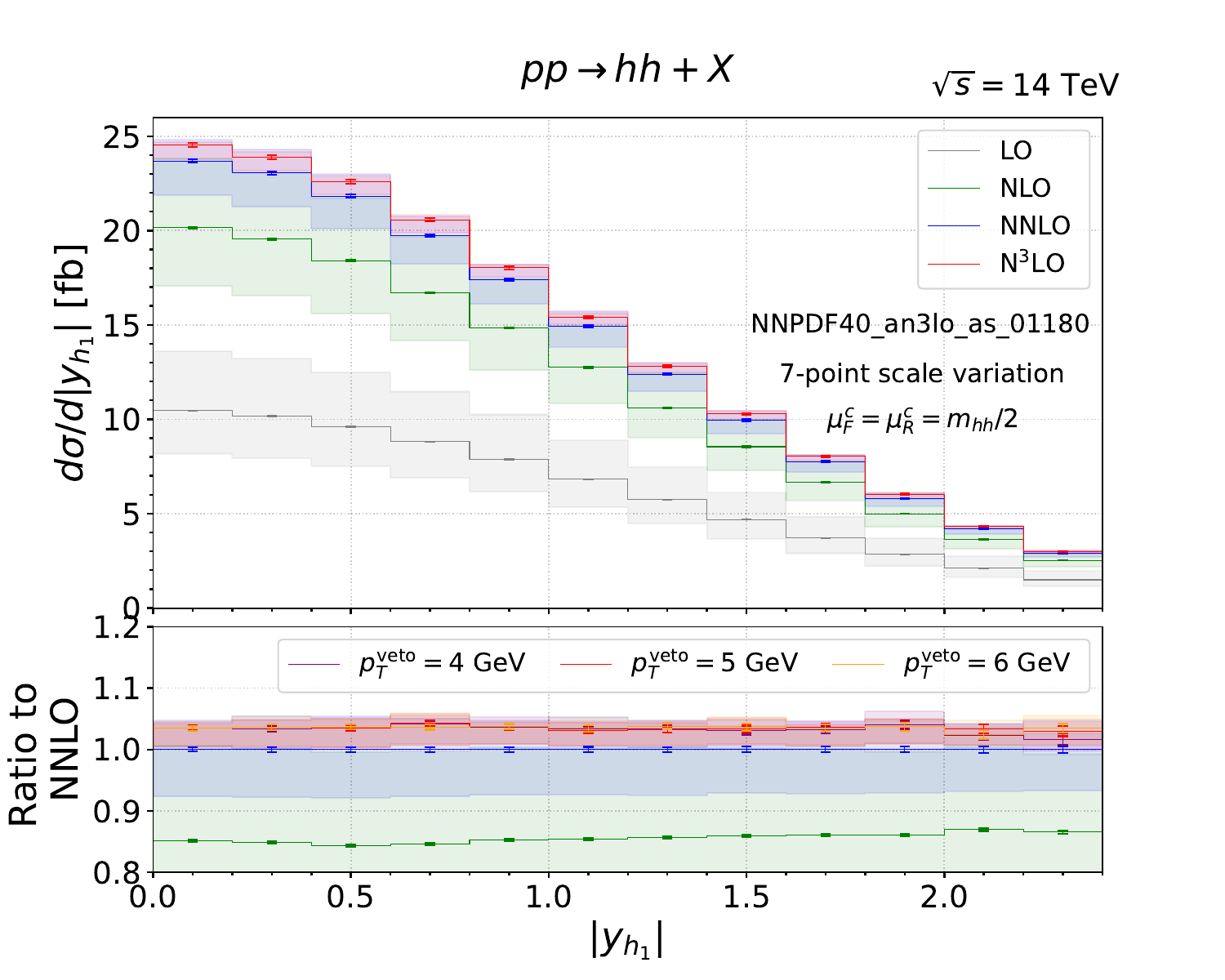}
    \caption{}
    \label{fig:abs(y_h1st)-a+b+c}
  \end{subfigure}%\hspace{-0.8cm}
  \begin{subfigure}[b]{0.5\textwidth}
    \includegraphics[width=\textwidth]{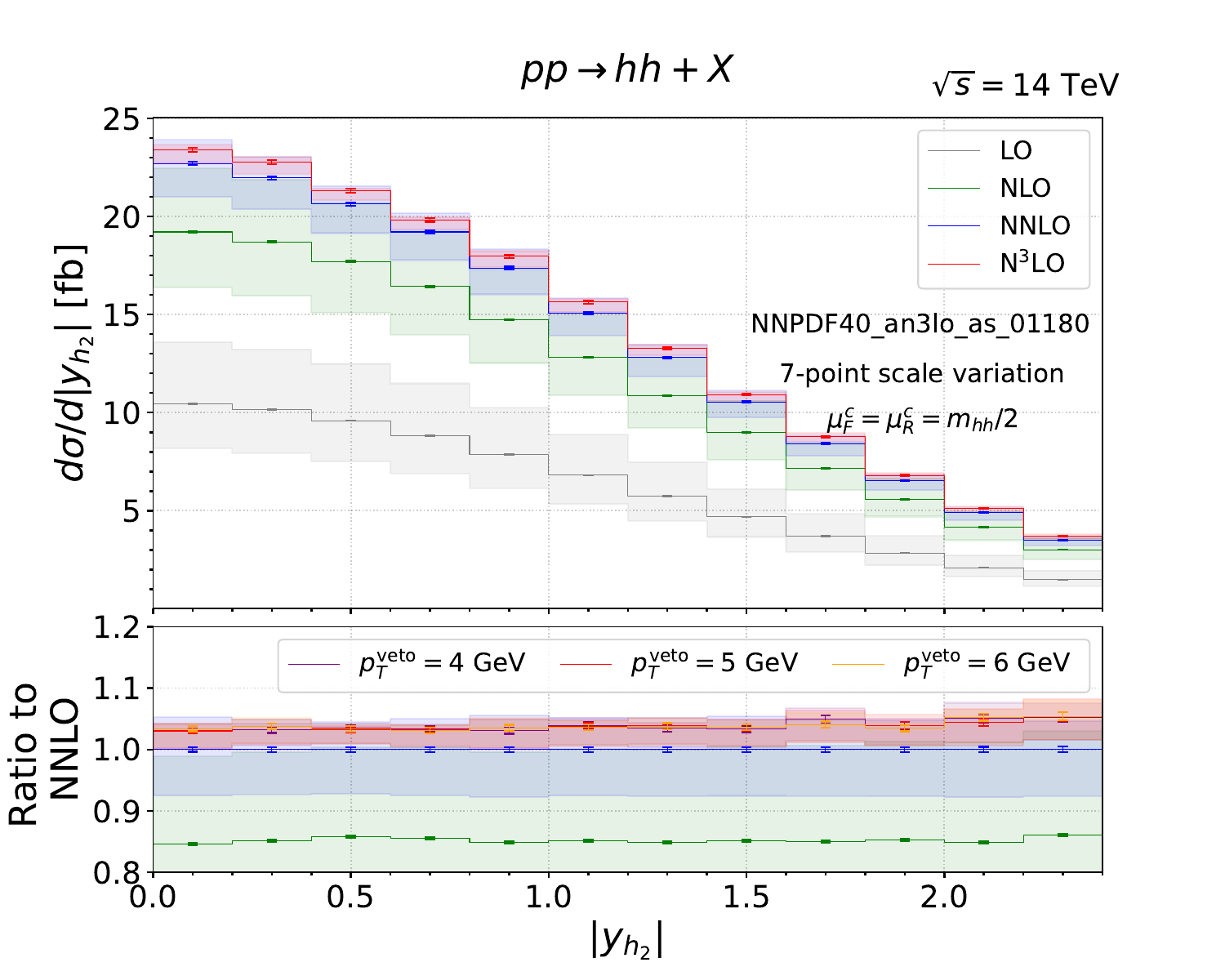}
    \caption{}
    \label{fig:abs(y_h2nd)-a+b+c}
  \end{subfigure}
  \caption{Differential distributions for Higgs-boson pair production in the HTL from LO to $\mathrm{N^{3}LO}$ accuracy: 
(a) The absolute rapidity of the Higgs-pair system, $|y_{hh}|$. 
(b) The absolute rapidity gap between the Higgs bosons, $|y_{h_1}-y_{h_2}|$. 
(c) The absolute rapidity of the leading-$p_{T}$ Higgs, $|y_{h_1}|$.
(d) The absolute rapidity of the subleading-$p_{T}$ Higgs, $|y_{h_2}|$. 
The colored bands represent theoretical uncertainties from $7$-point scale variations. The bottom panels show the ratios with respect to NNLO for three different values of $p_{T}^{\rm veto}$.}
  \label{fig:HH-a+b+c-2}
\end{figure}

We now turn to the discussion of the differential distributions. Figures~\ref{fig:HH-a+b+c-1} and \ref{fig:HH-a+b+c-2} show the LO to N$^3$LO differential cross sections in the fiducial region defined in eq.~\eqref{eq:fiducialcuts}. The distributions include:
\begin{itemize}
\item the invariant mass of the Higgs pair, $m_{hh}$ (figure~\ref{fig:m_hh-a+b+c}); 
\item the transverse momentum of the leading-$p_{T}$ Higgs, $p_{T,h_{1}}$ (figure~\ref{fig:pt_h1st-a+b+c}); \item the transverse momentum of the subleading-$p_{T}$ Higgs, $p_{T,h_{2}}$ (figure~\ref{fig:pt_h2nd-a+b+c}); 
\item the azimuthal angle difference between the two Higgs bosons normalized by \mbox{$\pi$, $|\phi_{h_{1}}-\phi_{h_{2}}|/\pi$} (figure~\ref{fig:abs(phi_h1st-phi_h2nd)_over_pi-a+b+c}); 
\item the absolute rapidity of the Higgs-pair system, $|y_{hh}|$ (figure~\ref{fig:abs(y_hh)-a+b+c});
\item the absolute rapidity gap between the two Higgs bosons, $|y_{h_{1}}-y_{h_{2}}|$ (figure~\ref{fig:abs(y_h1st-y_h2nd)-a+b+c});
\item the absolute rapidity of the leading-$p_{T}$ Higgs, $|y_{h_{1}}|$ (figure~\ref{fig:abs(y_h1st)-a+b+c}); \item and the absolute rapidity of the subleading-$p_{T}$ Higgs, $|y_{h_{2}}|$ (figure~\ref{fig:abs(y_h2nd)-a+b+c}). 
\end{itemize}
In each plot, 
the colored bands represent the $7$-point scale uncertainties. The bottom panels show the ratios with respect to the central values of the NNLO distributions. We also display the ratios of N$^3$LO over NNLO for three different values of $p_{T}^{\mathrm{veto}}$ in the bottom panels, while only the curves corresponding to $p_{T}^{\rm veto}=5~\mathrm{GeV}$ are shown in the upper panels.

We begin with the di-Higgs invariant mass $m_{hh}$ distribution, whose shape is particularly sensitive to the value of $\lambda_{3h}$ due to a delicate cancellation between the trilinear Higgs self-coupling–dependent and –independent terms in the SM. From figure~\ref{fig:m_hh-a+b+c}, we observe that the peak of the distribution is located around 500 GeV at LO, and higher-order QCD corrections do not shift the peak position. A similar significant reduction of scale uncertainties, as observed for the fiducial cross sections, is found. The N$^3$LO band lies almost entirely within the NNLO band, except in the threshold region ($m_{hh}\to 2m_h$), where the scale uncertainties are largest. 
This feature can be attributed to the impact of the fiducial cuts, which affect predominantly the low-mass region,~\footnote{For instance, linear power corrections~\cite{Salam:2021tbm} are more enhanced in the threshold region.} compared to the inclusive case shown in figure 12b of ref.~\cite{Chen:2019fhs}. 
Interestingly, the $K$ factor, $d\sigma^{\mathrm{N}^3\mathrm{LO}}_{hh}/d\sigma^{\mathrm{NNLO}}_{hh}$, is generally flat, except near threshold. The N$^3$LO QCD corrections enhance the NNLO invariant-mass differential cross section from about $8\%$ in the threshold region to about $4\%$ in the tail region. This behavior is slightly different from the inclusive case studied in ref.~\cite{Chen:2019fhs}.

On the other hand, the N$^3$LO QCD corrections significantly affect the shapes of the transverse momentum distributions of the leading and subleading Higgs bosons, as shown in figures~\ref{fig:pt_h1st-a+b+c} and \ref{fig:pt_h2nd-a+b+c}, respectively. In particular, in the small $p_{T}$ region, the fixed-order predictions are not stable: the N$^3$LO QCD corrections to the $p_{T,h_{1}}$ spectrum are negative with a magnitude larger than the NNLO ones, and the scale uncertainties at $\rm N^3LO$ exceed those at NNLO. These behaviors are caused by the additional parton emission that has a small transverse momentum (mainly from class-a contribution), and thus a $q_T$ resummation is necessary to stabilize the theoretical predictions in this region. In the intermediate and large $p_T$ regions above 100 GeV, the fixed-order predictions remain valid. The N$^3$LO corrections modify the shapes of the NNLO distributions, as can be clearly seen from the non-flat $K$ factors in the bottom panels of figures~\ref{fig:pt_h1st-a+b+c} and \ref{fig:pt_h2nd-a+b+c}.

The azimuthal angle difference distribution, $|\phi_{h_{1}}-\phi_{h_{2}}|/\pi$, shown in figure \ref{fig:abs(phi_h1st-phi_h2nd)_over_pi-a+b+c}, peaks in the back-to-back configuration ($|\phi_{h_{1}}-\phi_{h_{2}}|/\pi\to 1$). Since all LO events are located at $|\phi_{h_{1}}-\phi_{h_{2}}|/\pi=1$, an N$^k$LO calculation yields only N$^{k-1}$LO accuracy in bins away from this value. The $K$ factors are also non-uniform, with the largest N$^3$LO QCD corrections exceeding $10\%$ around $|\phi_{h_{1}}-\phi_{h_{2}}|/\pi\simeq 0.7$. Excellent convergence between NNLO and N$^3$LO is observed for the entire distribution. The exact N$^3$LO QCD corrections for both the $p_T$ and azimuthal-angle-difference distributions differ from the approximate N$^3$LO results given in ref.~\cite{Chen:2019fhs}, highlighting the importance of including the exact computation for accurate predictions of these observables.

All the absolute rapidity distributions shown in figure~\ref{fig:HH-a+b+c-2} peak around zero. The N$^3$LO $K$ factors are nearly independent of the rapidity bins, in good agreement with the approximate N$^3$LO results given in ref.~\cite{Chen:2019fhs}. As in the phase-space-integrated case,
the N$^3$LO corrections are around $4\%$ in all bins, and the N$^3$LO bands lie within the NNLO bands. This is not surprising, given the extremely flat N$^3$LO $K$ factors observed in the rapidity distributions of single Higgs production~\cite{Dulat:2018bfe}.

To estimate the theoretical uncertainty associated with $p_{T}^{\rm veto}$, we plot the $\mathrm{N^{3}LO}$ results for $p_{T}^{\mathrm{veto}}=4~\mathrm{GeV}$, $5~\mathrm{GeV}$, and $6~\mathrm{GeV}$ in the bottom panels. These panels confirm that the uncertainty from varying $p_{T}^{\rm veto}$ is negligible compared to the scale uncertainty at $\mathrm{N^{3}LO}$, further validating the $p_{T}^{\rm veto}$ plateau region.

\subsection{Top-quark mass effects\label{sec:n3lomt}}

As mentioned in section \ref{sec:introduction}, the effects of the finite top-quark mass cannot be neglected for Higgs-boson pair production through gluon fusion. We therefore combine our HTL results with the full-$m_t$-dependent NLO QCD results (denoted as ``NLO$_{m_t}$'') in this subsection. Ref.~\cite{Chen:2019fhs} proposes three schemes for this combination, corresponding to different working assumptions:
\begin{align}
\mathrm{N}^k \mathrm{LO} \oplus \mathrm{NLO}_{m_t}\quad :& \quad d \sigma_{hh}^{\mathrm{N}^k \mathrm{LO} \oplus \mathrm{NLO}_{m_{t}}}=d \sigma_{hh}^{\mathrm{NLO}_{m_t}}+d\sigma_{hh}^{\mathrm{N}^k\mathrm{LO}}-d\sigma_{hh}^{\mathrm{NLO}}\,,\nonumber\\
\mathrm{N}^k \mathrm{LO}_{\mathrm{B}-\mathrm{i}} \oplus \mathrm{NLO}_{m_t}\quad :& \quad d \sigma_{hh}^{\mathrm{N}^k\mathrm{LO}_{\mathrm{B}-\mathrm{i}} \oplus \mathrm{NLO}_{m_{t}}}=d \sigma_{hh}^{\mathrm{NLO}_{m_t}}+\left(d\sigma_{hh}^{\mathrm{N}^k\mathrm{LO}}-d\sigma_{hh}^{\mathrm{NLO}}\right)\frac{d \sigma_{hh}^{\mathrm{LO}_{m_t}}}{d \sigma_{hh}^{\mathrm{LO}}}\,,\nonumber\\
\mathrm{N}^k \mathrm{L O} \otimes \mathrm{NLO}_{m_t}\quad :& \quad d \sigma_{hh}^{\mathrm{N}^k \mathrm{LO} \otimes \mathrm{NLO}_{m_{t}}}=
d \sigma_{hh}^{\mathrm{N}^k \mathrm{LO}}
\frac{d \sigma_{hh}^{\mathrm{NLO}_{m_t}}}{d \sigma_{hh}^{\mathrm{NLO}}}\,,
\label{eq:mtsigma}
\end{align}
where $d \sigma_{hh}^{\mathrm{(N)LO}_{m_t}}$ denotes the (N)LO differential cross section with full top-quark mass dependence. 
In the first scheme,  labeled $\mathrm{N}^k \mathrm{LO} \oplus \mathrm{NLO}_{m_t}$, the QCD corrections beyond NLO computed in the HTL are combined additively with the full top-mass NLO result. 
In the second scheme, $\mathrm{N}^k \mathrm{LO}_{\mathrm{B}-\mathrm{i}} \oplus \mathrm{NLO}_{m_t}$, a differential reweighting of the HTL N$^k$LO contributions by the LO full-$m_t$-dependent results is performed; this is referred to as the Born-improved reweighting.
In the third scheme, $\mathrm{N}^k\mathrm{LO} \otimes \mathrm{NLO}_{m_t}$, the HTL predictions are reweighted by the NLO full-$m_t$ results \cite{Borowka:2016ypz,Borowka:2016ehy,Baglio:2018lrj,Davies:2019dfy,Baglio:2020wgt,Baglio:2020ini,Davies:2025qjr}.
In this study, the full $m_t$-dependent results are calculated using the public code \ggxy~\cite{Davies:2025qjr}, which implements the analytic expansions of two-loop amplitudes in various phase-space regions~\cite{Davies:2018ood,Davies:2018qvx,Davies:2023vmj}.~\footnote{Other analytic expansions of two-loop amplitudes with full $m_t$ dependence are provided in refs.~\cite{Bonciani:2018omm,Xu:2018eos,Wang:2020nnr,Bellafronte:2022jmo}.}
The scale variations of the combined cross sections are evaluated using eq. (3.4) in ref.~\cite{Chen:2019fhs}.

\subsubsection{Phase-space-integrated fiducial cross sections}

The phase-space-integrated fiducial cross sections for the three schemes introduced above are listed in table~\ref{tab:integrated-fiducial-cross-sections-reweighted}.
\begin{table}[hbt!]
    \centering
    \begin{tabular}{|c|c|c|c|}
    \hline
        &  $\sigma_{hh}^{\mathrm{N}^k \mathrm{LO} \oplus \mathrm{NLO}_{m_t}}$ [fb]  & $\sigma_{hh}^{\mathrm{N}^k \mathrm{LO}_{\mathrm{B}-\mathrm{i}} \oplus \mathrm{NLO}_{m_t}}$ [fb]  &   $\sigma_{hh}^{\mathrm{N}^k \mathrm{LO} \otimes \mathrm{NLO}_{m_t}}$ [fb]  \\
    \hline
    $k=1$  &  $28.44_{-12 \%}^{+14 \%}$  &   $28.44_{-12 \%}^{+14 \%}$ &   $28.44_{-12 \%}^{+14 \%}$  
     \\
    \hline
     $k=2$  &  $33.30_{-7.2 \%}^{+8.1 \%}$  &   $34.03_{-6.8 \%}^{+7.6 \%}$ &   $33.40_{-7.3 \%}^{+5.2 \%}$  
     \\
    \hline
     $k=3$   &  $34.56(4)_{-8.0\%}^{+6.8\%}$ &   $35.47(4)^{+6.6\%}_{-9.2\%}$ &   $34.68(4)^{+1.4\%}_{-2.9\%}$ 
     \\
    \hline
    \end{tabular}
    \caption{Phase-space-integrated fiducial cross sections (in units of fb) at the LHC with a center-of-mass energy of $\sqrt{s}=14~\rm TeV$ for three schemes combining finite top-quark-mass effects. The percentages indicate the systematic uncertainties arising from scale variations according to eq.~\eqref{scale}. Values in parentheses denote the Monte Carlo statistical uncertainties.}
    \label{tab:integrated-fiducial-cross-sections-reweighted}
\end{table}
The NNLO QCD corrections enhance the NLO cross section by approximately $17\%$-$20\%$, depending on the combination scheme. In contrast, in all three schemes, the N$^3$LO corrections increase the NNLO cross sections by around $4\%$.
However, the scale uncertainties strongly depend on the scheme. In particular, except for the N$^3$LO$\otimes$NLO$_{m_t}$ scheme, no evident reduction of scale uncertainties is observed when going from NNLO to N$^3$LO accuracy. For instance, in the $\mathrm{N}^k \mathrm{LO} \oplus \mathrm{NLO}_{m_t}$ scheme, the scale uncertainties are almost the same at NNLO and N$^3$LO. In the $\mathrm{N}^k \mathrm{LO}_{\mathrm{B}-\mathrm{i}} \oplus \mathrm{NLO}_{m_t}$ scheme, the scale uncertainties are even slightly larger at N$^3$LO than at NNLO. This behavior differs from the inclusive case shown in table 4 of ref.~\cite{Chen:2019fhs}, where a clear reduction of scale uncertainties is observed in all three schemes when going from NNLO to N$^3$LO at $\sqrt{s}=14$ TeV. This difference arises because the absence of finite top-quark-mass corrections at NNLO and N$^3$LO prevents the scale cancellations that occur in the first two schemes for fiducial cross sections. In the last scheme, since different perturbative orders in the HTL are assigned the same reweighting factor, the scale variations are, by construction, identical to those in the HTL case. In the following, when discussing differential distributions, we restrict ourselves to presenting results obtained using the $\mathrm{N}^k \mathrm{L O} \otimes \mathrm{NLO}_{m_t}$ scheme, which can be argued to be more perturbatively stable to the other two schemes.

\subsubsection{Differential distributions}

\begin{figure}[hbt!]
%  \centering
%\hspace{-0.3cm}
  \begin{subfigure}[b]{0.5\textwidth}
    \includegraphics[width=\textwidth]{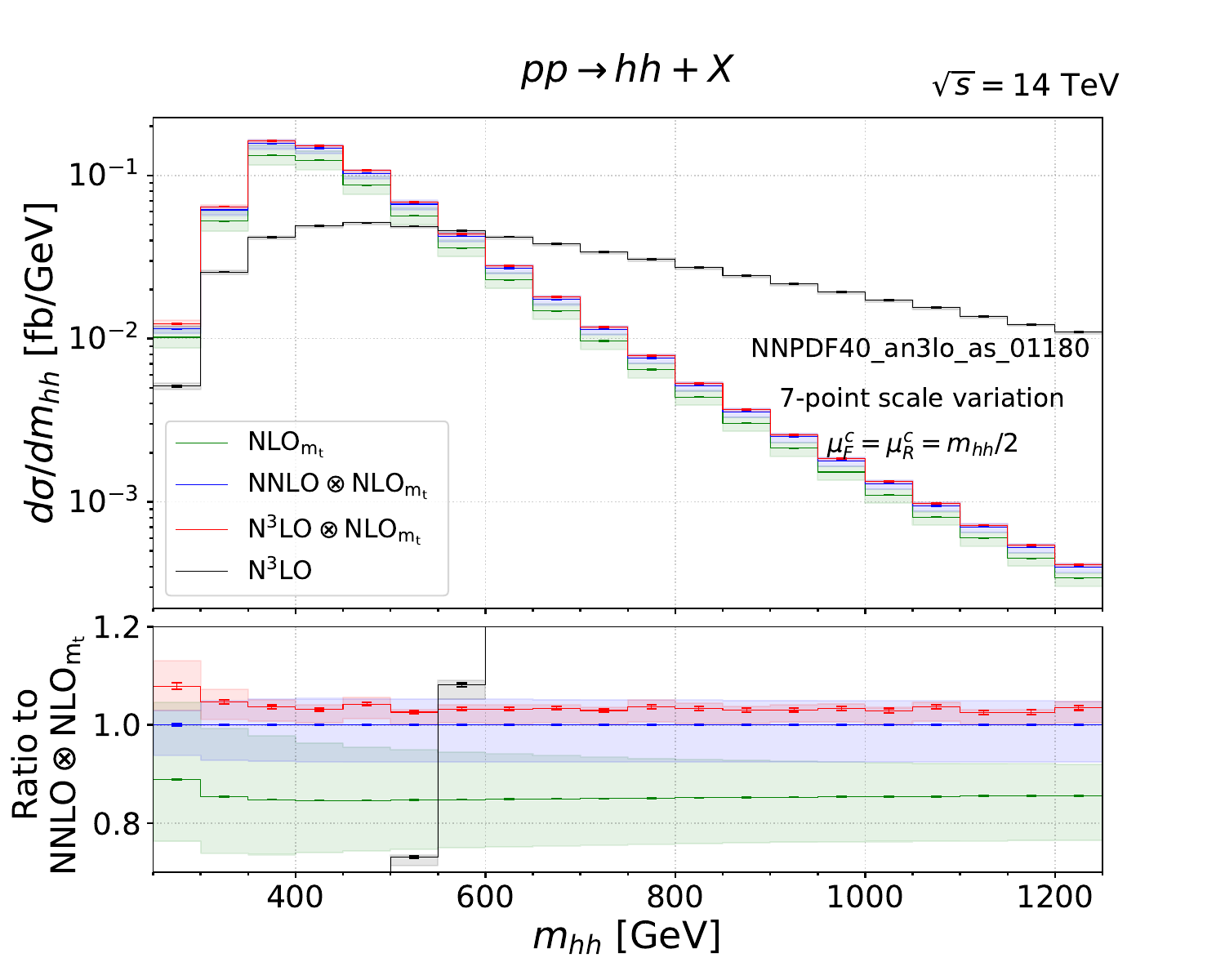}
    \caption{}
    \label{fig:m_hh-reweighted}
  \end{subfigure}%\hspace{-0.8cm}
  \begin{subfigure}[b]{0.5\textwidth}
    \includegraphics[width=\textwidth]{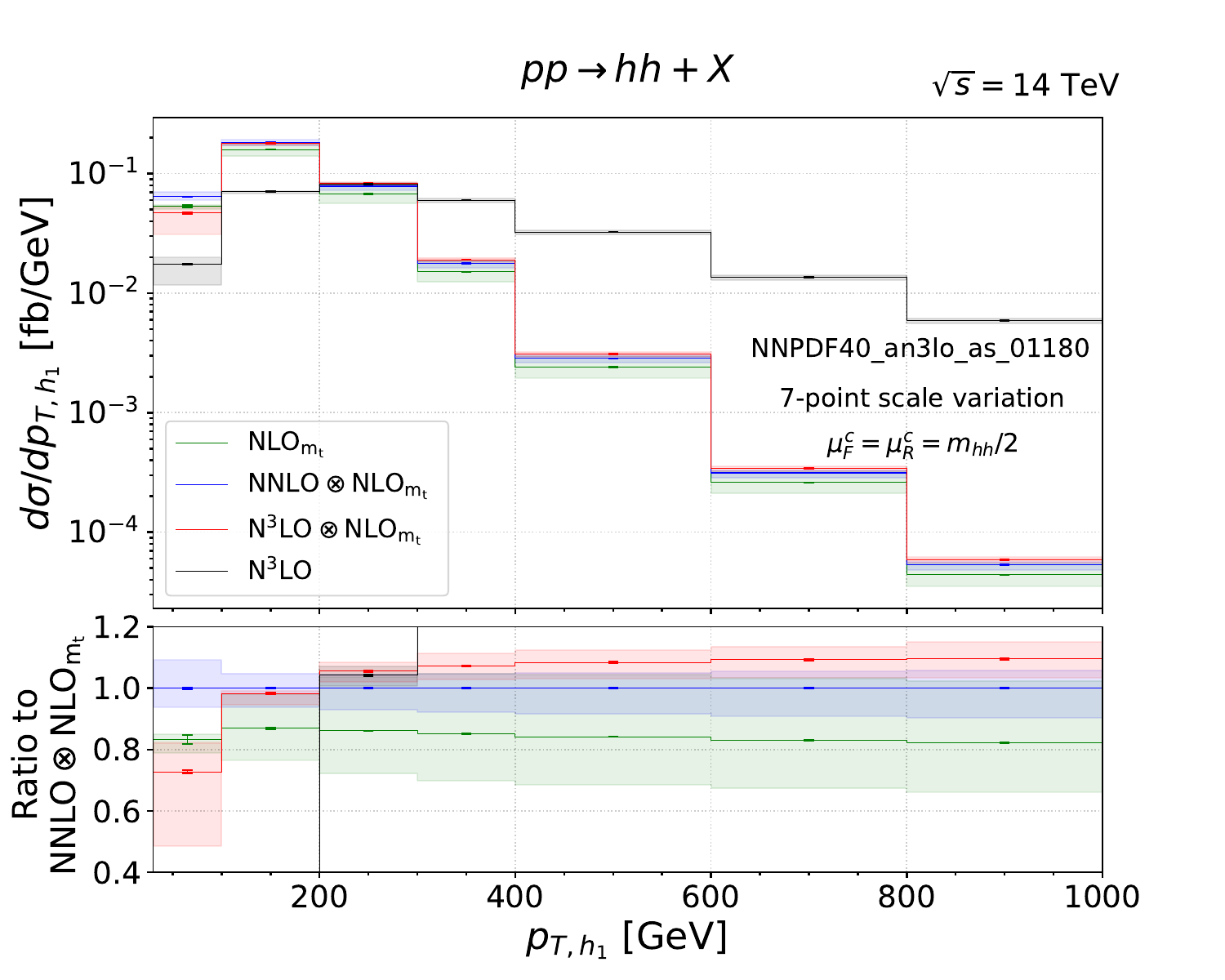}
    \caption{}
    \label{fig:pt_h1st-reweighted}
  \end{subfigure}
  \begin{subfigure}[b]{0.5\textwidth}
    \includegraphics[width=\textwidth]{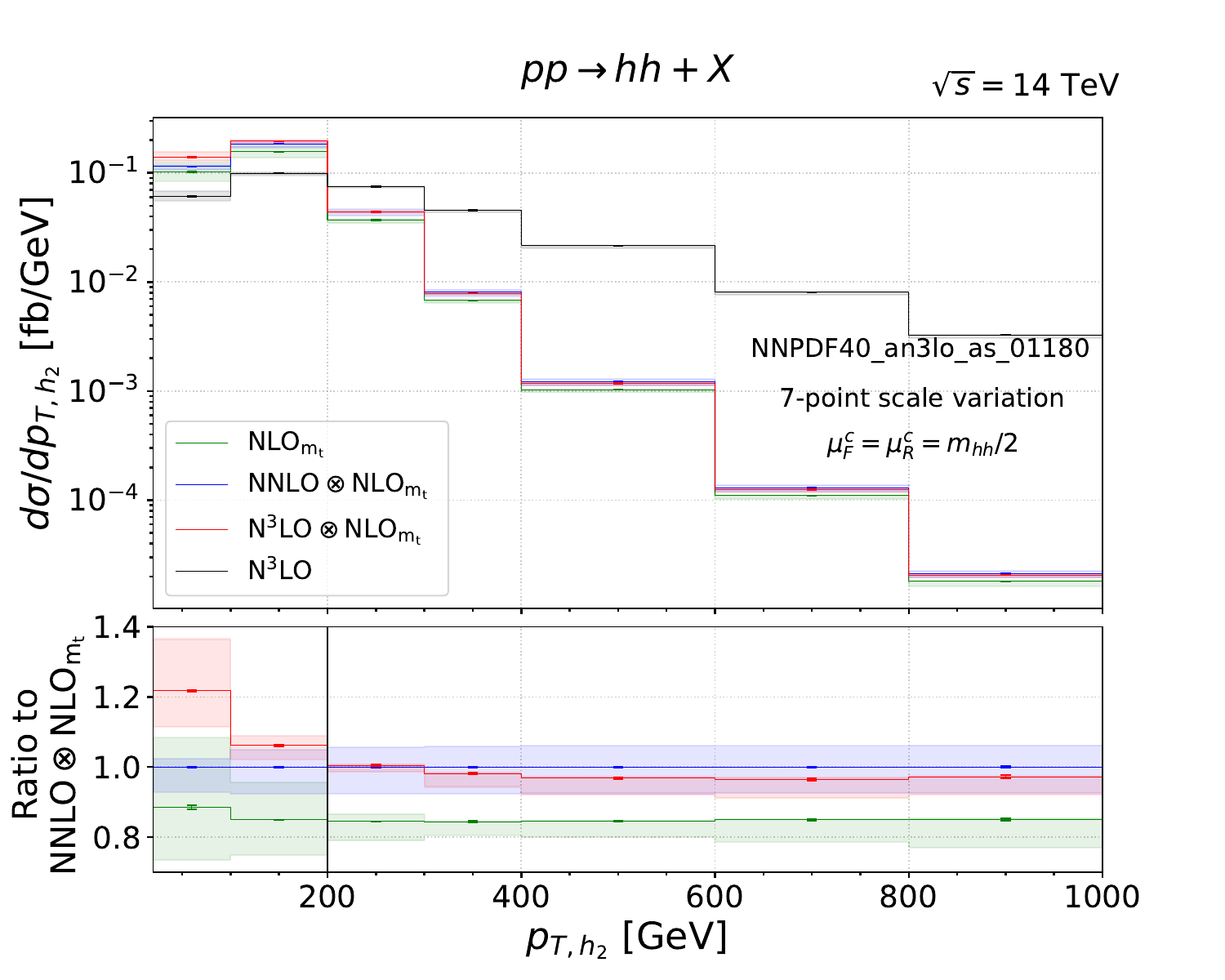}
    \caption{}
    \label{fig:pt_h2nd-reweighted}
  \end{subfigure}%\hspace{-0.8cm}
  \begin{subfigure}[b]{0.5\textwidth}
    \includegraphics[width=\textwidth]{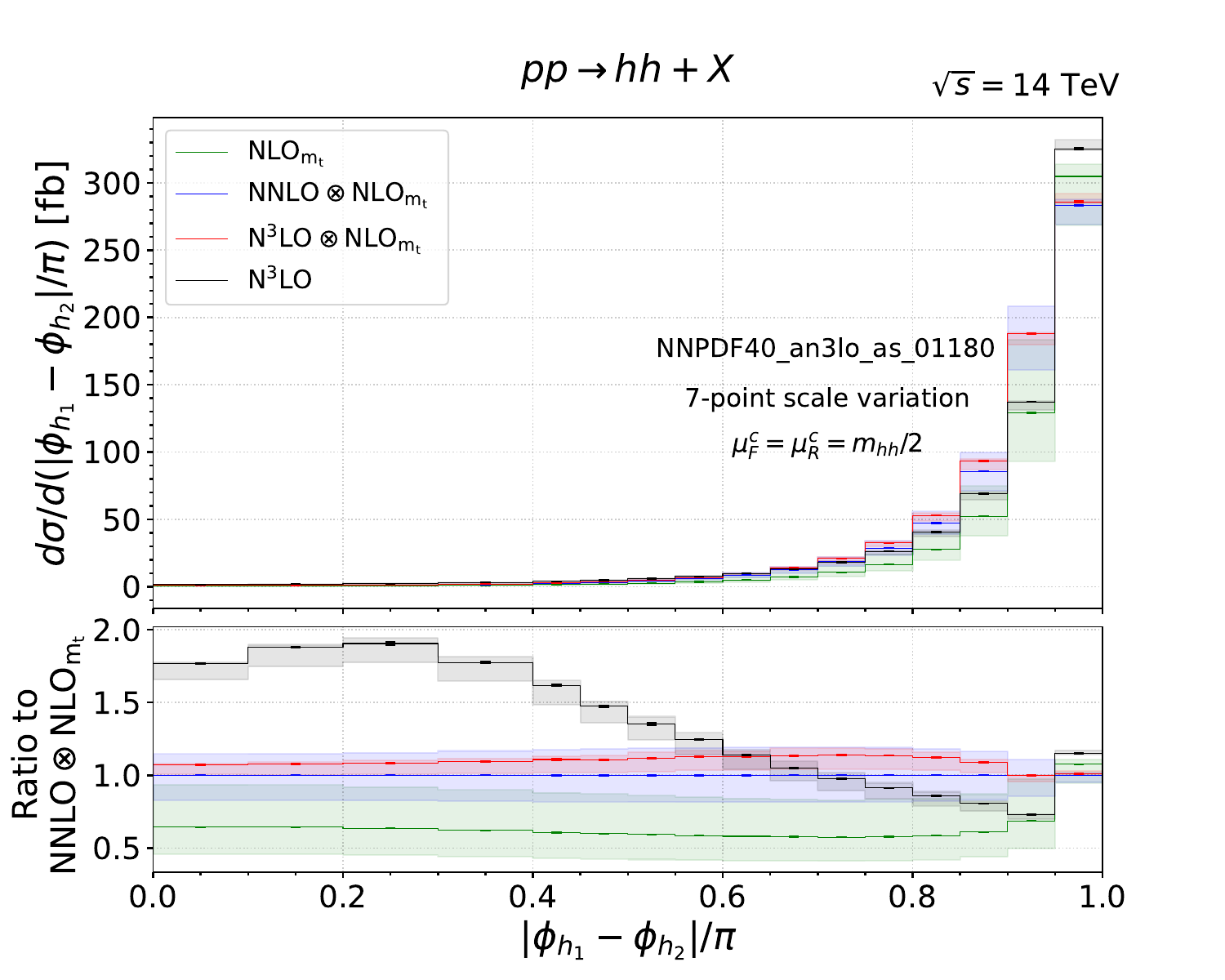}
    \caption{}
    \label{fig:abs(phi_h1st-phi_h2nd)_over_pi-reweighted}
  \end{subfigure}
  \caption{Differential distributions for Higgs-boson pair production in the $\mathrm{N}^k \mathrm{LO} \otimes \mathrm{NLO}_{m_t}$ scheme: 
(a) Invariant mass of the Higgs pair, $m_{hh}$;
(b) Transverse momentum of the leading-$p_{T}$ Higgs, $p_{T,h_1}$; 
(c) Transverse momentum of the subleading-$p_{T}$ Higgs, $p_{T,h_2}$;
(d) Azimuthal-angle difference between the Higgs bosons, $|\phi_{h_1}-\phi_{h_2}|/\pi$.
The colored bands represent theoretical uncertainties from $7$-point scale variations. The bottom panels show the ratios with respect to $\mathrm{NLO}_{m_{t}}$.}
  \label{fig:HH-reweighted-1}
\end{figure}

\begin{figure}[hbt!]
%  \centering
%\hspace{-0.3cm}
  \begin{subfigure}[b]{0.5\textwidth}
    \includegraphics[width=\textwidth]{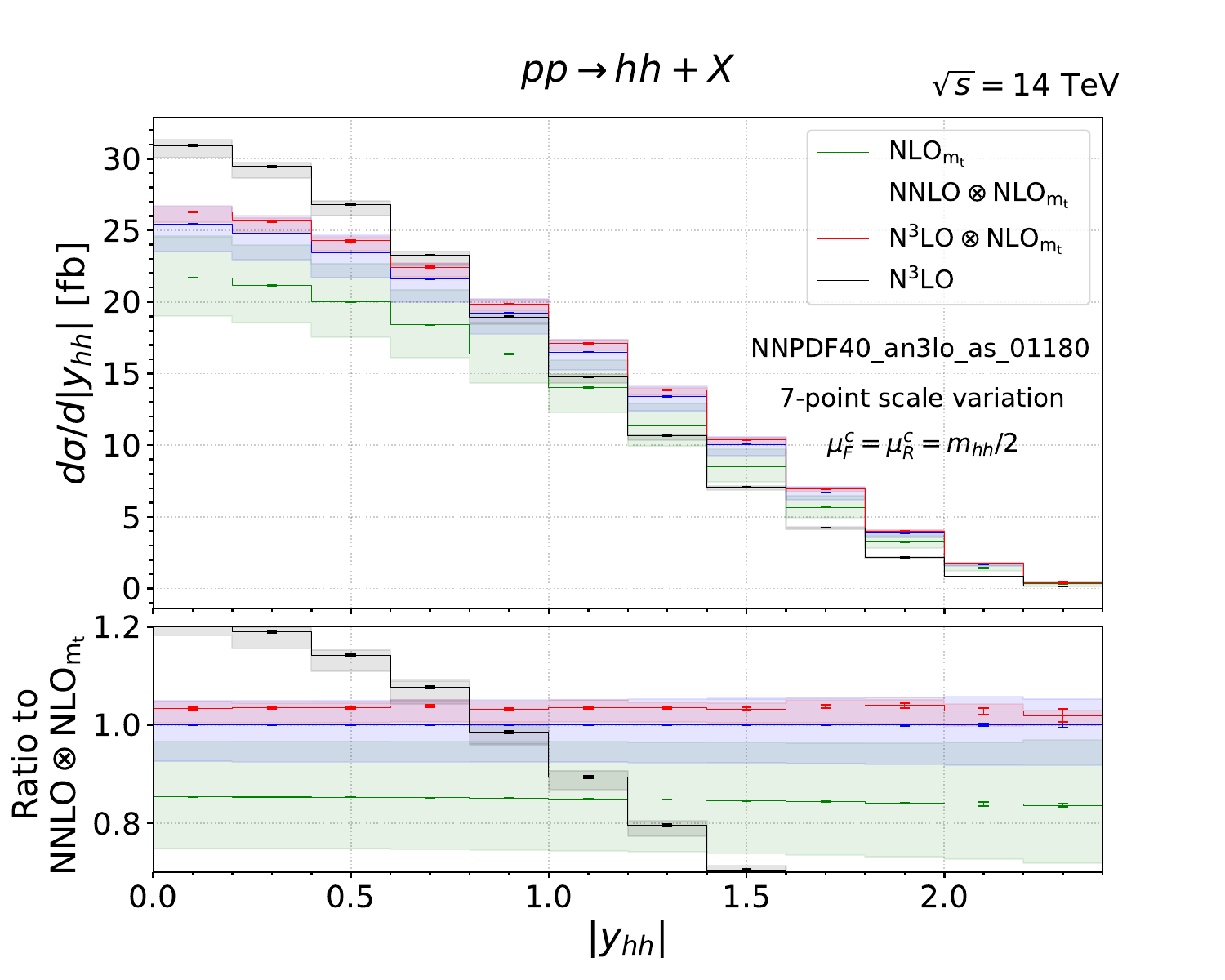}
    \caption{}
    \label{fig:abs(y_hh)-reweighted}
  \end{subfigure}%\hspace{-0.8cm}
  \begin{subfigure}[b]{0.5\textwidth}
    \includegraphics[width=\textwidth]{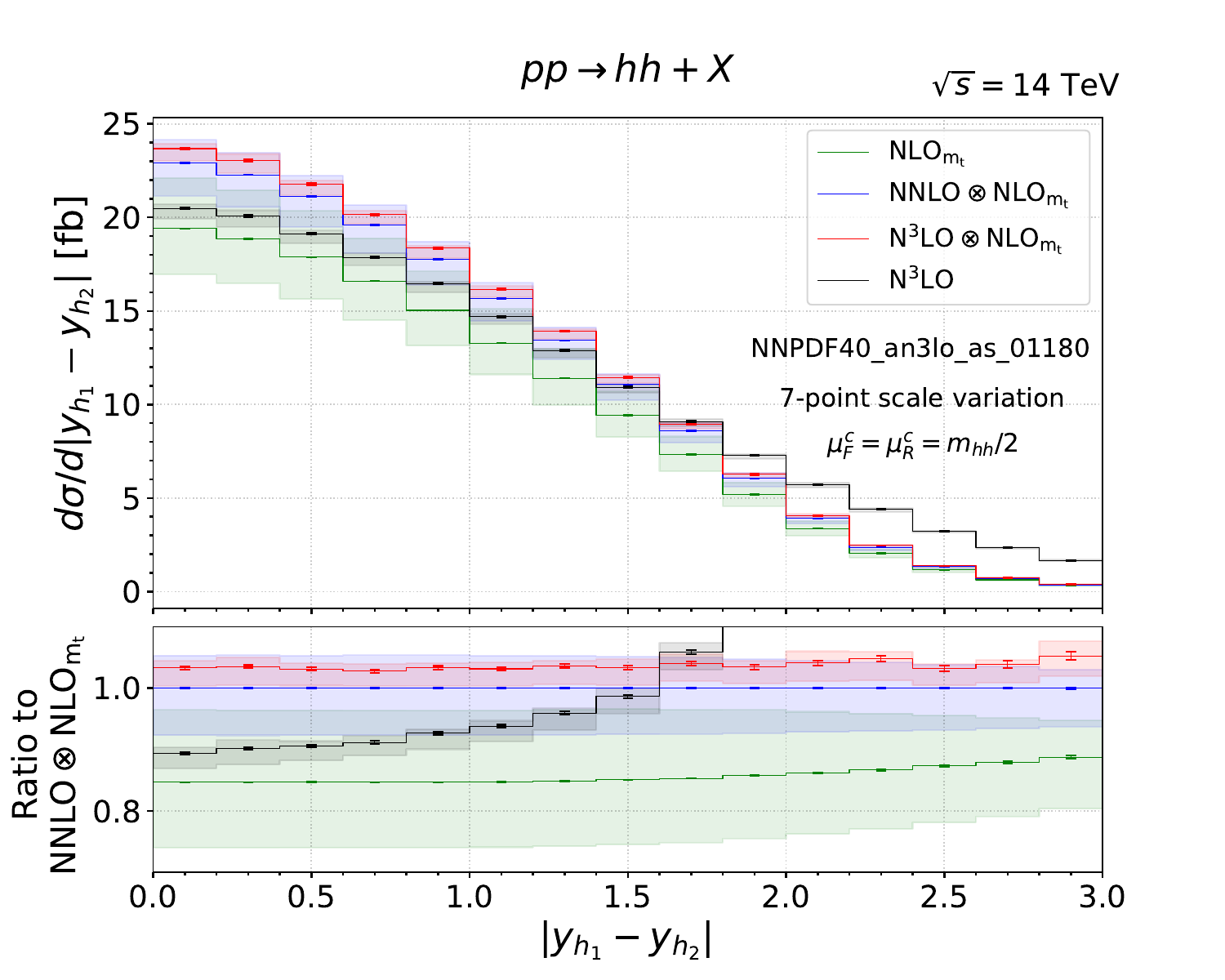}
    \caption{}
    \label{fig:abs(y_h1st-y_h2nd)-reweighted}
  \end{subfigure}
  \begin{subfigure}[b]{0.5\textwidth}
    \includegraphics[width=\textwidth]{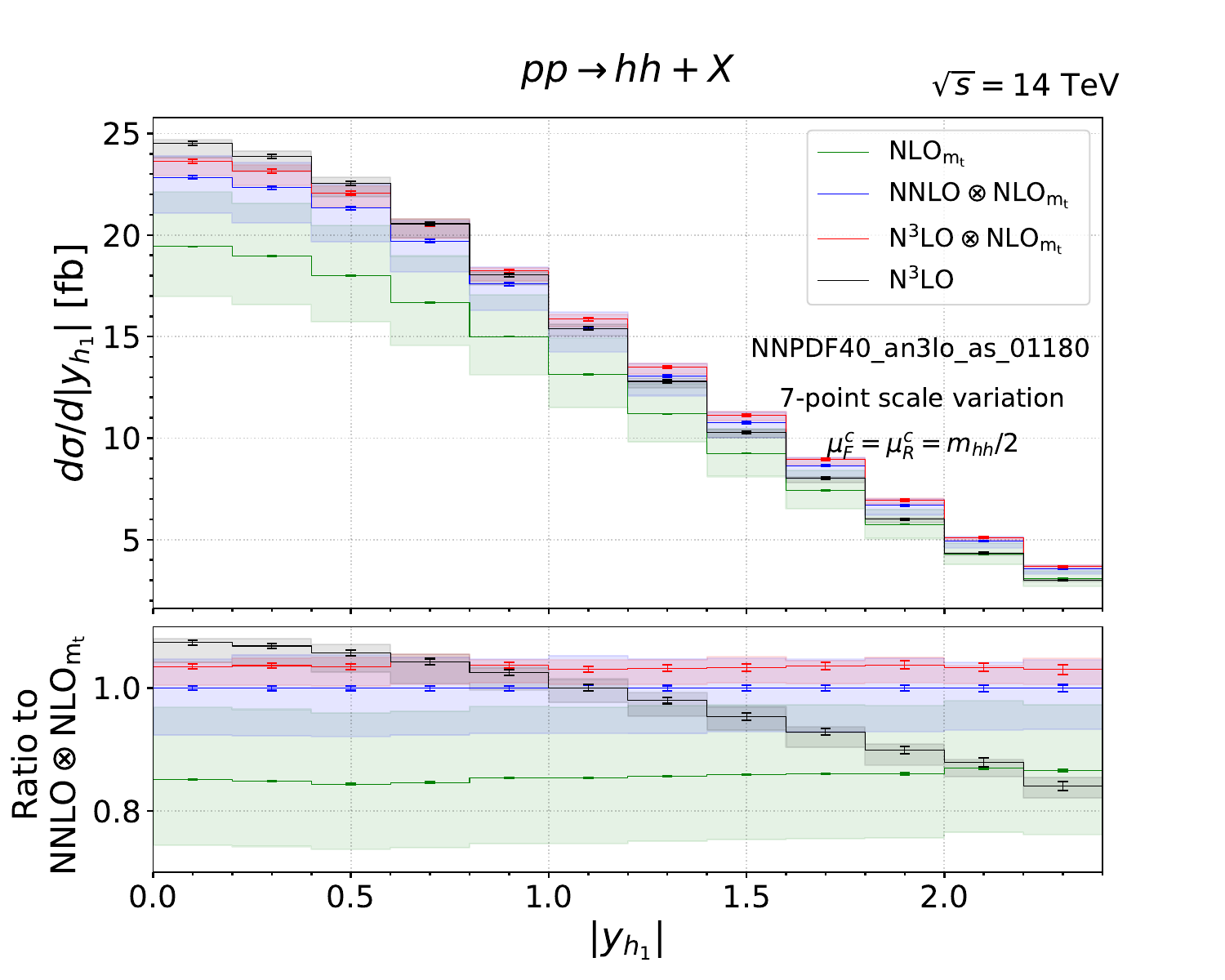}
    \caption{}
    \label{fig:abs(y_h1st)-reweighted}
  \end{subfigure}%\hspace{-0.8cm}
  \begin{subfigure}[b]{0.5\textwidth}
    \includegraphics[width=\textwidth]{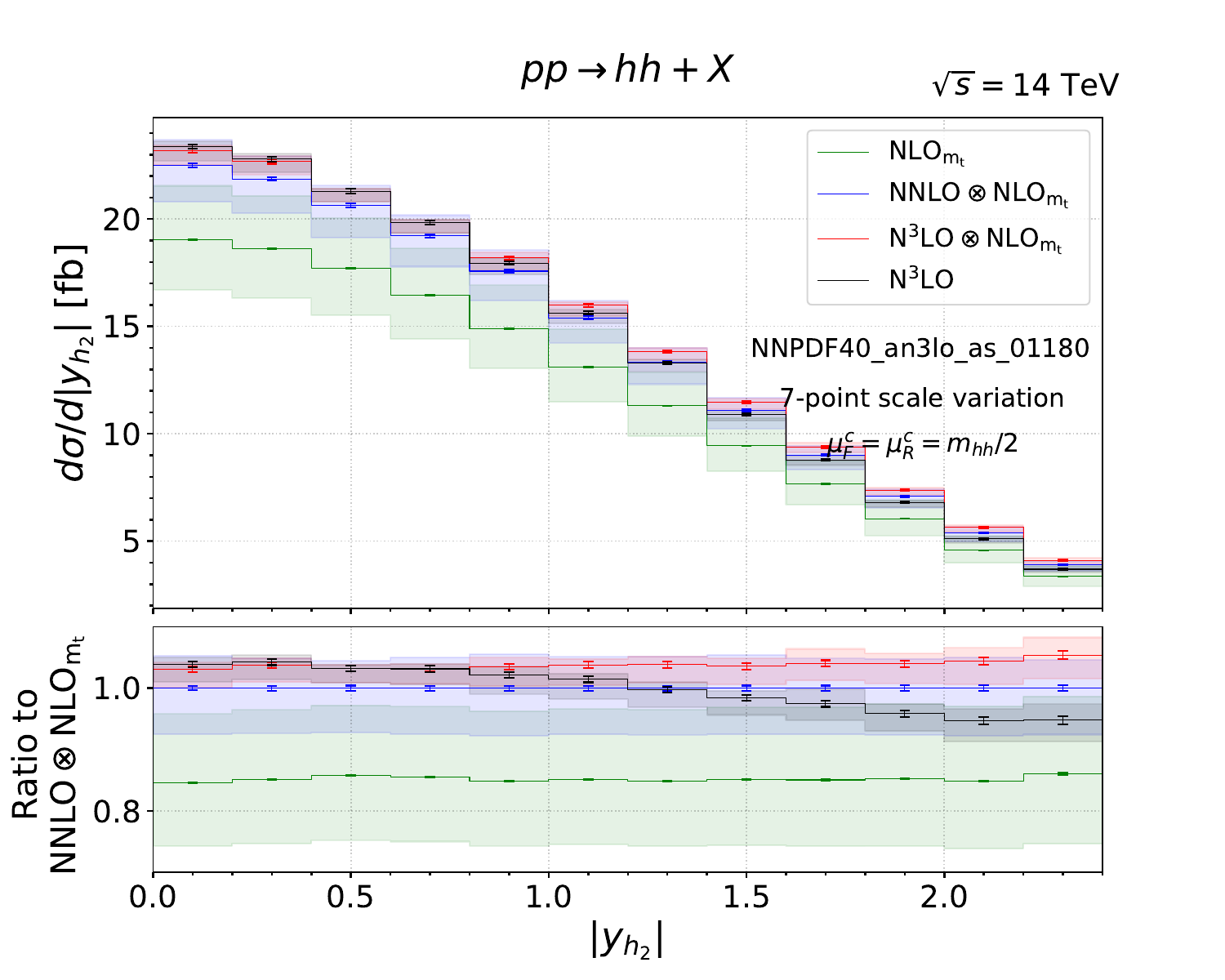}
    \caption{}
    \label{fig:abs(y_h2nd)-reweighted}
  \end{subfigure}
  \caption{Differential distributions for Higgs-boson pair production in the $\mathrm{N}^k \mathrm{LO} \otimes \mathrm{NLO}_{m_t}$ scheme: 
(a) Absolute rapidity of the Higgs-pair system, $|y_{hh}|$; 
(b) Absolute rapidity gap between the Higgs bosons, $|y_{h_1}-y_{h_2}|$; 
(c) Absolute rapidity of the leading-$p_{T}$ Higgs, $|y_{h_1}|$;
(d) Absolute rapidity of the subleading-$p_{T}$ Higgs, $|y_{h_2}|$. 
The colored bands represent theoretical uncertainties from $7$-point scale variations. The bottom panels show the ratios with respect to $\mathrm{NLO}_{m_{t}}$.}
  \label{fig:HH-reweighted-2}
\end{figure}

The differential distributions in the $\mathrm{N}^k \mathrm{LO} \otimes \mathrm{NLO}_{m_t}$ scheme are shown in figures~\ref{fig:HH-reweighted-1} and \ref{fig:HH-reweighted-2}. These correspond, respectively, to figures~\ref{fig:HH-a+b+c-1} and \ref{fig:HH-a+b+c-2} obtained in the HTL case (cf. section \ref{sec:mhhnomt}). For comparison, we also include the N$^3$LO HTL results, shown as black lines, in these figures.

For the invariant-mass distribution $m_{hh}$ shown in figure~\ref{fig:m_hh-reweighted}, we observe that, compared with the results obtained without finite top-quark-mass effects, the distribution in the threshold region ($m_{hh}\to 2m_h$) is enhanced by approximately a factor of two, while that in the large-$m_{hh}$ region is notably suppressed. The latter behavior is expected, since the contact interaction in eq.~\eqref{eq:effL} yields an overly hard spectrum once the condition $m_{hh}<2m_t$ is no longer satisfied. The former indicates that finite-$m_t$ effects are already essential at the lowest invariant mass, $m_{hh}=2m_h\simeq 250~\mathrm{GeV}$, which can be qualitatively understood from the sizable power corrections $m_h^2/m_t^2\simeq 0.5$. As a result, the overall shapes of the \mbox{$\mathrm{N}^k\mathrm{LO} \otimes \mathrm{NLO}_{m_t}$} distributions are significantly modified when full $m_t$ dependence is included at NLO. The corresponding $K$ factors are exactly the same as in the HTL case, as expected. The NLO$_{m_t}$ scale dependence, however, differs slightly from its HTL counterpart. In particular, the relative scale uncertainties in the former decrease mildly with increasing $m_{hh}$, which causes the $\mathrm{N}^3\mathrm{LO} \otimes \mathrm{NLO}_{m_t}$ uncertainty band to lie outside the $\mathrm{NLO}_{m_t}$ band over the entire kinematic range.

The $p_{T,h_{1}}$ and $p_{T,h_{2}}$ distributions shown in figures~\ref{fig:pt_h1st-reweighted} and \ref{fig:pt_h2nd-reweighted} exhibit behavior similar to that of the $m_{hh}$ distribution: a factor-of-two enhancement in the small-$p_T$ region, and overly hard HTL $p_T$ spectra in the tails compared to those including finite top-quark-mass corrections. At NLO$_{m_t}$, the scale uncertainties of the leading and subleading $p_T$ spectra show different trends. The fractional scale uncertainties in the $p_{T,h_{1}}$ distribution increase with $p_T$, whereas those in the $p_{T,h_{2}}$ distribution decrease with $p_T$.
This difference arises because events with a large leading-$p_T$ Higgs boson can originate from configurations involving a hard Higgs boson produced in association with one or more jets, accompanied by the bremsstrahlung of a soft Higgs boson.

For the azimuthal-angle difference distribution $|\phi_{h_{1}}-\phi_{h_{2}}|/\pi$ shown in figure~\ref{fig:abs(phi_h1st-phi_h2nd)_over_pi-reweighted}, finite-$m_t$ corrections suppress the distribution in the back-to-back region \mbox{($|\phi_{h_{1}}-\phi_{h_{2}}|/\pi\to 1$)} by more than $10\%$, and reduce the differential cross section in the collinear region ($|\phi_{h_{1}}-\phi_{h_{2}}|/\pi\to 0$) by about a factor of $1.7$. To qualitatively understand this behavior, we note that the dominant event configuration in the back-to-back region consists of a Higgs pair recoiling against soft jet(s), whereas in the collinear region the Higgs pair recoils against hard jet(s). The latter configuration pushes the partonic center-of-mass energy $\sqrt{\hat{s}}$ to values well above $2m_t$. In the intermediate region $0.6<|\phi_{h_{1}}-\phi_{h_{2}}|/\pi<0.95$, the N$^3$LO$\otimes$NLO$_{m_t}$ results become larger than the N$^3$LO HTL results, with the transition occurring around $|\phi_{h_{1}}-\phi_{h_{2}}|/\pi\simeq 0.6$.

For the $|y_{hh}|$ distribution shown in figure~\ref{fig:abs(y_hh)-reweighted}, we observe an enhancement due to finite-$m_t$ corrections in the large-rapidity region, which is mainly populated by events with small $m_{hh}$. As the rapidity decreases, this enhancement gradually turns into a suppression, with the transition occurring around $|y_{hh}|\simeq 1$. A similar, though much weaker, behavior is observed in the $|y_{h_{1}}|$ and $|y_{h_{2}}|$ distributions shown in figures~\ref{fig:abs(y_h1st)-reweighted} and \ref{fig:abs(y_h2nd)-reweighted}, respectively. On the other hand, for the rapidity-gap distribution $|y_{h_{1}}-y_{h_{2}}|$ shown in figure~\ref{fig:abs(y_h1st-y_h2nd)-reweighted}, a pronounced suppression is observed at large rapidity gaps, while an enhancement appears at small rapidity separations. This behavior can be understood as follows. At $p_{T,hh}=0$, where the bulk of the cross section resides, the Higgs-pair invariant mass $m_{hh}$ is closely related to the rapidity gap $|y_{h_{1}}-y_{h_{2}}|$ through
\begin{equation}
m_{hh}=2\sqrt{m_h^2+p_{T,h}^2}\cosh{\left(\frac{|y_{h_{1}}-y_{h_{2}}|}{2}\right)}\,,
\end{equation}
and therefore encodes similar kinematic information.

Overall, given the small scale uncertainties of the N$^3$LO$\otimes$NLO$_{m_t}$ predictions, the red bands in figures~\ref{fig:HH-reweighted-1} and \ref{fig:HH-reweighted-2} lie almost entirely outside the NLO$_{m_t}$ green bands, while remaining well within the NNLO$\otimes$NLO$_{m_t}$ blue bands. The only minor exception is the leading-Higgs $p_T$ distribution shown in figure~\ref{fig:pt_h1st-reweighted}. Other assumptions to include finite top-quark mass effects defined in eq.\eqref{eq:mtsigma} result in similar corrections to the shapes of the distributions.

\subsubsection{Assessment of other theoretical uncertainties}

In the discussion above on theoretical uncertainties, we have mainly focused on scale uncertainties, which are the dominant ones captured by our N$^3$LO calculations in the HTL. However, there are additional sources of theoretical uncertainties in modeling the di-Higgs (differential) cross sections. Besides the parametric uncertainties stemming from the PDFs, $\alpha_s$, $m_t$, and $m_h$, the main theoretical uncertainties are:
\begin{itemize}
\item \textbf{Missing finite-$m_t$ corrections at NNLO and beyond}: According to ref.~\cite{Grazzini:2018bsd}, this uncertainty can be estimated by comparing the NNLO$\otimes$NLO$_{m_t}$ results with the more advanced NNLO finite-top (FT) approximation. For the inclusive total cross section at $\sqrt{s}=14$ TeV, this procedure yields an uncertainty of about $5\%$. This uncertainty can be further reduced by combining the NNLO FT approximation with our N$^3$LO calculations in the HTL.
\item \textbf{Top-quark-mass scheme uncertainty}: NLO$_{m_t}$ calculations indicate that the remaining uncertainty due to the top-quark-mass scheme (OS versus $\overline{\mathrm{MS}}$) is quite significant~\cite{Baglio:2018lrj}, comparable to the scale uncertainties at NLO$_{m_t}$. A first attempt at understanding the top-quark-mass scheme uncertainty in the high-energy or Sudakov region for $gg\to hh$ has been reported recently in ref.~\cite{Jaskiewicz:2024xkd}. Using \ggxy, we show the NLO$_{m_t}$ results with different schemes in appendix~\ref{app:top-quark-mass-scheme}. Across most kinematic regions, the systematic uncertainty induced by different $m_t$ schemes is indeed significantly larger than the scale uncertainty at N$^3$LO. However, a partial study~\cite{Davies:2025ghl} suggests that if full-$m_t$-dependent calculations are extended to NNLO, the top-quark-mass scheme uncertainty can be significantly reduced.
\item \textbf{Missing higher-order EW corrections}: The complete NLO EW corrections with full-$m_t$ dependence are known from ref.~\cite{Bi:2023bnq}. At the total cross section level, the NLO EW corrections amount to approximately $-4\%$ of the LO prediction, and their impact is more pronounced in differential distributions. Our N$^3$LO calculations can be combined with NLO EW corrections in the same way as done here with NLO QCD corrections. Missing NNLO mixed QCD$\times$EW corrections can be estimated by comparing additive and multiplicative schemes for combining NLO QCD and NLO EW $K$ factors, which is valid in the non-Sudakov region. For the total cross section, this leads to an uncertainty below $2.8\%$ at LHC energies. Missing NNLO EW corrections, on the other hand, can be estimated by varying the EW renormalization scheme.
\item \textbf{Bottom-quark loop contributions}: Given the small bottom Yukawa coupling, the bottom-quark loop contribution to $gg\to hh$ induces only a few-percent-level correction in the low invariant-mass region (cf. table 1 of ref.~\cite{Hu:2025aeo}). The LO amplitude with full bottom-quark mass dependence can be readily computed using modern automated event generators such as \mgshort~\cite{Hirschi:2015iia}. Studies based on \mgshort\ indicate that bottom-quark loop contributions enhance the LO predictions obtained from top-quark loops alone by more than $5\%$ for $m_{hh}<300$ GeV at $\sqrt{s}=13$ TeV, mainly due to top-bottom interference effects, 
while remaining negligible (well below $1\%$) for the inclusive cross section. A first effort toward computing the corresponding two-loop bottom-quark amplitudes has appeared recently in ref.~\cite{Hu:2025aeo}.~\footnote{Recently, an NLO QCD calculation~\cite{Bonetti:2026cih} of the EW production of a Higgs-boson pair in the massless quark-antiquark channel ($q\bar{q}\to hh$) shows that this contribution can be as large as $10\%$ relative to the LO ggF channel for $m_{hh}<270$ GeV at LHC energies. This channel was previously thought to be negligible. From the mixed-coupling expansion point of view at the matrix-element level, the LO ($\mathcal{O}(\alpha^4)$) and NLO QCD ($\mathcal{O}(\alpha_s\alpha^4)$) contributions constitute subleading LO and NLO quantum corrections~\cite{Frederix:2018nkq,Shao:2025fwd} to the inclusive reaction $pp\to hh+X$. The $q\bar{q}$ channel is further suppressed by PDFs compared to the ggF channel. At NLO ($\mathcal{O}(\alpha_s\alpha^4)$), the real emission contributions, in particular those involving the $qg$ initial state with its enhanced parton luminosity, lead to sizable effects in the low-$m_{hh}$ region. Therefore, to accurately model this kinematic regime, the $q\bar{q}\to hh$ channel at NLO QCD should also be taken into account.}
\end{itemize} 

\section{Conclusions\label{sec:conclusion}}

In this paper, for the first time, we report a fully differential N$^3$LO calculation of the cross sections for Higgs-boson pair hadroproduction via gluon-fusion in the HTL. The N$^3$LO QCD corrections enhance the NNLO fiducial cross sections by about $4\%$. The scale uncertainties in the phase-space-integrated fiducial cross sections are reduced by roughly a factor of three at N$^3$LO compared to NNLO. In differential distributions, the N$^3$LO results generally lie within the NNLO scale-uncertainty bands. However, there are N$^3$LO QCD corrections that modify the shapes of certain distributions, such as the leading- and subleading-Higgs transverse momentum spectra and the azimuthal-angle difference between the two Higgs bosons. This highlights the necessity of including N$^3$LO QCD corrections in precise predictions of Higgs-boson pair production at the LHC. 

To account for top-quark-mass effects, we investigate three schemes of combining our HTL results with the full top-quark-mass-dependent results at NLO QCD accuracy, computed using the \ggxy\ generator~\cite{Davies:2025qjr}. Some differences between the fiducial and inclusive~\cite{Chen:2019fhs} cases are observed. The differential distributions in the $\mathrm{N}^k \mathrm{LO} \otimes \mathrm{NLO}_{m_t}$ scheme are analyzed. Finite-top-quark-mass effects are generally indispensable and should be included at the highest available order. Finally, we comment on four remaining intrinsic theoretical uncertainties in Higgs-boson pair production, in addition to the scale uncertainties.

\section*{Acknowledgements} 
The authors gratefully acknowledge the valuable discussions and insights provided by Alexander Huss, Matteo Marcoli, Kay Schönwald, Dingyu Shao, and the members of the Collaboration of Precision Testing and New Physics.
This work is supported in part by the National Natural Science Foundation of China under grants No.12275156, No.12275157, No.12321005, No.12375076, and No.12475085. 
The work of HSS is supported by the European Research Council (grant No.101041109, ``BOSON'') and the French National Research Agency  (grant ANR-20-CE31-0015, ``PrecisOnium''). Views and opinions expressed are however those of the authors only and do not necessarily reflect those of the European Union or the European Research Council Executive Agency. Neither the European Union nor the granting authority can be held responsible for them.

%For the purpose of Open Access, a CC-BY public copyright license has been applied by the authors to the present document and will be applied to all subsequent versions up to the Author Accepted Manuscript arising from this submission.

%\newpage

\section*{Appendix}
%{\color{red} delete appendix}
\appendix
\section{NLO$_{m_{t}}$ results with different top-quark-mass schemes}
\label{app:top-quark-mass-scheme}
\begin{figure}[hbt!]
%  \centering
  \begin{subfigure}[b]{0.5\textwidth}
    \includegraphics[width=\textwidth]{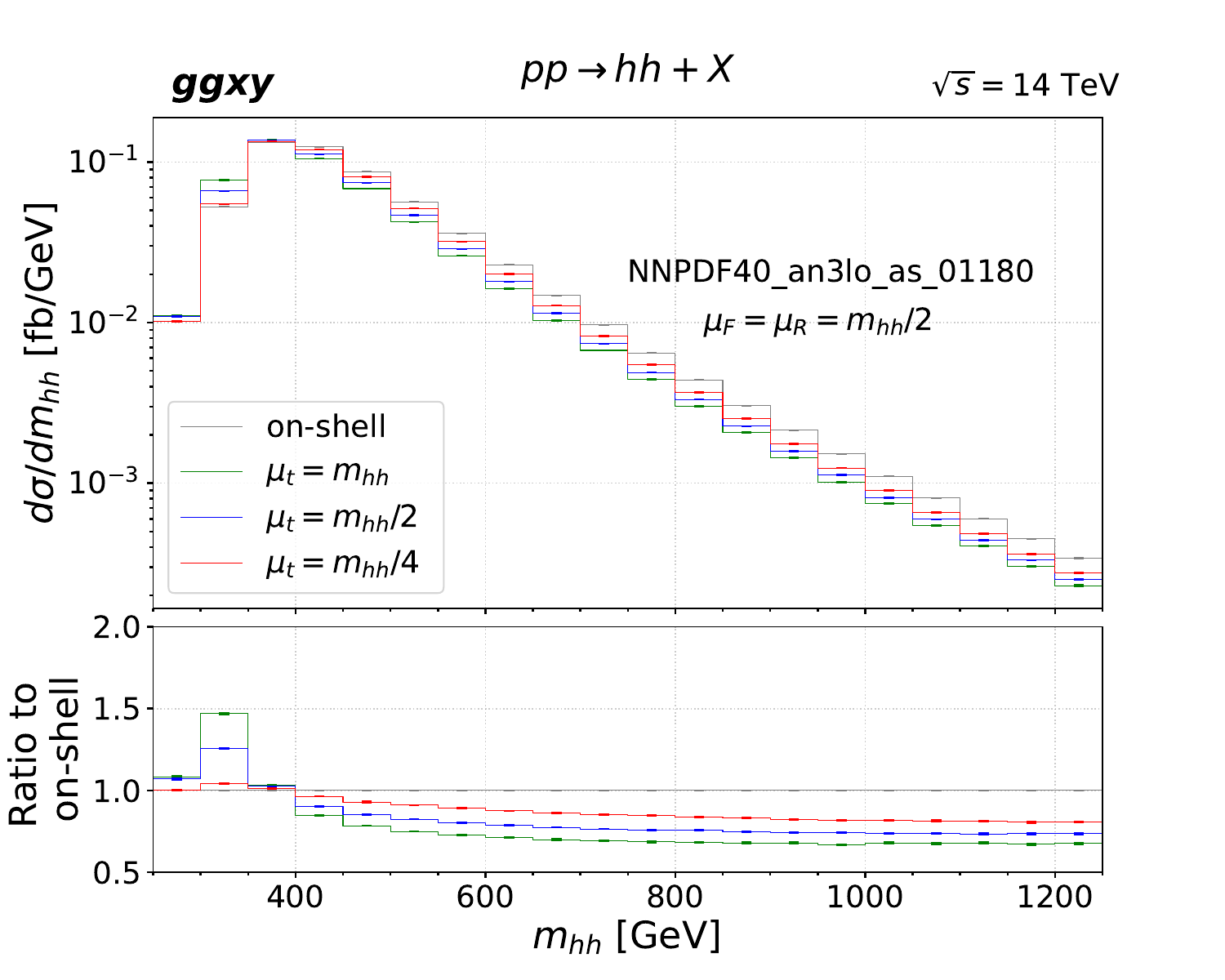}
    \caption{}
    \label{fig:m_hh-compare}
  \end{subfigure}%\hspace{-0.8cm}
  \begin{subfigure}[b]{0.5\textwidth}
    \includegraphics[width=\textwidth]{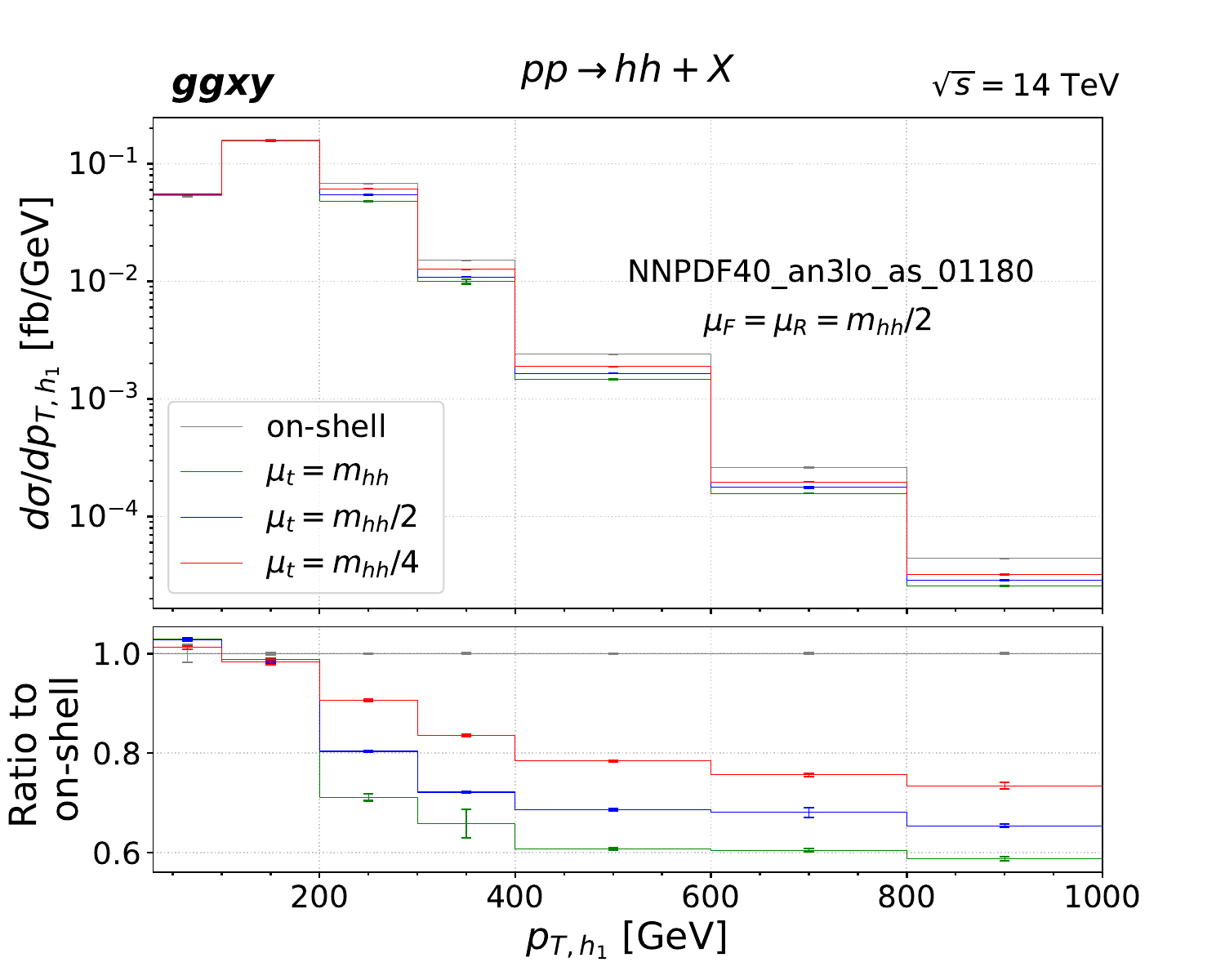}
    \caption{}
    \label{fig:pt_h1st-compare}
  \end{subfigure}%\hspace{-0.8cm}
  \\
  \begin{subfigure}[b]{0.5\textwidth}
    \includegraphics[width=\textwidth]{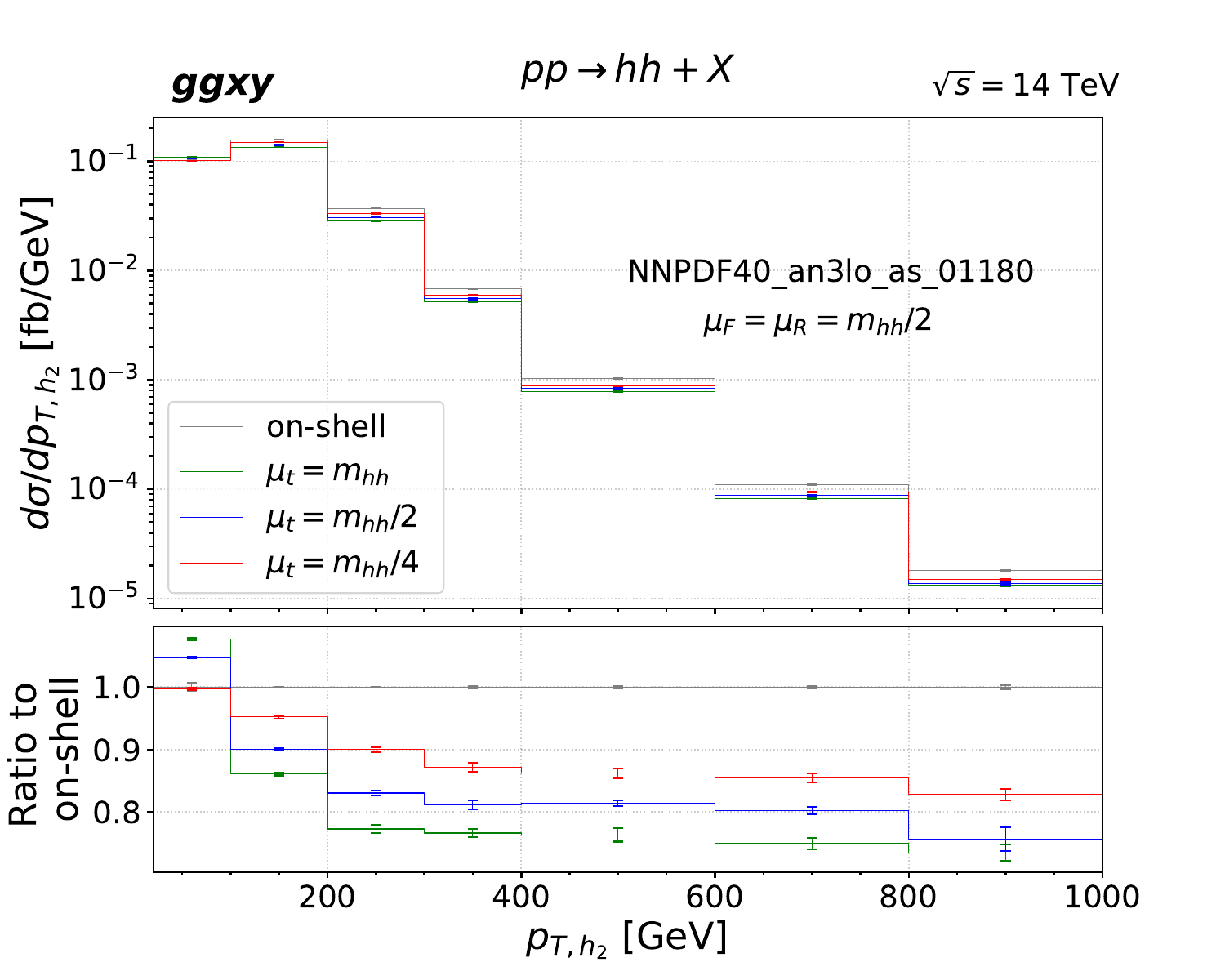}
    \caption{}
    \label{fig:pt_h2nd-compare}
  \end{subfigure}%\hspace{-0.8cm}
  \begin{subfigure}[b]{0.5\textwidth}
    \includegraphics[width=\textwidth]{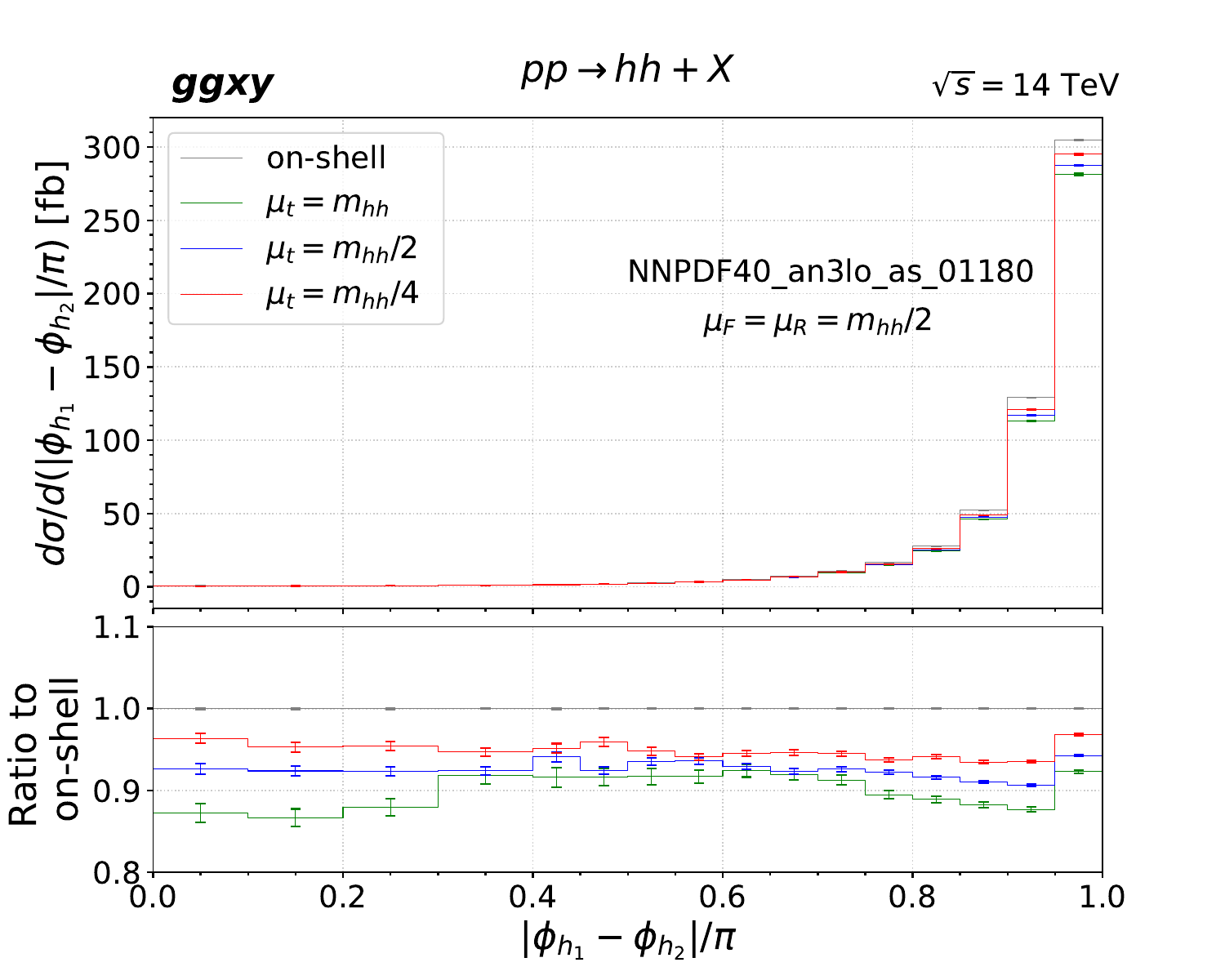}
    \caption{}
    \label{fig:abs(phi_h1st-phi_h2nd)_over_pi-compare}
  \end{subfigure}
  \caption{Differential distributions for Higgs-boson pair production at ${\rm NLO}_{m_{t}}$ in different top-quark-mass schemes: 
(a) Invariant mass of the Higgs pair, $m_{hh}$;
(b) Transverse momentum of the leading-$p_{T}$ Higgs, $p_{T,h_1}$; 
(c) Transverse momentum of the subleading-$p_{T}$ Higgs, $p_{T,h_2}$;
(d) Azimuthal-angle difference between the Higgs bosons, $|\phi_{h_1}-\phi_{h_2}|/\pi$.
The bottom panels show the ratios with respect to the OS scheme result.}
  \label{fig:HH-compare-1}
\end{figure}

\begin{figure}[hbt!]
%  \centering
  \begin{subfigure}[b]{0.5\textwidth}
    \includegraphics[width=\textwidth]{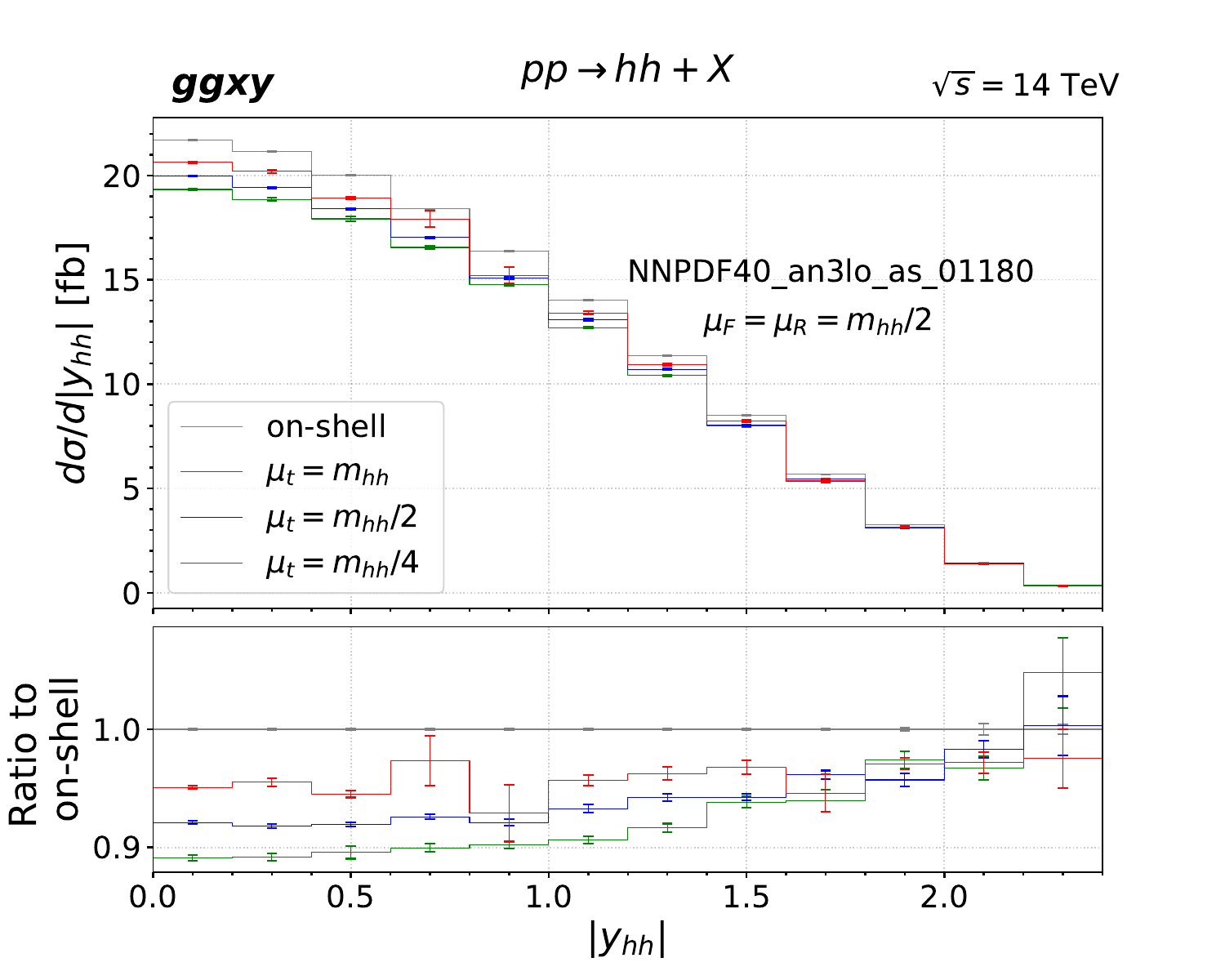}
    \caption{}
    \label{fig:abs(y_hh)-compare}
  \end{subfigure}%\hspace{-0.8cm}
  \begin{subfigure}[b]{0.5\textwidth}
    \includegraphics[width=\textwidth]{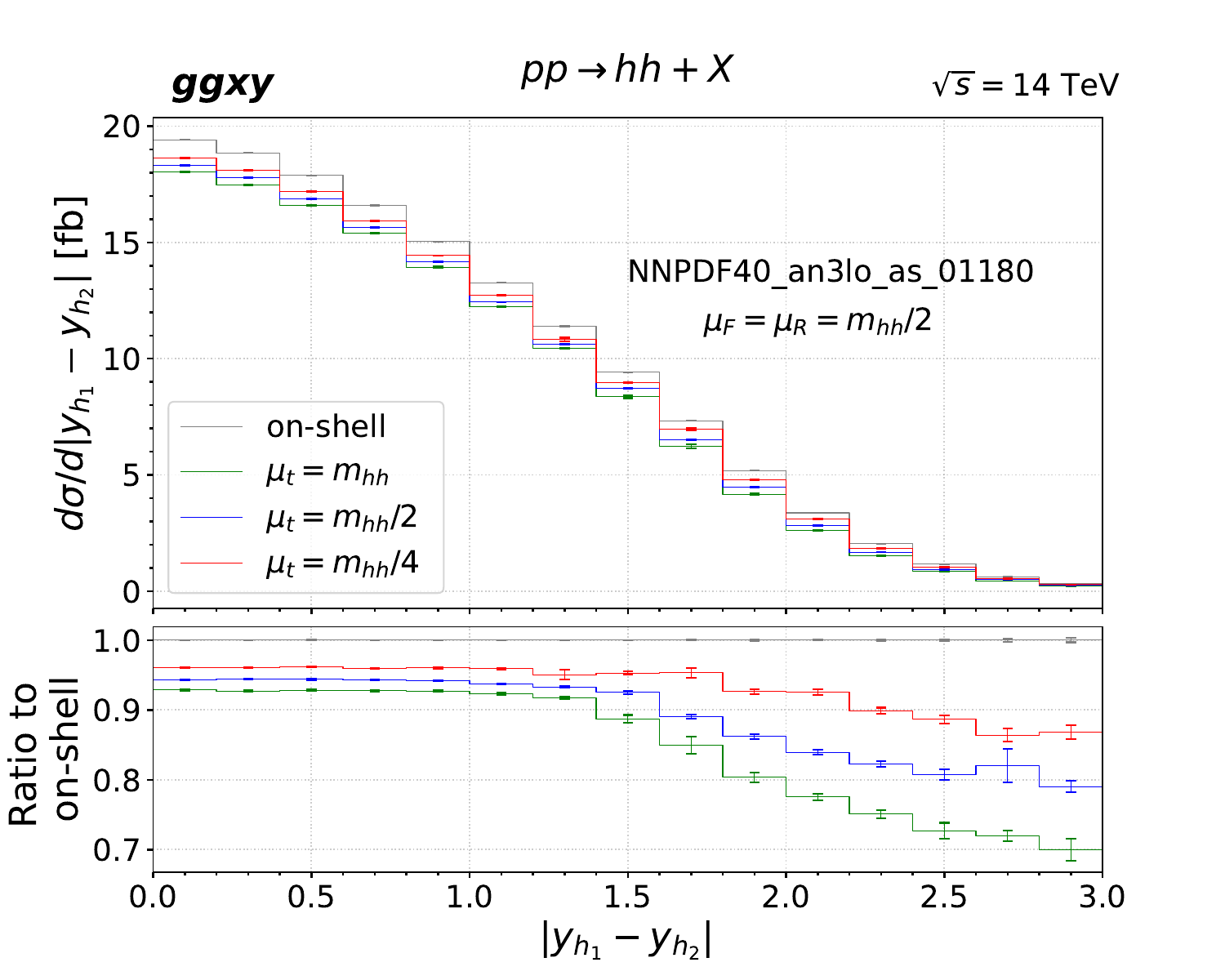}
    \caption{}
    \label{fig:abs(y_h1st-y_h2nd)-compare}
  \end{subfigure}
  \\
  \begin{subfigure}[b]{0.5\textwidth}
    \includegraphics[width=\textwidth]{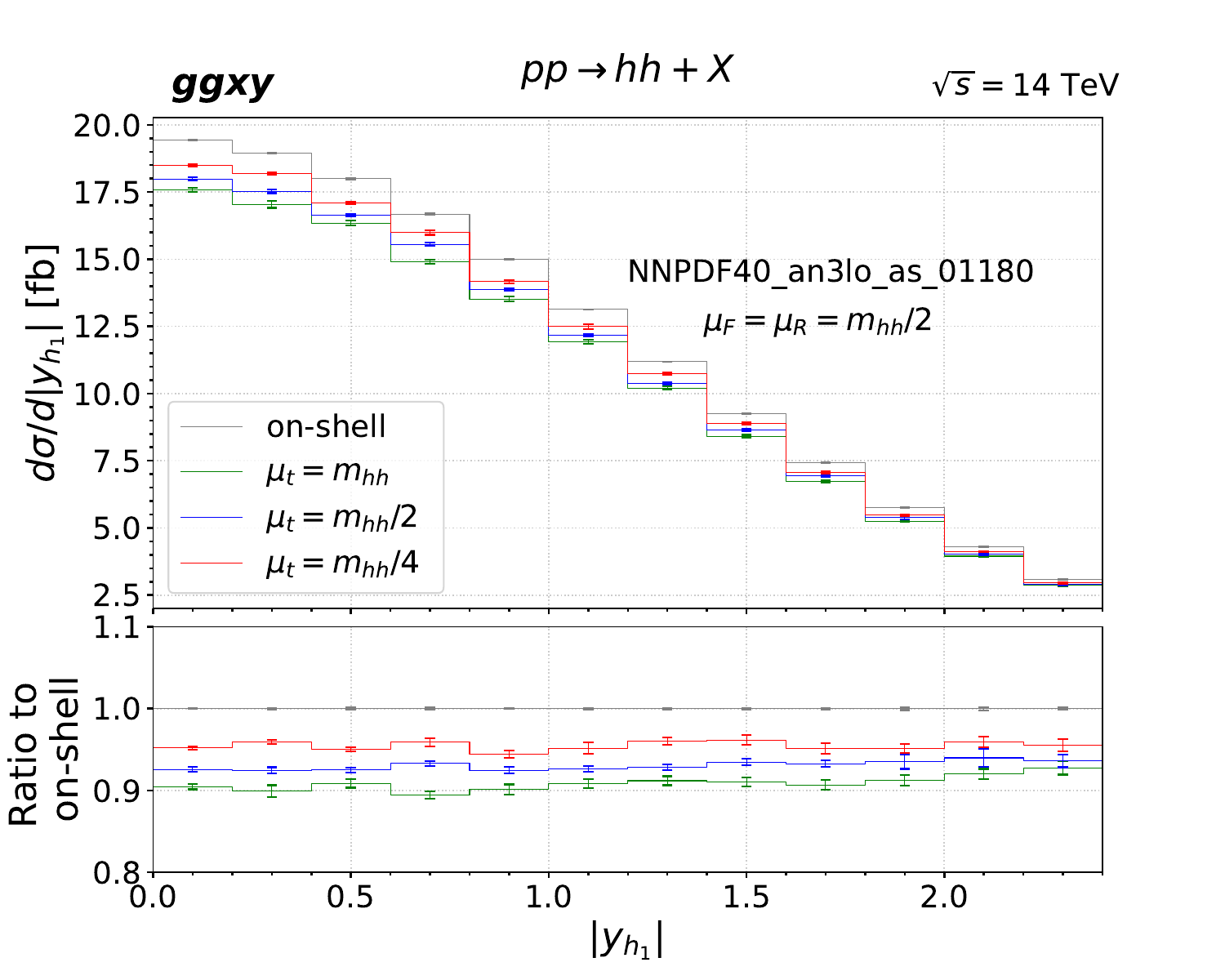}
    \caption{}
    \label{fig:abs(y_h1st)-compare}
  \end{subfigure}%\hspace{-0.8cm}
  \begin{subfigure}[b]{0.5\textwidth}
    \includegraphics[width=\textwidth]{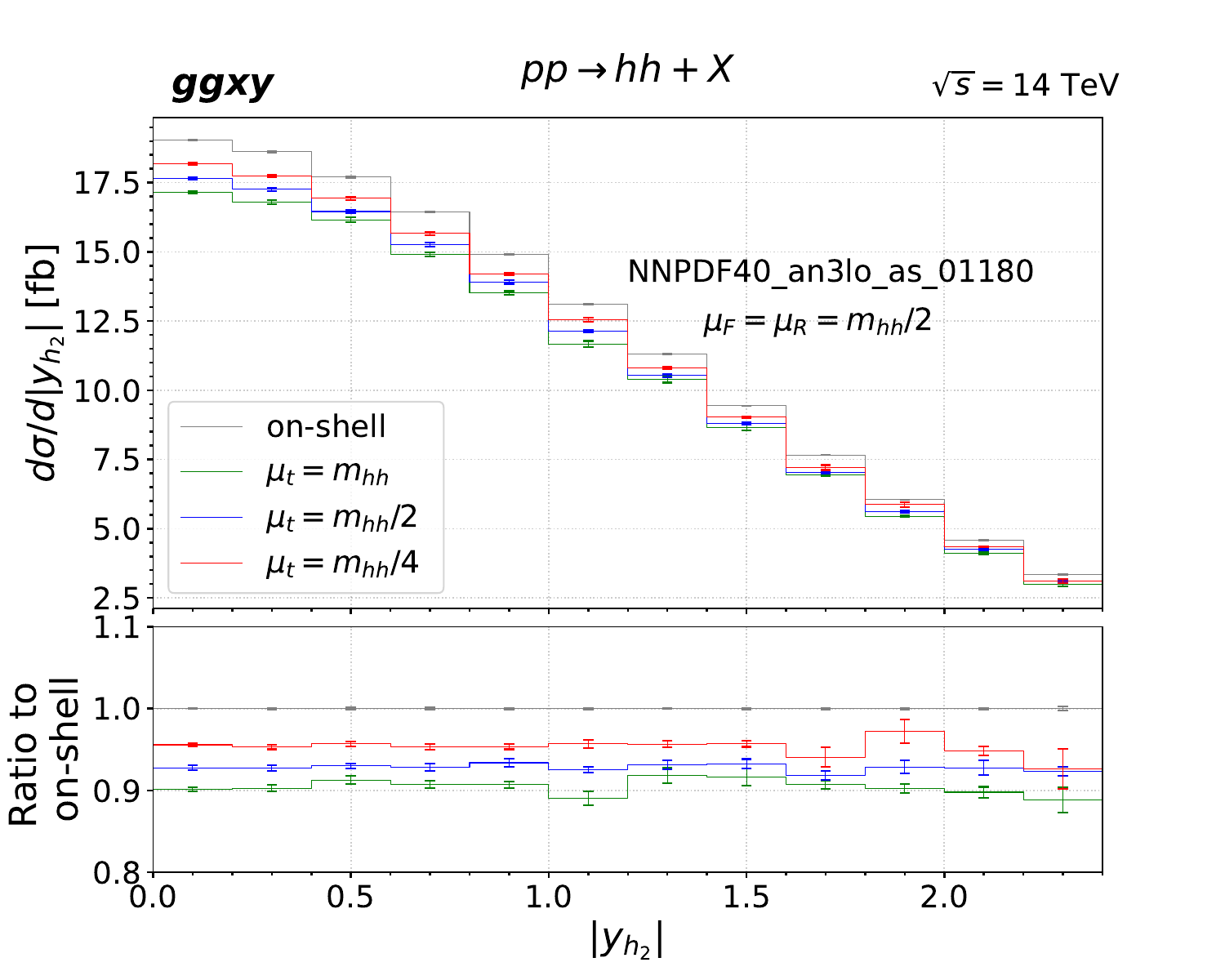}
    \caption{}
    \label{fig:abs(y_h2nd)-compare}
  \end{subfigure}
  \caption{Differential distributions for Higgs-boson pair production at ${\rm NLO}_{m_{t}}$ in different top-quark-mass schemes: 
(a) Absolute rapidity of the Higgs-pair system, $|y_{hh}|$; 
(b) Absolute rapidity gap between the Higgs bosons, $|y_{h_1}-y_{h_2}|$; 
(c) Absolute rapidity of the leading-$p_{T}$ Higgs, $|y_{h_1}|$;
(d) Absolute rapidity of the subleading-$p_{T}$ Higgs, $|y_{h_2}|$. 
The bottom panels show the ratios with respect to the OS scheme result.}
  \label{fig:HH-compare-2}
\end{figure}

Figures~\ref{fig:HH-compare-1} and \ref{fig:HH-compare-2} show the NLO distributions with full top-quark mass dependence computed in four top-quark-mass schemes: OS (black lines), $\overline{\rm MS}~(\mu_{t}=m_{hh})$ (green lines), $\overline{\rm MS}~(\mu_{t}=m_{hh}/2)$ (blue lines), and $\overline{\rm MS}~(\mu_{t}=m_{hh}/4)$ (red lines) using \ggxy~\cite{Davies:2025qjr}, where $\mu_t$ denotes the renormalization scale of the $\overline{\rm MS}$ top-quark mass. 

We find that the uncertainty associated with the choice of the top-quark mass scheme (OS and $\overline{\rm MS}$ variants) can reach up to $40\%$ when comparing the four mass schemes considered, 
with the exception of the second bin of the $m_{hh}$ distribution, where it increases to $47\%$. This is a well-known effect near the top-quark-pair production threshold, where the system should be treated non-relativistically and the $\overline{\rm MS}$ mass scheme is not appropriate. The findings of ref.~\cite{Davies:2025ghl} suggest to adopt the $\overline{\rm MS}$ scheme to estimate this uncertainty, since in this scheme the uncertainty band of the higher-order predictions overlaps with that of the lower-order ones, as illustrated in figure~7 of ref.~\cite{Davies:2025ghl}. Restricting to the $\overline{\rm MS}$ scheme with three choices of $\mu_t$, we further find that the corresponding uncertainty reaches up to $20\%$, again except for the second bin of the $m_{hh}$ distribution, where it rises to $43\%$.

\newpage
\bibliographystyle{JHEP}
\bibliography{bib}

%\clearpage

%\onecolumngrid

%\setcounter{equation}{0}
%\setcounter{figure}{0}

%\section*{Supplemental material}
%\section*{Appendix}
%\appendix
%\counterwithin{figure}{section}

\end{document}